\documentclass[twocolumn]{aastex631}

\usepackage{amsmath,bm}

\DeclareMathOperator\arctanh{arctanh}

\newcommand{\beq}{\begin{equation}}           % begin equation
\newcommand{\eeq}{\end{equation}}             % end equation
\newcommand{\bfi}{\begin{figure}}           % begin figure
\newcommand{\efi}{\end{figure}}             % end figure

\defcitealias{karamazov_etal21}{Paper~I}

\shortauthors{Karamazov \& Heyrovsk\'y}

\begin{document}

\title{Gravitational Lensing By a Massive Object in a Dark Matter Halo.\\ II. Shear, Phase, and Image Geometry}
	
\author[0000-0002-7919-499X]{Michal Karamazov}
\affiliation{Institute of Theoretical Physics, Faculty of Mathematics and Physics, Charles University,\\ V Hole\v{s}ovi\v{c}k\'ach 2, 18000~Praha 8, Czech Republic}
\author[0000-0002-5198-5343]{David Heyrovsk\'y}
\affiliation{Institute of Theoretical Physics, Faculty of Mathematics and Physics, Charles University,\\ V Hole\v{s}ovi\v{c}k\'ach 2, 18000~Praha 8, Czech Republic}
\email{michal.karamazov@gmail.com, david.heyrovsky@mff.cuni.cz}

\begin{abstract}

We study the gravitational lensing influence of a massive object in a dark matter halo, using a simple model of a point mass embedded in a spherical Navarro--Frenk--White halo. Building on the analysis of critical curves and caustics presented in the first part of this work, we proceed to explore the geometry of images formed by the lens. First, we analyze several lensing quantities including shear, phase, and their weak-lensing approximations, illustrating the results with image-plane maps. We derive formulae and present a geometric interpretation for the shear and phase of a combination of two axially symmetric mass distributions. In the case of our lens model, we describe the occurrence of zero-shear points and specify the conditions under which they become umbilic points. Second, we use the eigenvalue decomposition of the inverse of the lens-equation Jacobian matrix to compute the magnification and flattening of lensed images. Based on this, we introduce the convergence--shear diagram, a novel and compact way of visualizing the properties of images formed by a particular gravitational lens. We inspect relative deviations of the analyzed lensing quantities in order to evaluate the perturbing effect of the point mass and the applicability of the weak-lensing approximation. We explore the dependence of the results on the point-mass parameters by studying grids of plots for different combinations of its position and mass. We provide analytical explanations for important patterns arising in these plots and discuss the implications for the lensing influence of isolated compact bodies in dark matter halos.

\end{abstract}

\section{Introduction}
\label{sec:Intro}

Dark matter halos form the basic building blocks in the bottom-up structure formation of $\Lambda$CDM cosmology. They constitute the dominant matter component of the astrophysical objects they are associated with: the largest halos with galaxy clusters; smaller halos with individual galaxies and dwarf galaxies. Their properties can be studied by ``observations'' in the virtual universes arising from large-scale-structure formation simulations  \citep[e.g.,][]{zandanel_etal18}. In our universe, main observational constraints on galaxy-cluster dark matter halos come from the study of the kinematics of cluster galaxies \citep[starting from][]{zwicky33}, from measurements of X-ray emission from intracluster baryonic gas \citep[e.g.,][]{ettori_etal13}, and from analyses of weak and strong gravitational lensing of background galaxies \citep[e.g.,][]{limousin_etal07,okabe_etal13}. Gravitational lensing analyses are particularly useful as tools for studying the finer-scale substructure of cluster halos, such as subhalos of individual cluster galaxies, local clumps or other inhomogeneities. A detailed analysis of 11 galaxy clusters by \cite{meneghetti_etal20} revealed a surprisingly high efficiency of substructure lensing, more than an order of magnitude higher than expected from CDM simulations. This result indicates the need for a better understanding of the lensing effects of individual bodies within the cluster.

The goal of our work is to study the gravitational lensing influence of a compact massive body in a dark matter halo. For this purpose, we use a simple model consisting of a point mass embedded in a spherical Navarro--Frenk--White (NFW) density profile \citep{navarro_etal96}. In the first part of this work \citep[][hereafter \citetalias{karamazov_etal21}]{karamazov_etal21}, we studied the critical curves and caustics of the lens model as a function of the mass and position of the point mass. We discovered that the model exhibited a rich diversity of critical-curve topologies and caustic geometries. We mapped the boundaries separating the corresponding lensing regimes in the point-mass parameter space and identified the accompanying caustic metamorphoses. Among other findings, we demonstrated the existence of a critical value of the mass parameter. For centrally positioned lighter (sub-critical) point masses, the lens has two radial critical curves. Heavier (super-critical) point masses are strong enough to fully eliminate the radial critical curves. For critical point masses the lens has a single radial critical curve with peculiar properties, which are described in Appendix~B of \citetalias{karamazov_etal21}. We discussed the relevance of the model to the lensing by galaxies in galaxy-cluster halos as well as other astrophysical scenarios, such as the lensing influence of a satellite galaxy or a (super-)massive black hole in a galactic dark matter halo.

In this sequel to \citetalias{karamazov_etal21}, we explore other lensing properties of the model. Here we concentrate on the shear and phase and their relation to the geometric distortions of images formed by the lens. In the weak-lensing regime the relation is tight, with the shear specifying the semi-axis ratio and the phase specifying the orientation of the major axis of the image. However, this will not be the case in the regions with high convergence (near the halo center) or high shear (near the point mass and near the halo center). In Section~\ref{sec:NFW} we describe the shear, phase, and image geometry for a NFW-halo-only lens. We study the images in Section~\ref{sec:NFW-images}, starting from the eigenvalue decomposition of the inverse of the Jacobian matrix and utilizing the convergence--shear diagram, a new tool described in detail in Appendix~\ref{sec:Appendix-images}. In Section~\ref{sec:NFW-weak} we introduce the weak shear and weak phase and compare these weak-lensing estimates with the shear and phase.

In Section~\ref{sec:NFWP} we proceed with the analysis of the NFW halo + point-mass lens in a similar manner. More specifically, in Section~\ref{sec:NFWP-csp} we derive formulae for the shear and phase of the combined mass distribution. We describe the emergence and occurrence of points with zero shear, which may constitute umbilic points under conditions discussed in Section~\ref{sec:NFWP-Jacobian}. In Section~\ref{sec:NFWP-images} we study the properties of images using the convergence--shear diagram. We present the main results in Section~\ref{sec:NFWP-plots} in the form of grids of image-plane maps of different lens characteristics and convergence--shear diagrams, utilizing the same point-mass parameter grid as in \citetalias{karamazov_etal21}. We discuss the results and their broader relevance in Section~\ref{sec:discussion} and summarize our findings in Section~\ref{sec:summary}.

\section{Lensing by a NFW Halo}
\label{sec:NFW}

\subsection{Convergence, Shear, and Phase}
\label{sec:NFW-csp}

The surface density of a halo with a spherical NFW profile expressed in units of the critical surface density $\Sigma_{\text{cr}}$ yields the dimensionless convergence profile (\citealt{bartelmann96}; \citetalias{karamazov_etal21}),
\beq
\kappa_{\text{\tiny NFW}}(x)=2\,\kappa_{\text{s}}\;\frac{1-\mathcal{F}(x)}{x^2-1}\,,
\label{eq:NFW_kappa}
\eeq
where $x$ is the plane-of-the-sky distance from the halo center expressed in units of the halo scale radius $r_{\text{s}}$, and $\kappa_{\text{s}}$ is the halo convergence parameter. The function
\beq
\mathcal{F}(x)=\begin{cases}
\frac{\displaystyle\arctanh{\sqrt{1-x^2}}}{\displaystyle\sqrt{1-x^2}} & \text{for $x<1$}\,,\\
\hfil 1 & \text{for $x=1$}\,,\\
\frac{\displaystyle\arctan{\sqrt{x^2-1}}}{\displaystyle\sqrt{x^2-1}} & \text{for $x>1$}\,,
\end{cases}
\label{eq:f(x)}
\eeq
has a similar radial behavior to the convergence $\kappa_{\text{\tiny NFW}}(x)$: both decrease monotonically from $\infty$ at the halo center to $0$ for $x\gg 1$ (see \citetalias{karamazov_etal21} for details). The radius $x_0$ at which the convergence is equal to $1$ can be computed numerically from
\beq
\frac{1-x_0^2}{\mathcal{F}(x_0)-1}=2\,\kappa_{\text{s}}\,.
\label{eq:unit-convergence_radius}
\eeq
This unit-convergence radius and the circle that it defines play a key role when studying the geometry of images formed by a lens. For the NFW halo $x_0$ increases monotonically with the convergence parameter $\kappa_{\text{s}}$, as illustrated in \citetalias{karamazov_etal21}. We note here that for $\kappa_{\text{s}}=3/2$ the unit-convergence radius is equal to the scale radius, $x_0=1$. For lower values of $\kappa_{\text{s}}$ the unit-convergence circle lies inside the scale-radius circle; for higher values outside.

A light ray passing through the halo at a position $\boldsymbol x$ in the plane of the sky is deflected by an angle
\beq
\boldsymbol \alpha(\boldsymbol x)=\frac{4\,\kappa_{\text{s}}\,r_{\text{s}}\,D_{\text{s}}}{D_{\text{l}}\,D_{\text{ls}}}\, \left[\ln{\frac{x}{2}}+\mathcal{F}(x)\right]\,\frac{\boldsymbol x}{x^2}\,,
\label{eq:NFW_deflection}
\eeq
where $D_{\text{l}}$, $D_{\text{s}}$, and $D_{\text{ls}}$ are the angular-diameter distances from the observer to the lens, from the observer to the source, and from the lens to the source, respectively. Expressed in units of the angular scale radius, the position of a source $\boldsymbol y$ and the position of its image $\boldsymbol x$ formed by the gravitational field of the NFW halo are connected by the lens equation
\beq
\boldsymbol y=\boldsymbol x - 4\,\kappa_{\text{s}}\,
\left[\ln{\frac{x}{2}}+\mathcal{F}(x)\right]\,\frac{\boldsymbol x}{x^2}\,.
\label{eq:NFW_lens_equation}
\eeq

For illustration, in the top row of Figure~\ref{fig:images} we show the lensing of a circular source by a NFW halo with convergence parameter $\kappa_{\text{s}}\approx 0.239035$, the fiducial value used in \citetalias{karamazov_etal21}. As seen in the top left panel, in this example the black circular source centered at \mbox{$\bm{y_{\text{c}}}=(0.015,\,-0.005)$} with radius $y_{\text{r}}=0.005$ lies inside the radial caustic without overlapping the central point-like tangential caustic. Solving the lens Equation~(\ref{eq:NFW_lens_equation}) numerically for a point $\boldsymbol y$ on the circumference of the source yields three points $\boldsymbol x$ on the boundaries of the three black images shown in the top right panel. One image lies outside the tangential critical curve, a second image lies between the tangential and radial critical curves, and the smallest third image lies inside the radial critical curve. The dashed circle with radius $x_0\approx0.0936$ is the unit-convergence circle of this halo. In this case, the first two images are elongated in the tangential (azimuthal) direction, while the third image is elongated in the radial direction.

\bfi
{\centering
\vspace{0cm}
\hspace{0cm}
\includegraphics[width=8.5 cm]{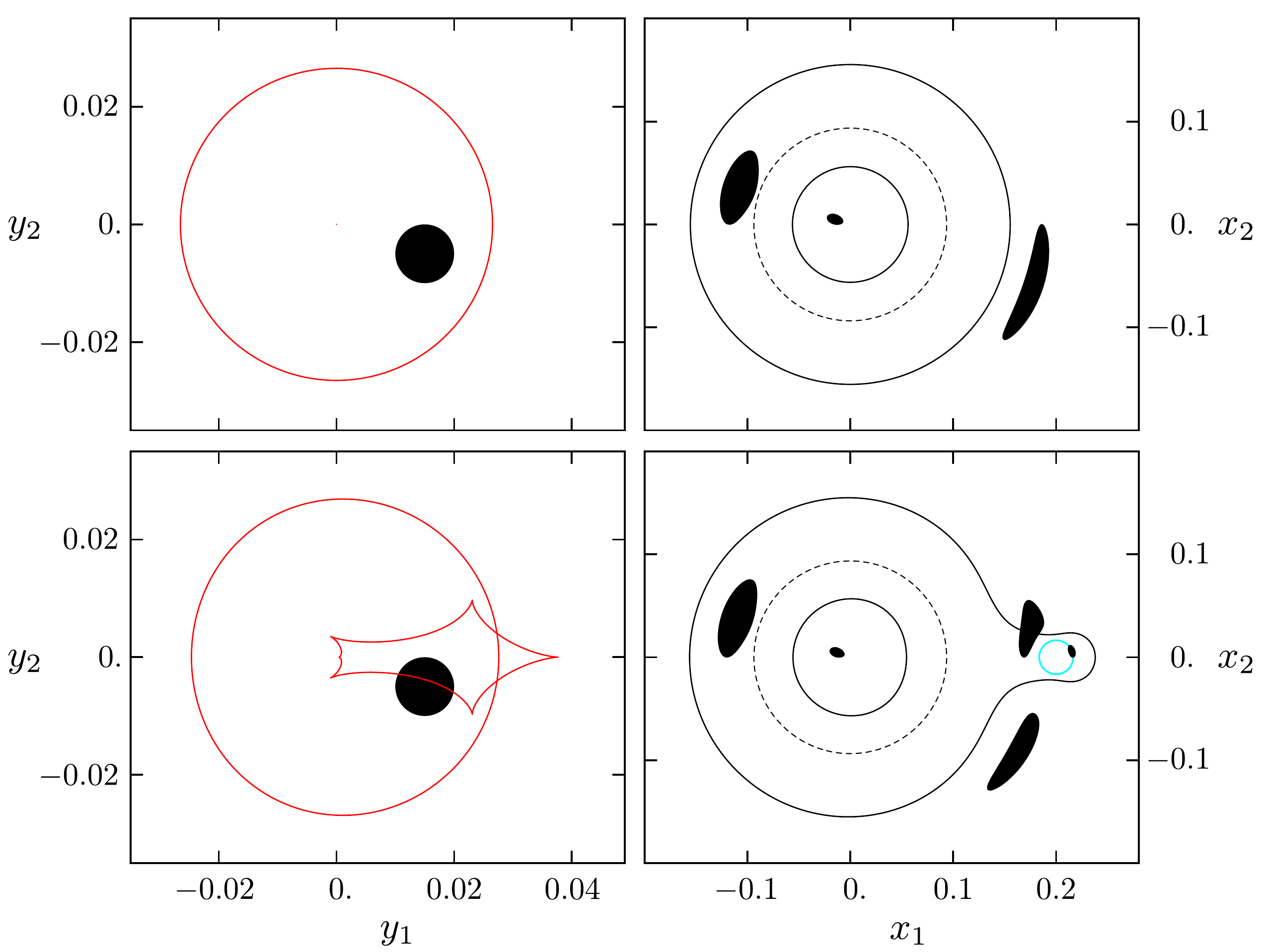}
\caption{Gravitational lensing of a circular source. Top row: lensing by a NFW halo with convergence parameter $\kappa_{\text{s}}\approx 0.239035$. Bottom row: lensing by the same halo with an additional point mass with mass parameter $\kappa_{\text{P}}\approx 2.714\cdot 10^{-4}$ positioned at $\bm{x_{\text{P}}}=(0.2,\,0)$. Left column: source-plane plots indicating the position of the source (black circle) with respect to the lens caustic (red lines). Right column: image-plane plots indicating the positions of images (black patches) with respect to the critical curve (solid black lines). The cyan circle in the bottom right panel marks the Einstein circle of the point mass; the dashed black lines in the right panels mark the unit-convergence circle.\label{fig:images}}}
\efi

For a source lying inside the radial caustic and overlapping the central tangential caustic, the first two images would merge along the tangential critical curve, forming an Einstein ring. For a smaller source positioned close to the inner side of the radial caustic, the second image would lie inside the unit-convergence circle and be elongated in the radial direction. For a source positioned on the radial caustic, the second and third images would merge at the radial critical curve. For a source lying outside the radial caustic, these two images would vanish, leaving only the first image.

The deformations and orientations of the images are best studied by computing the lens shear and its phase, quantities that may be introduced by means of the lens potential. The deflection angle can be written in terms of the gradient of the lens potential $\psi(x)$, which in this case is circularly symmetric,
\beq
\boldsymbol \alpha(\boldsymbol x)=\frac{D_{\text{s}}}{D_{\text{ls}}}\,\nabla_{\boldsymbol \theta}\,\psi(x)=\frac{D_{\text{s}}\,D_{\text{l}}}{D_{\text{ls}}\,r_{\text{s}}}\,\frac{\boldsymbol x}{x}\, \frac{{\rm d}\psi}{{\rm d} x}\,,
\label{eq:NFW_gradient}
\eeq
where we converted the angular position in radians $\boldsymbol \theta$ to the angular position in scale-radius units, $\boldsymbol x = \boldsymbol \theta \, D_{\text{l}}/r_{\text{s}}$. By substituting for the deflection angle from Equation~(\ref{eq:NFW_deflection}) we can express the lens-potential derivative
\beq
\frac{{\rm d}\psi_{\text{\tiny NFW}}}{{\rm d} x}= \frac{4\,\kappa_{\text{s}}\,r_{\text{s}}^2}{D_{\text{l}}^2}\,\frac{1}{x}\, \left[\ln{\frac{x}{2}}+\mathcal{F}(x)\right]\,.
\label{eq:NFW_potential,x}
\eeq
Integration yields the following expressions for the NFW halo lens potential \citep{meneghetti_etal03,golse_kneib02}:
\beq
\psi_{\text{\tiny NFW}}(x)=\begin{cases}
\frac{2\,\kappa_{\text{s}}\,r_{\text{s}}^2}{D_{\text{l}}^2}\left[\ln{x}\ln{\frac{x}{4}}- (\displaystyle\arctanh{\sqrt{1-x^2}}\,)^2\right]\\
\hfil 0\\
\frac{2\,\kappa_{\text{s}}\,r_{\text{s}}^2}{D_{\text{l}}^2}\left[\ln{x}\ln{\frac{x}{4}}+ (\displaystyle\arctan{\sqrt{x^2-1}}\,)^2\right]\,,
\end{cases}
\label{eq:NFW_psi}
\eeq
where the expression in the first row holds for $x<1$, in the second row for $x=1$, and in the third row for $x>1$. For $x\ll 1$, the potential close to the halo center
\beq
\psi_{\text{\tiny NFW}}(x)= -\,\frac{2\,\kappa_{\text{s}}\,r_{\text{s}}^2}{D_{\text{l}}^2} \left(\ln^2{2}+\frac{x^2}{2}\,\ln{\frac{x}{2}}\right) +\mathcal{O}(x^4\,\ln{x})\,,
\label{eq:NFW_psi_origin}
\eeq
starting from a finite negative value and increasing monotonically outward, crossing zero at the scale radius. Note that the expressions for the potential in \cite{meneghetti_etal03} and \cite{golse_kneib02} are higher by the constant $-\psi_{\text{\tiny NFW}}(0)$ and thus they start at zero.

\begin{figure*}
{\centering
\vspace{0cm}
\hspace{0cm}
\includegraphics[width=15.5 cm]{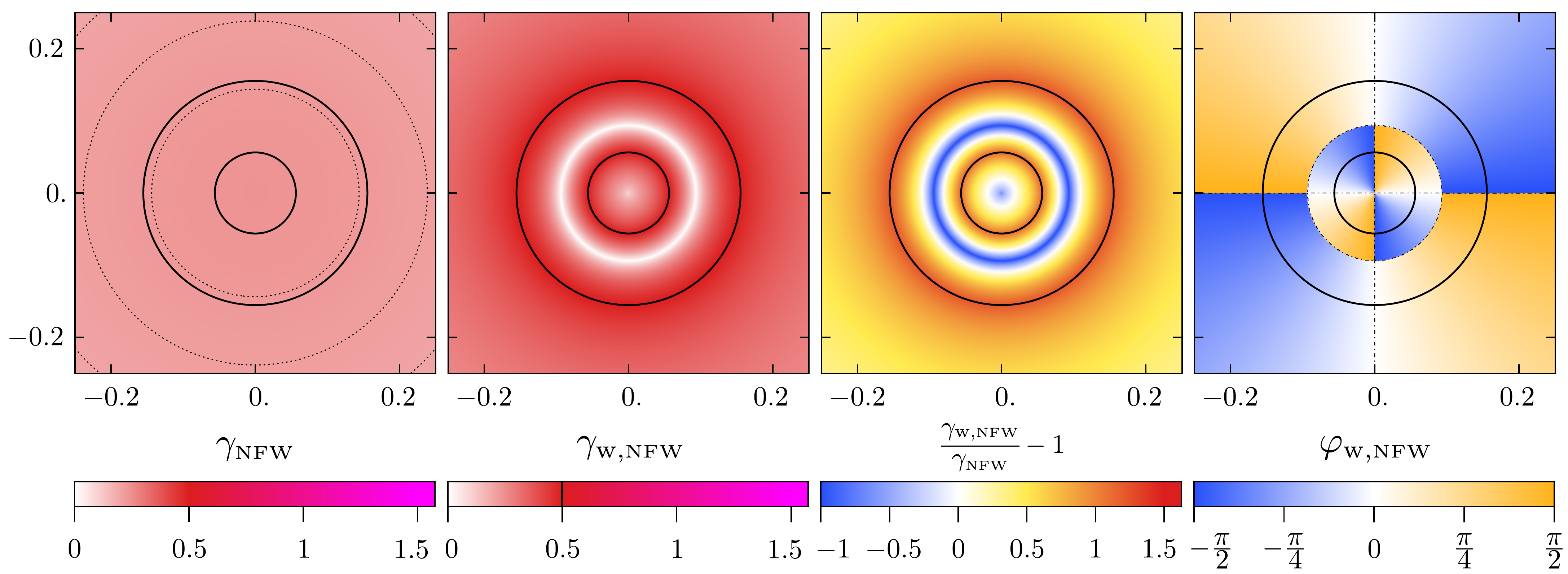}
\caption{Image-plane color maps of lensing characteristics of the NFW halo from the top row of Figure~\ref{fig:images}. For orientation, solid black circles mark the critical curves in all panels. First panel: shear $\gamma_{\text{\tiny NFW}}(\boldsymbol x)$ with dotted contours from inside to outside marking $95\%, 90\%,$ and $85\%$ of the central shear $\gamma_{\text{\tiny NFW}}(0)=\kappa_{\text{s}}\approx 0.239035$. Second panel: weak shear $\gamma_{\text{w,\tiny NFW}}(\boldsymbol x)$ computed from image flattening, using the same color bar as in the first panel. Third panel: relative weak-shear deviation from the shear, $\gamma_{\text{w,\tiny NFW}}(\boldsymbol x)/\gamma_{\text{\tiny NFW}}(\boldsymbol x)-1$. Fourth panel: weak phase $\varphi_{\text{w,\tiny{NFW}}}(\boldsymbol x)$ defined by image orientation, with orange corresponding to images oriented counterclockwise, and blue to images oriented clockwise from the horizontal. Dot-dashed lines mark the unit-convergence circle, and locations of horizontal (white background) and vertical images (high-saturation blue/orange boundary).\label{fig:NFW-line}}}
\end{figure*}

For a circularly symmetric lens potential the lens shear $\gamma$ can be computed as
\beq
\gamma = \frac{D_{\text{l}}^2}{2\,r_{\text{s}}^2} \left|\frac{{\rm d}^2\psi}{{\rm d} x^2}-\frac{1}{x}\frac{{\rm d}\psi}{{\rm d} x}\right|\,.
\label{eq:shear_sym}
\eeq
For the NFW profile the second derivative of the lens potential,
\begin{multline}
\frac{{\rm d}^2\psi_{\text{\tiny NFW}}}{{\rm d} x^2}= \frac{4\,\kappa_{\text{s}}\,r_{\text{s}}^2}{D_{\text{l}}^2}\\ \times \left[\frac{1}{x^2-1}+\frac{2x^2-1}{x^2(1-x^2)}\,\mathcal{F}(x)- \frac{1}{x^2}\ln{\frac{x}{2}}\right]\,,
\label{eq:NFW_potential,xx}
\end{multline}
can be used together with the first derivative from Equation~(\ref{eq:NFW_potential,x}) in Equation~(\ref{eq:shear_sym}) to yield the shear of the NFW halo,
\begin{multline}
\gamma_{\text{\tiny NFW}}(x)=2\,\kappa_{\text{s}}\left[\frac{2}{x^2}\ln{\frac{x}{2}} +\frac{1}{1-x^2}+ \frac{2-3x^2}{x^2(1-x^2)}\,\mathcal{F}(x)\right]\,,\\
\label{eq:NFW_gamma}
\end{multline}
as shown by \cite{wright_brainerd00}. In order to understand its behavior close to the origin, we expand Equation~(\ref{eq:NFW_gamma}) for $x\ll 1$ and obtain
\beq
\gamma_{\text{\tiny NFW}}(x)= \kappa_{\text{s}}\,\left[\,1+\frac{3}{2}\,x^2\,\ln{\frac{x}{2}}+\frac{13}{8}\,x^2\,\right] +\mathcal{O}(x^4\,\ln{x})\,.
\label{eq:NFW_gamma_origin}
\eeq
By setting $x=0$ we see that the NFW shear at the center is equal to the convergence parameter of the halo, $\gamma_{\text{\tiny NFW}}(0)=\kappa_{\text{s}}$. From this value the shear decreases outward monotonically. An expansion close to the scale radius shows that for $x\to 1$
\begin{multline}
\gamma_{\text{\tiny NFW}}(x)= 2\,\kappa_{\text{s}}\,\left[\,\frac{5}{3}-2\,\ln{2}-4\,\left(\frac{11}{15}-\ln{2}\right)\, (x-1)\,\right]\\[1ex] +\mathcal{O}((x-1)^2)\,,
\label{eq:NFW_gamma_radius}
\end{multline}
which yields $\gamma_{\text{\tiny NFW}}(1)\approx 0.561\,\kappa_{\text{s}}$. The NFW shear decreases for $x\gg 1$ to zero,
\begin{multline}
\gamma_{\text{\tiny NFW}}(x)=2\,\kappa_{\text{s}}\,\left[\,\frac{2}{x^2}\,\ln{\frac{x}{2}}-x^{-2}+\frac{3\pi}{2}\,x^{-3} -4\,x^{-4}\,\right]\\ +\mathcal{O}(x^{-5})\,.
\label{eq:NFW_gamma_infty}
\end{multline}
The first panel in Figure~\ref{fig:NFW-line} shows a contour plot of the shear $\gamma_{\text{\tiny NFW}}(\boldsymbol x)$ in the central part of a NFW halo. Going outward from the center, the dotted contours correspond to 95\%, 90\%, and 85\% of the central shear $\gamma_{\text{\tiny NFW}}(0)=\kappa_{\text{s}}$. Clearly, the shear changes very slowly on this scale, as indicated also by the practically homogeneous color, with the color bar set for comparison with further figures. The solid black circles mark the radial (smaller) and tangential (larger) critical curve for the fiducial halo convergence parameter $\kappa_{\text{s}}\approx 0.239035$.

The NFW shear can be written in terms of its two components, defined as
\beq
(\gamma_{\text{\tiny NFW}1},\gamma_{\text{\tiny NFW}2})=\gamma_{\text{\tiny NFW}}\, (\cos{2\varphi_{\text{\tiny NFW}}},\sin{2\varphi_{\text{\tiny NFW}}})\,,
\label{eq:NFW_gamma12}
\eeq
where the trigonometric functions of the phase $\varphi_{\text{\tiny NFW}}$ can be computed for a point $\boldsymbol x = (x_1,x_2)= x(\cos{\phi},\sin{\phi})$ in the image plane as
\begin{multline}
(\cos{2\varphi_{\text{\tiny NFW}}},\sin{2\varphi_{\text{\tiny NFW}}})=x^{-2}(x_2^2-x_1^2,-2x_1 x_2)\\=-(\cos{2\phi},\sin{2\phi})\,.
\label{eq:NFW_phi}
\end{multline}
The negative sign in front of the last parentheses indicates that the phase $\varphi_{\text{\tiny NFW}}=\phi+\pi/2+k\pi$, i.e., its orientation is always perpendicular to the position vector of the point. Note that this also means that the phase and the shear components are undefined at $x=0$, since the phase depends on the direction of approach to the center.

\subsection{Jacobian}
\label{sec:NFW-Jacobian}

\begin{figure*}
{\centering
\vspace{0cm}
\hspace{0cm}
\includegraphics[width=18 cm]{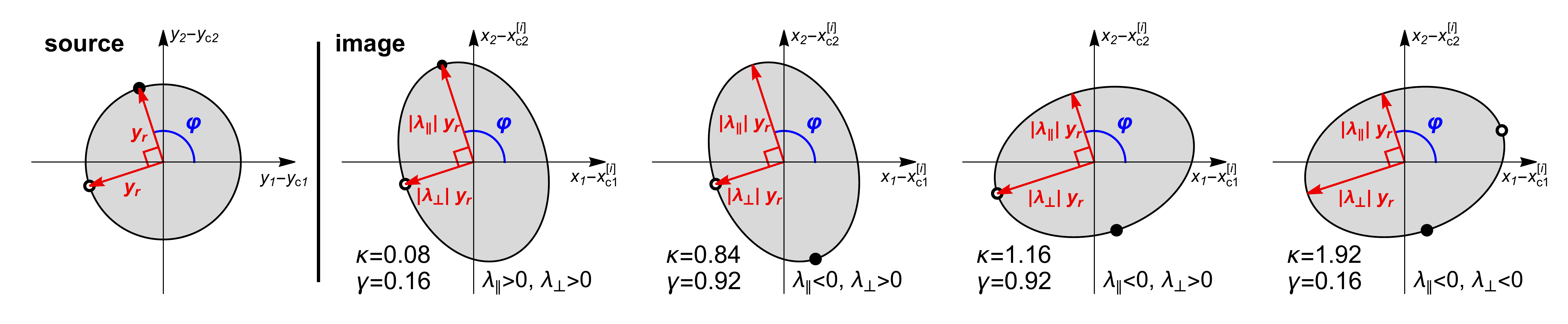}
\caption{Geometry of an image of a small circular source. Left panel: source with radius $y_\text{r}$; the black and white points lie in the directions of the eigenvectors of the inverse Jacobian matrix $\mathbb{A}$ parallel and perpendicular to the phase $\varphi$. Right panels: elliptical image for $(\kappa,\gamma)=(0.08,0.16)$ and the three other combinations producing an ellipse of the same shape and size, as marked at the bottom left of each panel. The sizes of the semi-axes are marked in red; the positions of the images of the two points marked on the circumference are determined by the signs of the eigenvalues marked at the bottom right of each panel.\label{fig:small-source-images}}}
\end{figure*}

The Jacobian matrix of a general lens equation expressed in terms of the convergence $\kappa(\boldsymbol x)$, shear $\gamma(\boldsymbol x)$, and phase $\varphi(\boldsymbol x)$ has the form \citep[e.g.,][]{schneider_etal92}
\beq
J(\boldsymbol x)=\frac{\partial\,\boldsymbol y}{\partial\,\boldsymbol x}=
\begin{pmatrix}
  1-\kappa-\gamma\cos{2\varphi} & -\gamma\sin{2\varphi} \\
  -\gamma\sin{2\varphi} & 1-\kappa+\gamma\cos{2\varphi}
\end{pmatrix}\,.
\label{eq:Jacobi_matrix}
\eeq
Its determinant, the Jacobian, can be computed from the convergence and the shear:
\beq
\mathrm{det}\,J(\boldsymbol x)=\left[\,1-\kappa(\boldsymbol x)-\gamma(\boldsymbol x)\,\right]\,\left[\,1-\kappa(\boldsymbol x)+\gamma(\boldsymbol x)\,\right]\,.
\label{eq:Jacobian}
\eeq
The critical curves, explored in detail in \citetalias{karamazov_etal21}, can be obtained by setting $\mathrm{det}\,J(\boldsymbol x)=0$. We note here merely that for an axially symmetric lens such as the studied NFW halo, the first term in Equation~(\ref{eq:Jacobian}) yields the tangential critical curve and the second term yields the radial critical curve. Hence, the shear is related to the convergence by $\gamma(x_{\text{T}})=1-\kappa(x_{\text{T}})$ at the tangential critical curve, and by $\gamma(x_{\text{R}})=\kappa(x_{\text{R}})-1$ at the radial critical curve, with the two critical curves separated by the unit-convergence circle.

\subsection{Geometry of Images}
\label{sec:NFW-images}

To study the geometry of the images, we invert the Jacobian matrix to obtain the mapping from the source plane to the image plane. We can write the inverse matrix in terms of its two eigenvalues,
\beq
\lambda_\parallel(\boldsymbol x)=(1-\kappa-\gamma)^{-1},\qquad \lambda_\perp(\boldsymbol x)=(1-\kappa+\gamma)^{-1}\,,
\label{eq:A_eigenvalues}
\eeq
as
\begin{multline}
\mathbb{A}(\boldsymbol x)=\lambda_\parallel
\begin{pmatrix}
  \cos^2{\varphi} & \cos{\varphi}\sin{\varphi} \\
  \cos{\varphi}\sin{\varphi} & \sin^2{\varphi}
\end{pmatrix}\\
+\lambda_\perp
\begin{pmatrix}
  \sin^2{\varphi} & -\cos{\varphi}\sin{\varphi} \\
  -\cos{\varphi}\sin{\varphi} & \cos^2{\varphi}
\end{pmatrix}\,,
\label{eq:A_matrix}
\end{multline}
where $\kappa$, $\gamma$, and $\varphi$ are functions of the image-plane position $\boldsymbol x$. The matrix accompanying $\lambda_\parallel$ is a projection matrix onto the eigenvector $(\cos{\varphi},\sin{\varphi})$; the matrix accompanying $\lambda_\perp$ is a projection matrix onto the eigenvector $(-\sin{\varphi},\cos{\varphi})$. Equation~(\ref{eq:A_matrix}) shows that an image at position $\boldsymbol x$ is scaled by a factor $\lambda_\parallel$ in the direction parallel to the phase $\varphi$, and by a factor $\lambda_\perp$ in the direction  perpendicular to the phase, $\varphi+\pi/2$.

A small circular source with radius $y_{\text{r}}$ centered at $\bm y_\text{c}$ is thus portrayed by the lens as a set of $n$ small elliptical images with semiaxes $y_{\text{r}}/|1-\kappa(\bm x^{[i]}_\text{c})-\gamma(\bm x^{[i]}_\text{c})|$ and $y_{\text{r}}/|1-\kappa(\bm x^{[i]}_\text{c})+\gamma(\bm x^{[i]}_\text{c})|$. Their positions $\bm x^{[i]}_\text{c}(\bm y_\text{c})$, $i=1\ldots n$, can be found by solving the lens equation, i.e., Equation~(\ref{eq:NFW_lens_equation}) for the NFW halo. We illustrate the geometry of one such image in the second panel of Figure~\ref{fig:small-source-images} for convergence $\kappa(\bm x^{[i]}_\text{c})=\kappa=0.08$ and shear $\gamma(\bm x^{[i]}_\text{c})=\gamma=0.16$. In the three right panels we include all other $(\kappa, \gamma)$ combinations that lead to the same combination of semiaxes $|\lambda_\parallel|\,y_{\text{r}}$ and $|\lambda_\perp|\,y_{\text{r}}$, i.e., they generate an elliptical image of the same shape and size. The images differ in their orientation and parity. For the combinations in the two right panels with $\kappa>1$ the major axis is oriented perpendicular to the phase $\varphi$ rather than parallel to it. The images in the third and fourth panels have negative parity, as indicated by the positions of the images of the black and white points on the circumference of the source. The signs of the eigenvalues are marked in each panel, with negative values indicating mirroring along the corresponding eigenvector.

For a general source, the distortion of its image can be quantified by the dimensionless flattening,
\beq
f(\kappa,\gamma)=1-\min\left({\left|\frac{1-\kappa-\gamma}{1-\kappa+\gamma}\right|, \left|\frac{1-\kappa+\gamma}{1-\kappa-\gamma}\right|}\right)\,,
\label{eq:flattening}
\eeq
defined here using the ratio of the smaller to larger eigenvalues, with their definitions taken from Equation~(\ref{eq:A_eigenvalues}). For an elliptical image of a circular source, $f$ is equal to its ellipticity. For the sample images in Figure~\ref{fig:small-source-images}, $f\approx0.30$. While the absolute value of the ratio of the eigenvalues determines the distortion of the image, their product determines its magnification and parity. Hence, it is sufficient to know the convergence $\kappa$ and shear $\gamma$ at the position of an image of a small source in order to fully determine its distortion, magnification, parity, and orientation with respect to the phase.

\bfi
{\centering
\vspace{0cm}
\hspace{0cm}
\includegraphics[width=8.5 cm]{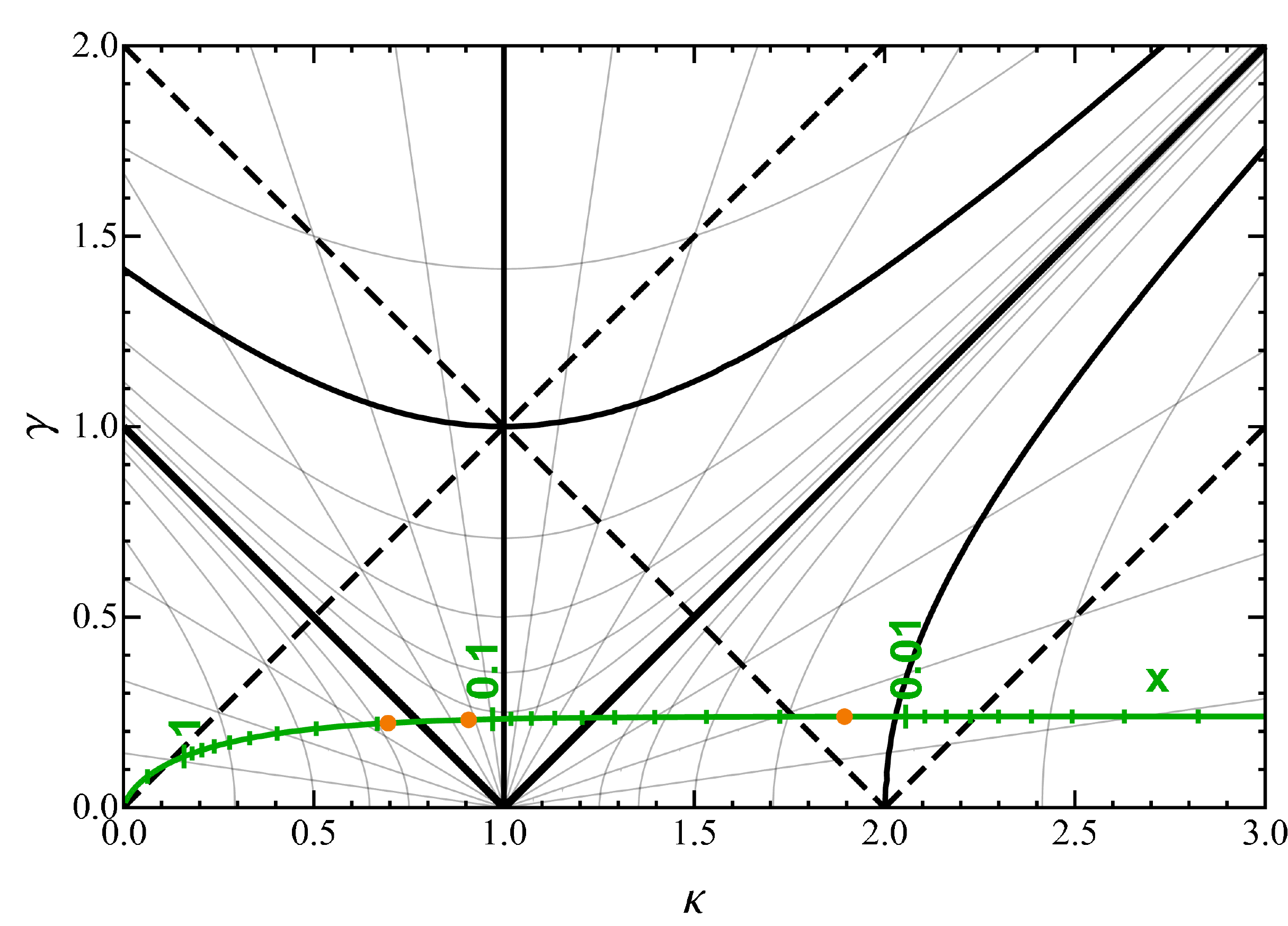}
\caption{CS diagram illustrating the geometry of images formed by the NFW halo from the top row of Figure~\ref{fig:images} with $\kappa_{\text{s}}\approx 0.239035$. The green line marks all $\left(\kappa_{\text{\tiny NFW}}(x),\gamma_{\text{\tiny NFW}}(x)\right)$ combinations of the lens, with radial positions $x$ marked by tick marks, starting from $x=0.002$ near the right edge and ending with $x=2$ close to the origin of the diagram. The orange dots mark the $(\kappa,\gamma)$ combinations at the positions of the three images in the top right panel of Figure~\ref{fig:images}. For more details on the diagram and its interpretation see Figure~\ref{fig:CS-diagram}.\label{fig:CS-NFW}}}
\efi

For a given gravitational lens, this information can be condensed into the convergence--shear (hereafter CS) diagram, introduced in Appendix~\ref{sec:Appendix-images}. The geometry of images formed by a NFW halo lens as a function of their radial position $x$ can be easily identified from its CS diagram, shown in Figure~\ref{fig:CS-NFW} for the fiducial convergence parameter $\kappa_{\text{s}}\approx 0.239035$. The green curve connecting the $\left(\kappa_{\text{\tiny NFW}}(x),\gamma_{\text{\tiny NFW}}(x)\right)$ points is obtained by using the NFW halo convergence from Equation~(\ref{eq:NFW_kappa}) and the NFW halo shear from Equation~(\ref{eq:NFW_gamma}). The center of the halo corresponds to position $(\kappa,\gamma)=(\infty,\kappa_{\text{s}})$ in the diagram, while for $x\to\infty$ the green curve reaches the origin, $(\kappa,\gamma)\to(0,0)$. The tick marks along the curve correspond to radii (from the right side of the plot) $x\in\{0.002,0.003,\ldots,0.01,0.02,\ldots,0.1,0.2,\ldots,1,2\}$.

In order to interpret the image geometries in a NFW halo from Figure~\ref{fig:CS-NFW}, we follow the green curve, starting from the origin for very distant images and progressing toward the halo center. We first recall the discussion following Equation~(\ref{eq:NFW_phi}), which implies that for the NFW profile the direction of the phase corresponds in polar coordinates to the tangential, and the direction perpendicular to the phase to the radial direction, respectively. Initially, the image has positive parity with magnification increasing from $1$ and the flattening increasing from $0$. The image is expanded tangentially ($|\lambda_\parallel|>1$) but contracted radially ($|\lambda_\perp|<1$), as indicated by the position above the dashed line.

Between $x=2$ and $x=1$ the green line crosses the diagonal and the image becomes expanded also radially ($|\lambda_\perp|>1$). For lower $x$, the green curve approaches the solid black tangential-critical-curve line, along which the magnification and $|\lambda_\parallel|$ become infinite, the flattening increases to $1$, and $|\lambda_\perp|$ remains finite. After crossing the tangential critical curve at $x=x_{\text{T}}\approx0.155$, the image parity changes to negative (due to the sign of $\lambda_\parallel$ changing to negative) and the magnification and flattening decrease. The maximum distortion stays oriented tangentially until reaching the unit-convergence circle, $x=x_0\approx0.0936$, corresponding to the vertical solid black line at $\kappa=1$. To the right of this line the maximum distortion is oriented radially. At the unit-convergence radius the negative-parity image is close to its lowest magnification (though it stays higher than $16$ in this case, as indicated by the hyperbolic contours), and with zero flattening its shape is undistorted.

Proceeding further toward the halo center, the magnification and flattening increase again, with the radial expansion $|\lambda_\perp|$ growing rapidly while the tangential expansion $|\lambda_\parallel|$ decreases. At the solid black radial-critical-curve line, which corresponds to $x=x_{\text{R}}\approx0.056$, the magnification and $|\lambda_\perp|$ become infinite, the flattening increases to $1$, and $|\lambda_\parallel|$ remains finite. For lower radii, image parity changes back to positive (due to the sign of $\lambda_\perp$ changing to negative), and the magnification and flattening decrease. The expansion in both perpendicular directions decreases, until the intersection with the dashed black line at $x\approx 0.019$. Images lying closer to the halo center are tangentially contracted rather than expanded ($|\lambda_\parallel|<1$). After crossing the solid black unit-magnification hyperbola at $x\approx 0.0105$, all images are demagnified. The last important intersection occurs at $x\approx 0.0068$; images to the right of the last dashed black line are contracted even radially ($|\lambda_\perp|<1$), and their magnification and flattening decrease to $0$ at the halo center.

The three orange points marked along the green curve in Figure~\ref{fig:CS-NFW} correspond to the radial positions of the centers of the three images in the top right panel of Figure~\ref{fig:images}. The first point at $x\approx 0.188$ corresponds to the image outside the tangential critical curve at the right side of the panel. The second point at $x\approx 0.115$ corresponds to the image between the tangential critical curve and the unit-convergence circle at the left side of the panel. The third point at $x\approx 0.014$ corresponds to the smallest image inside the radial critical curve. The magnification, parity, flattening, orientation, and the two scaling factors of the corresponding images can be determined from the positions of these points in the diagram in Figure~\ref{fig:CS-NFW}. Note that the values obtained from the diagram are technically valid at the positions of the source-center images and thus correspond to the local ``point-source'' values. Taking into account the radial extent of each image, their position (and the relevant range of their properties) should be marked by line segments along the green line in Figure~\ref{fig:CS-NFW} rather than by points.

\subsection{Weak Shear and Phase}
\label{sec:NFW-weak}

Weak-lensing cluster-mass reconstructions are based on statistical analyses of the images of background galaxies \citep{kaiser_squires93}. The convergence map is computed from maps of the shear components. These are constructed from the shear and the phase, which are in turn determined from the shapes and orientations of the images. In the weak-lensing limit, the geometry of the elliptical image of a small circular source corresponds to the second panel of Figure~\ref{fig:small-source-images}. Its axis ratio yields the shear and the orientation angle of its major axis is equal to the phase. The semi-minor to semi-major axis ratio $b/a$ is obtained by expanding the ratio $\lambda_\perp/\lambda_\parallel$ of the eigenvalues from Equation~(\ref{eq:A_eigenvalues}) to first order in $\kappa$ and $\gamma$. The ellipticity, which is equal to the flattening of the image defined in Equation~(\ref{eq:flattening}), reduces in this limit to
\beq
f = 1-b/a\simeq2\,\gamma\,,
\label{eq:ellipticity}
\eeq
i.e., double the value of the local shear. For illustration, for the image in the second panel of Figure~\ref{fig:small-source-images} Equation~(\ref{eq:ellipticity}) yields  an approximate flattening $2\,\gamma=0.32$, which is an $8\%$ overestimate of the actual value $f\approx0.30$ from Equation~(\ref{eq:flattening}).

Based on the weak-lensing regime, we introduce the weak shear and weak phase, which are computed from the images using the weak-lensing relations from the previous paragraph. We define the weak shear as
\beq
\gamma_{\text{w}}(\boldsymbol x)=\frac{1}{2}\,f\left(\kappa(\boldsymbol x),\gamma(\boldsymbol x)\right)\,,
\label{eq:weak_shear}
\eeq
where the flattening $f$ is computed from Equation~(\ref{eq:flattening}) using lens-specific convergence and shear functions. Note that $\gamma_{\text{w}}$ by its definition attains only values from the interval $\left[0,\,0.5\right]$. In the weak-lensing regime $\gamma_{\text{w}}\approx \gamma$, but as $\gamma$ and $\kappa$ increase, the weak shear computed from the image flattening deviates from the shear.

For the NFW halo we compute the weak shear $\gamma_{\text{w,\tiny NFW}}(\boldsymbol x)$ from Equation~(\ref{eq:weak_shear}) using its convergence from Equation~(\ref{eq:NFW_kappa}) and shear from Equation~(\ref{eq:NFW_gamma}). The second panel in Figure~\ref{fig:NFW-line} shows a plot of $\gamma_{\text{w,\tiny NFW}}(\boldsymbol x)$ in the central part of a NFW halo, using the same color scale as for the shear in the first panel of the figure. Due to its relation to the flattening in Equation~(\ref{eq:weak_shear}), the plot can be interpreted following the $f$ values along the green curve in the diagram in Figure~\ref{fig:CS-NFW}. At the halo center the flattening and thus also the weak shear are equal to zero. Going outward from the center, the weak shear increases to its maximum value of $0.5$ at the radial critical curve, marked by the inner black circle. From there it drops to $0$ at the unit-convergence radius and increases back to $0.5$ at the tangential critical curve, marked by the outer black circle. Beyond the tangential critical curve the weak shear drops asymptotically to $0$. Comparison with the left panel shows the substantial difference between the shears in the central region of the NFW halo, in terms of amplitude as well as radial pattern. Here the variations in image distortion are primarily driven by the convergence rather than by the shear.

In the third panel of Figure~\ref{fig:NFW-line} we illustrate the difference between the first two panels by plotting the relative deviation of the weak shear from the shear, $\gamma_{\text{w,\tiny NFW}}/\gamma_{\text{\tiny NFW}}-1$. The blue regions in which the weak shear underestimates the shear are limited to the vicinity of the origin and the vicinity of the unit-convergence circle. In both cases the weak shear drops to zero and the relative deviation thus reaches $-1$, its minimum possible value. Everywhere else the weak shear overestimates the shear, with the positive deviation peaking at the critical curves and dropping to $0$ asymptotically. Note that the maxima at the critical curves may be negative for NFW halos with a sufficiently high convergence parameter $\kappa_{\text{s}}$, for which $\gamma_{\text{\tiny NFW}}(x_{\text{R}})$ or even $\gamma_{\text{\tiny NFW}}(x_{\text{T}})$ exceeds the weak-shear value at critical curves (i.e., $0.5$).

In addition to the weak shear, we define the image-based weak phase $\varphi_{\text{w}}$ as the angle between the major axis of the image of a small circular source and the horizontal ($x_1$) axis of the image plane. Taking into account Figure~\ref{fig:small-source-images} and the discussion preceding Equation~(\ref{eq:flattening}), in the case of the NFW halo it is related to the phase $\varphi_{\text{\tiny NFW}}$ as follows:
\beq
\varphi_{\text{w,\tiny{NFW}}}(\boldsymbol x)=\Bigg\{
\begin{array}{ll}
\,\varphi_{\text{\tiny NFW}}(\boldsymbol x) & \,\text{for $\kappa_{\text{\tiny NFW}}(\boldsymbol x)<1$}\,,\\
\,\varphi_{\text{\tiny NFW}}(\boldsymbol x)+\pi/2 & \,\text{for $\kappa_{\text{\tiny NFW}}(\boldsymbol x)>1$}\,.
\end{array}
\label{eq:weak_phase}
\eeq
We use values from the interval $\left[-\pi/2,\,\pi/2\right]$ for both $\varphi_{\text{\tiny NFW}}$ and $\varphi_{\text{w,\tiny{NFW}}}$.

The fourth panel of Figure~\ref{fig:NFW-line} shows a color map of the weak phase $\varphi_{\text{w,\tiny{NFW}}}$ of the NFW halo. The white regions with $\varphi_{\text{w,\tiny{NFW}}}=0$ correspond to horizontally elongated images, in the orange regions with $\varphi_{\text{w,\tiny{NFW}}}>0$ the images are oriented counterclockwise and in the blue regions with $\varphi_{\text{w,\tiny{NFW}}}<0$ the images are oriented clockwise from the horizontal. The bright orange/blue boundaries correspond to images elongated exactly vertically, with $|\varphi_{\text{w,\tiny{NFW}}}|=\pi/2$. The weak phase flips by $\pi/2$ along the unit-convergence circle, separating the inner radially oriented from the outer tangentially oriented images. The weak phase is undefined along this circle as well as at the origin, which corresponds to zero flattening. The dot-dashed lines added for orientation mark the positions of all images with exactly horizontal, exactly vertical, or undefined orientation.

Since the weak phase differs from the phase only by the $\pi/2$ flip inside the unit-convergence circle, a similar color map of the phase $\varphi_{\text{\tiny NFW}}(\boldsymbol x)$ would differ merely by having inverted color and saturation inside the circle. In other words, the color and saturation outside the circle in the fourth panel of Figure~\ref{fig:NFW-line} would be radially extended to the halo center. Hence, the first and third quadrants would be entirely blue and the second and fourth quadrants would be entirely orange.

The plots of the different quantities in Figure~\ref{fig:NFW-line} are presented as reference plots to aid the interpretation of the results for the NFW halo + point-mass lens model presented in Section~\ref{sec:NFWP-plots}.

\section{Lensing by a NFW Halo + Point Mass}
\label{sec:NFWP}

\subsection{Convergence, Shear, and Phase}
\label{sec:NFWP-csp}

Adding a compact massive object modeled by a point mass positioned at $\bm{x_{\text{P}}}$ changes the convergence to
\beq
\kappa(\boldsymbol x)=2\,\kappa_{\text{s}}\;\frac{1-\mathcal{F}(x)}{x^2-1} +\pi\,\kappa_{\text{P}}\,\delta(\boldsymbol x-\bm{x_{\text{P}}})\,,
\label{eq:NFWP_kappa}
\eeq
where the mass parameter $\kappa_{\text{P}}$ corresponds to the ratio of the solid angles subtended by the point-mass Einstein circle and by the halo scale-radius circle \citepalias[for more details, see][]{karamazov_etal21}. Hence, $\sqrt{\kappa_{\text{P}}}$ is the point-mass Einstein radius in units of the halo scale radius. The convergence in Equation~(\ref{eq:NFWP_kappa}) is identical to the NFW halo convergence from Equation~(\ref{eq:NFW_kappa}), except exactly at the position of the added point mass. The lens equation can be written as
\beq
\boldsymbol y=\boldsymbol x - 4\,\kappa_{\text{s}}\,
\left[\ln{\frac{x}{2}}+\mathcal{F}(x)\right]\,\frac{\boldsymbol x}{x^2}-\kappa_{\text{P}}\,\frac{\boldsymbol x-\bm{x_{\text{P}}}}{|\boldsymbol x - \bm{x_{\text{P}}}|^2}\,,
\label{eq:NFWP_lens_equation}
\eeq
in the form used in \citetalias{karamazov_etal21}.

For illustration, in the bottom row of Figure~\ref{fig:images} we show the lensing of the same circular source as in the top row, by the same NFW halo with an additional point mass with mass parameter $\kappa_{\text{P}}\approx 2.714\cdot 10^{-4}$ positioned at $\bm{x_{\text{P}}}=(0.2,\,0)$. This parameter combination is selected from the parameter-space grid used in \citetalias{karamazov_etal21}. The position of the point mass is indicated by its Einstein circle (cyan) in the bottom right panel. As seen in the bottom left panel, the black circular source lies inside the weakly perturbed radial caustic, with its upper part lying also inside the strongly perturbed tangential caustic. For a source not lying on the caustic, lens Equation~(\ref{eq:NFWP_lens_equation}) yields $2, 4,$ or $6$ images. For a source lying on the caustic, several images appear combined into a lower number of macro-images. In the example shown in the bottom right panel there are five macro-images. Four of them are images of the full source; the fifth macro-image to the top left of the point mass consists of two additional images of the upper part of the source joined along the critical curve. Comparing the images with those in the top right panel, we see that the left and central images are affected only weakly by the point mass. The right image is affected more strongly, plus there are two new images closer to the point mass. For these images in particular, their distortion cannot be simply classified as tangential or radial.

In order to compute the shear we start from the lens potential, which has an additional term due to the point mass,
\beq
\psi(\boldsymbol x)=\psi_{\text{\tiny NFW}}(x)+\frac{r_{\text{s}}^2}{D_{\text{l}}^2}\,\kappa_{\text{P}}\, \ln{|\boldsymbol x-\bm{x_{\text{P}}}|}\,,
\label{eq:NFWP_potential}
\eeq
where $\psi_{\text{\tiny NFW}}(x)$ is given by Equation~(\ref{eq:NFW_psi}).

The point-mass shear has the simple form
\beq
\gamma_{\text{P}}(\boldsymbol x)=\frac{\kappa_{\text{P}}}{|\boldsymbol x - \bm{x_{\text{P}}}|^2}\,,
\label{eq:P_gamma}
\eeq
divergent at the point-mass position and dropping rapidly outward. Since the NFW halo shear peaks at its central value $\kappa_{\text{s}}$, the added point mass will dominate the lens shear in its vicinity, wherever it may be positioned. The shear can be generally computed from the second derivatives of the lens potential, namely
\beq
\gamma=\frac{D_{\text{l}}^2}{r_{\text{s}}^2}\,\sqrt{ \frac{1}{4}\left(\,\psi,_{\scriptscriptstyle 11}-\psi,_{\scriptscriptstyle 22}\right)^2+\left(\psi,_{\scriptscriptstyle 12}\right)^2}\,,
\label{eq:shear}
\eeq
where the commas denote partial derivatives with respect to image-plane coordinates $(x_1,x_2)$. For our combined lens we compute the derivatives of the lens potential from Equation~(\ref{eq:NFWP_potential}) and get
\beq
\gamma(\boldsymbol x)=\sqrt{\left[\gamma_{\text{\tiny{NFW}}}(x)-\gamma_{\text{P}}(\boldsymbol x)\right]^2+ 4\,\gamma_{\text{\tiny{NFW}}}(x)\,\gamma_{\text{P}}(\boldsymbol x)\,\cos^2{\omega(\boldsymbol x)}}\,,
\label{eq:NFWP_gamma}
\eeq
where $\gamma_{\text{\tiny{NFW}}}(x)$ and $\gamma_{\text{P}}(\boldsymbol x)$ are given by Equation~(\ref{eq:NFW_gamma}) and Equation~(\ref{eq:P_gamma}), respectively, and
\beq
\cos{\omega(\boldsymbol x)}=\frac{{\bm x}\cdot(\boldsymbol x - \bm{x_{\text{P}}})}{x\,|\boldsymbol x - \bm{x_{\text{P}}}|}
\label{eq:cos}
\eeq

\bfi
{\centering
\vspace{0cm}
\hspace{0cm}
\includegraphics[width=6 cm]{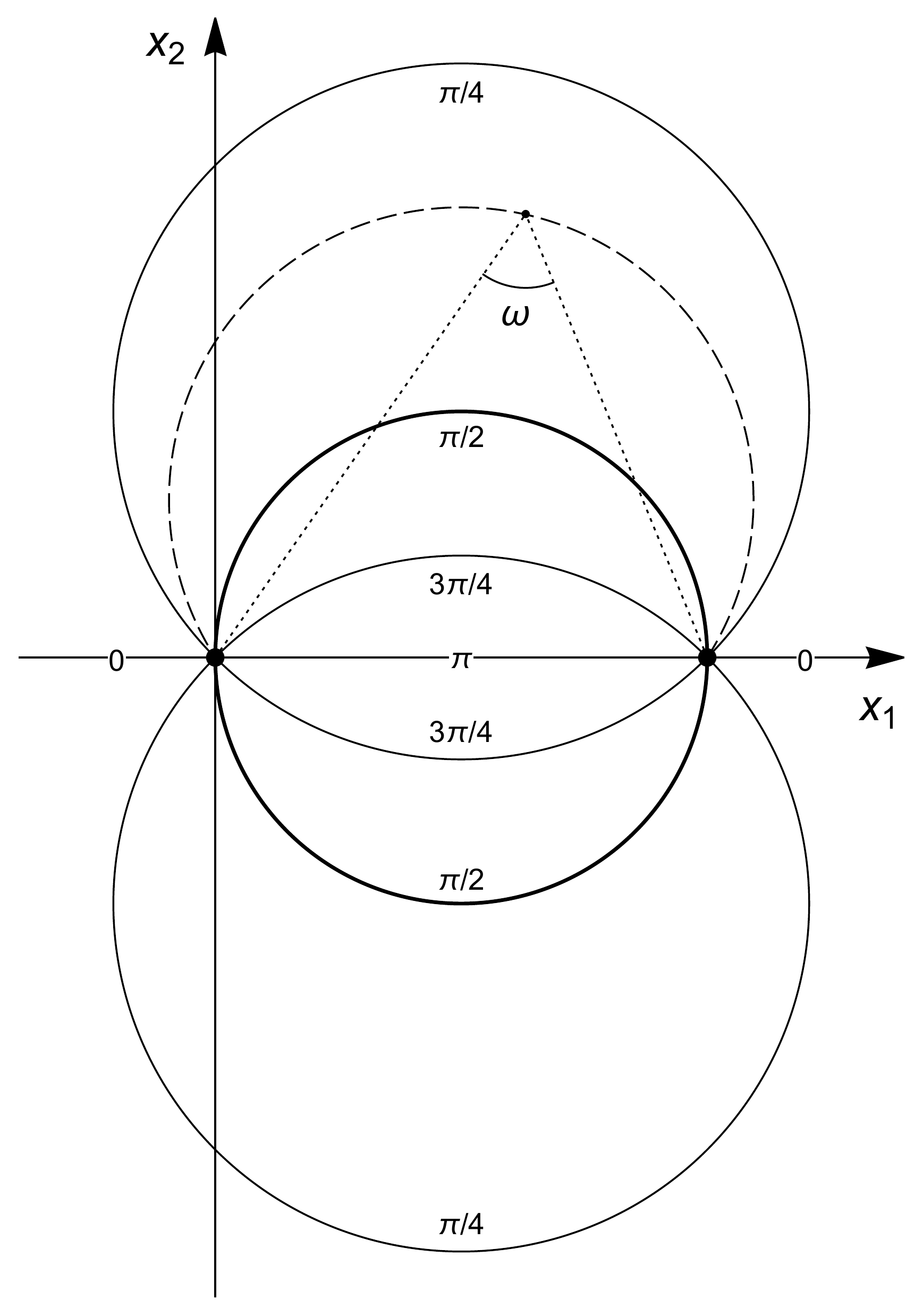}
\caption{Viewing angle $\omega(\boldsymbol x)$ of the line segment from the halo center to the point mass, appearing in Equation~(\ref{eq:NFWP_gamma}) for computing the shear $\gamma(\boldsymbol x)$ of the combined lens. Points on the horizontal axis mark the halo center and the point-mass position $\bm{x_{\text{P}}}$; the dot at the vertex of the angle marks the position $\boldsymbol x$. Contours of constant $\omega$ are symmetric pairs of circular arcs connecting the centers of the two lens components, with the dashed line marking the arc passing through position $\boldsymbol x$. The values of $\omega$ range from $\pi$ along the line segment to $0$ along the rest of the horizontal axis.\label{fig:omega}}}
\efi

is the dot product of the unit vectors pointing to $\bm x$ from the halo center and from the point-mass position. In terms of image-plane geometry, $\omega$ is the viewing angle from point $\boldsymbol x$ of the line segment connecting the halo center and the point-mass position. As shown in Figure~\ref{fig:omega}, curves of constant $\omega$ are circular arcs connecting symmetrically the halo center and the point-mass position. Note that $\omega$ is also equal to the angle
between the tangent to the arc at either of its end points and the outward horizontal direction, as follows from the tangent--chord theorem (alternate segment theorem).

Along the line segment connecting the center and the point mass the viewing angle reaches its maximum, $\omega=\pi$, while along the rest of the horizontal axis it reaches its minimum, $\omega=0$. In both cases, $\cos^2{\omega}=1$ and the total shear from Equation~(\ref{eq:NFWP_gamma}) is $\gamma(\boldsymbol x)=\gamma_{\text{\tiny{NFW}}}(x)+\gamma_{\text{P}}(\boldsymbol x)$. In this case both shears act in the same orientation, so that their combination is maximal. Along the circle bisected by the line segment, we find $\omega=\pi/2$ according to Thales's theorem and the total shear is $\gamma(\boldsymbol x)=|\gamma_{\text{\tiny{NFW}}}(x)-\gamma_{\text{P}}(\boldsymbol x)|$. In this case the two shears act in perpendicular directions, so that their combination is minimal. For the pair of small arcs in Figure~\ref{fig:omega} with $\omega=3\pi/4$ and for the pair of large arcs with $\omega=\pi/4$ we get $\cos^2{\omega}=1/2$ and the total shear is $\gamma(\boldsymbol x)=\sqrt{\gamma_{\text{\tiny{NFW}}}^2(x)+\gamma_{\text{P}}^2(\boldsymbol x)}$.

As discussed in Section~\ref{sec:NFW-csp}, for the NFW halo the central shear is defined, $\gamma_{\text{\tiny{NFW}}}(0)=\kappa_\text{s}$, while the phase and shear components are undefined. The same holds for the central properties of the point-mass lens. However, for the combined lens even the shear at the halo center is undefined. For $x\to 0$ Equation~(\ref{eq:NFWP_gamma}) yields
\beq
\gamma(\boldsymbol x)\to\sqrt{\left[\kappa_{\text{s}}-\kappa_{\text{P}}\,x_{\text{P}}^{-2}\right]^2+ 4\,\kappa_{\text{s}}\kappa_{\text{P}}\,x_{\text{P}}^{-2}\,\cos^2{\omega}}\,,
\label{eq:NFWP_gamma_center}
\eeq
a value that depends on the direction of approach to the center, due to the directional dependence of $\omega$. As seen from Figure~\ref{fig:omega} and as explained in the discussion above, approaching the center along the horizontal axis leads to the highest value (the sum of the two shears) while an approach along the vertical axis leads to the lowest value (the absolute value of the difference of the two shears). The situation at the position of the point mass is similar, though here the angular differences are suppressed by the divergence of $\gamma_{\text{P}}$.

The range of values of the shear occurring in the studied central region of the image plane is larger than for the halo or the point mass separately. Its upper limit is $\infty$, due to the divergence of $\gamma_{\text{P}}$ at the point-mass position $\bm{x_{\text{P}}}$. Its lower limit may reach $0$. As seen from the form of Equation~(\ref{eq:NFWP_gamma}), this may occur only along the $\omega=\pi/2$ circle at points where $\gamma_{\text{\tiny{NFW}}}(x)=\gamma_{\text{P}}(\boldsymbol x)$. Following the circle from the halo center to the point mass, $\gamma_{\text{\tiny{NFW}}}$ decreases while $\gamma_{\text{P}}$ increases. Zero-shear points thus exist only if $\gamma_{\text{\tiny{NFW}}}$ is equal to or larger than $\gamma_{\text{P}}$ at the halo center. Hence, for $x_{\text{P}}<\sqrt{\kappa_{\text{P}}/\kappa_{\text{s}}}$ there are no zero-shear points. In this range, for point masses closest to the halo center, the minimum shear $\gamma=\kappa_{\text{P}}\,x_{\text{P}}^{-2}-\kappa_{\text{s}}$ occurs at the halo center when approached along the vertical axis.

Zero-shear points exist for all larger point-mass distances from the halo center. For $x_{\text{P}}=\sqrt{\kappa_{\text{P}}/\kappa_{\text{s}}}$ there is one zero-shear point located directly at the halo center. For any $x_{\text{P}}>\sqrt{\kappa_{\text{P}}/\kappa_{\text{s}}}$ there are two zero-shear points lying symmetrically above and below the horizontal axis on the $\omega=\pi/2$ circle. With increasing distance of the point mass from the halo center the zero-shear points shift along the circle toward the position of the point mass, so that for larger distances they lie nearly vertically above and below the point mass. Their separation from the point mass, which is approximately $\sqrt{\kappa_{\text{P}}/\gamma_{\text{\tiny{NFW}}}(x_\text{P})}$ in this regime, increases with distance as the halo shear decreases.

Image-plane maps of the shear $\gamma(\boldsymbol x)$ for different masses and positions of the point mass are presented and discussed in Section~\ref{sec:plots-shear}.

The shear components are defined by
\beq
(\gamma_1,\gamma_2)=\gamma\, (\cos{2\varphi},\sin{2\varphi})\,,
\label{eq:NFWP_gamma12}
\eeq
where the trigonometric functions of the phase $\varphi$ can be computed for a point $\boldsymbol x = (x_1,x_2)$ in the image plane as
\begin{eqnarray}
\label{eq:NFWP_phi}
 \nonumber \cos{2\varphi} &=& \frac{1}{\gamma(\boldsymbol x)} \left[\frac{x_2^2-x_1^2}{x^2}\, \gamma_{\text{\tiny{NFW}}}(x)\right.\\
 \nonumber & & \left.\hspace{1cm}+\frac{(x_2-x_{\text{P}2})^2-(x_1-x_{\text{P}1})^2}{|\boldsymbol x - \bm{x_{\text{P}}}|^2}\, \gamma_{\text{P}}(\boldsymbol x)\right]\\[1ex]
 & & \\
 \nonumber \sin{2\varphi} &=& \frac{-2}{\gamma(\boldsymbol x)}\left[\frac{x_1 x_2}{x^2}\, \gamma_{\text{\tiny{NFW}}}(x)\right.\\
 \nonumber & & \left.\hspace{1cm}+\frac{(x_1-x_{\text{P}1})(x_2-x_{\text{P}2})}{|\boldsymbol x - \bm{x_{\text{P}}}|^2}\, \gamma_{\text{P}}(\boldsymbol x)\right]\,,
\end{eqnarray}
where the shears $\gamma(\boldsymbol x),\gamma_{\text{\tiny{NFW}}}(x),$ and $\gamma_{\text{P}}(\boldsymbol x)$ are given by Equations~(\ref{eq:NFWP_gamma}), (\ref{eq:NFW_gamma}), and (\ref{eq:P_gamma}), respectively. Note that in this case the phase and the shear components are undefined at the halo center and at the point-mass position, since the phase as well as the shear $\gamma(\boldsymbol x)$ depend on the direction of approach to these points.

We would like to point out that Equation~(\ref{eq:NFWP_gamma}) is a special case of the more general formula
\beq
\gamma=\sqrt{(\gamma_\text{A}-\gamma_\text{B})^2+ 4\,\gamma_\text{A}\gamma_\text{B}\cos^2{(\varphi_\text{A}-\varphi_\text{B})}}
\label{eq:shear_combination}
\eeq
for the shear of a combination of two mass distributions with shears $\gamma_\text{A}, \gamma_\text{B}$ and phases $\varphi_\text{A}, \varphi_\text{B}$. For two circularly symmetric mass distributions with the same sign of the expression $\psi''-x^{-1}\psi'$ that appears in Equation~(\ref{eq:shear_sym}), the absolute value of the phase difference in Equation~(\ref{eq:shear_combination}) is equal to the viewing angle $\omega$.

The expression $\psi''-x^{-1}\psi'$ is globally negative for a range of mass distributions, such as for the NFW profile, for a point mass, for a singular or a non-singular (cored) isothermal sphere. For a combination of two such distributions, the formula for the shear in Equation~(\ref{eq:NFWP_gamma}), the following discussion, and the formulae for the phase in Equation~(\ref{eq:NFWP_phi}) are valid. For example, the case of two point masses was studied by \cite{schneider_weiss86}, and the case of two isothermal spheres was studied by \cite{shin_evans08}.

\subsection{Jacobian and Umbilic Points}
\label{sec:NFWP-Jacobian}

The Jacobian of the lens equation can be computed from Equation~(\ref{eq:Jacobian}) using the convergence from Equation~(\ref{eq:NFWP_kappa}) and the shear from Equation~(\ref{eq:NFWP_gamma}). Its explicit form is presented in \citetalias{karamazov_etal21}, together with a detailed analysis of the critical curves which are obtained by setting the Jacobian equal to zero. The parts of the critical curve lying outside the unit-convergence circle ($x>x_0$) satisfy the equation
\beq
\gamma(\boldsymbol x)=1-\kappa(\boldsymbol x)\,,
\label{eq:NFWP_cc-tangential}
\eeq
which yields the tangential critical curve in absence of the point mass. The parts lying inside the unit-convergence circle ($x<x_0$) satisfy the equation
\beq
\gamma(\boldsymbol x)=\kappa(\boldsymbol x)-1\,,
\label{eq:NFWP_cc-radial}
\eeq
which yields the radial critical curve in absence of the point mass.

Equation~(\ref{eq:Jacobian}) also indicates that for the Jacobian to be equal to zero at a point lying directly on the unit-convergence circle, the shear must be zero at such a point. From the properties of zero-shear points discussed in Section~\ref{sec:NFWP-csp} it follows that such critical-curve points must lie at the intersections of the unit-convergence circle and the $\omega=\pi/2$ circle extending from the halo center to the point-mass position.

For $x_{\text{P}}<x_0$ these circles have no intersection and, thus, there are no critical-curve points along the unit-convergence circle. For $x_{\text{P}}=x_0$ these circles have an intersection exactly at the position of the point mass. However, at this point the shear is not zero, so that even in this case there is no critical-curve point along the unit-convergence circle. For any $x_{\text{P}}>x_0$ these circles have two intersections. In this case the requirement of zero shear leads to the condition
\beq
x_{\text{P}}=x_0\,\sqrt{1+\kappa_{\text{P}}\,/ \left[4\kappa_{\text{s}}\left(1+\ln{\frac{x_0}{2}}\right)+2-3 x_0^2\right]}\,.
\label{eq:umbilic_condition}
\eeq
We conclude that for any value of the mass parameter $\kappa_{\text{P}}$ Equation~(\ref{eq:umbilic_condition}) yields a single corresponding distance of the point mass from the halo center, for which the critical curve has points lying on the unit-convergence circle.

If we place the point mass along the horizontal axis in the image plane at $\bm{x_{\text{P}}}=(x_{\text{P}},0)$, the positions of these critical-curve points are
\beq
\boldsymbol x=\left(\frac{x_0}{x_{\text{P}}}, \pm\sqrt{1-\frac{x_0^2}{x_{\text{P}}^2}}\right)\,x_0\,,
\label{eq:umbilic_points}
\eeq
with the value of $x_{\text{P}}$ given by Equation~(\ref{eq:umbilic_condition}). These points with $\kappa=1$ and $\gamma=0$ have special significance. As discussed in Appendix~\ref{sec:Appendix-images}, such critical-curve points correspond to umbilics. Equation~(\ref{eq:umbilic_condition}) thus presents a condition for the existence of umbilics in the studied lens system. In the $(\kappa_{\text{P}},x_{\text{P}})$ parameter-space plots in Figure~6 of \citetalias{karamazov_etal21}, Equation~(\ref{eq:umbilic_condition}) describes the green and violet umbilic boundary, starting at $x_{\text{P}}=x_0$ at the $\kappa_{\text{P}}=0$ vertical axis and increasing monotonically for higher $\kappa_{\text{P}}$. In the image plane, the umbilic points lie along the unit-convergence circle. For $\kappa_{\text{P}}\ll 1$ they are located close to the horizontal axis in the direction of the point mass. Their displacement from the axis increases with increasing $\kappa_{\text{P}}$.

\subsection{Geometry of Images}
\label{sec:NFWP-images}

The geometry of the images can be studied using the eigenvalue decomposition of the inverse of the Jacobian matrix given by Equation~(\ref{eq:A_matrix}) and the convergence--shear (CS) diagram introduced in Appendix~\ref{sec:Appendix-images}. For a point mass placed at the center of the halo, the convergence and shear are purely radial functions. Hence, the range of their possible combinations is limited to the $\left(\kappa(x),\gamma(x)\right)$ curve in the CS diagram. In this case the analysis of possible image geometries can directly follow the example presented in Section~\ref{sec:NFW-images} for images formed by the NFW halo.

Even when the point mass is positioned away from the halo center, the convergence given by Equation~(\ref{eq:NFWP_kappa}) preserves its circular symmetry (with the exception of the single point at the position of the point mass). This means that any convergence value $\kappa$ can be one-to-one translated to the corresponding radial distance from the halo center $x$. However, the shear given by Equation~(\ref{eq:NFWP_gamma}) loses circular symmetry. In the CS diagram this results in the range of possible $\left(\kappa(\boldsymbol x),\gamma(\boldsymbol x)\right)$ combinations covering a two-dimensional region. In terms of image distortions and orientations, the lack of symmetry means that instead of the terms ``tangential'' and ``radial'' we revert to the more general ``in the direction of the phase'' and ``perpendicular to the phase'', respectively.

\bfi
{\centering
\vspace{0cm}
\hspace{0cm}
\includegraphics[width=8.5 cm]{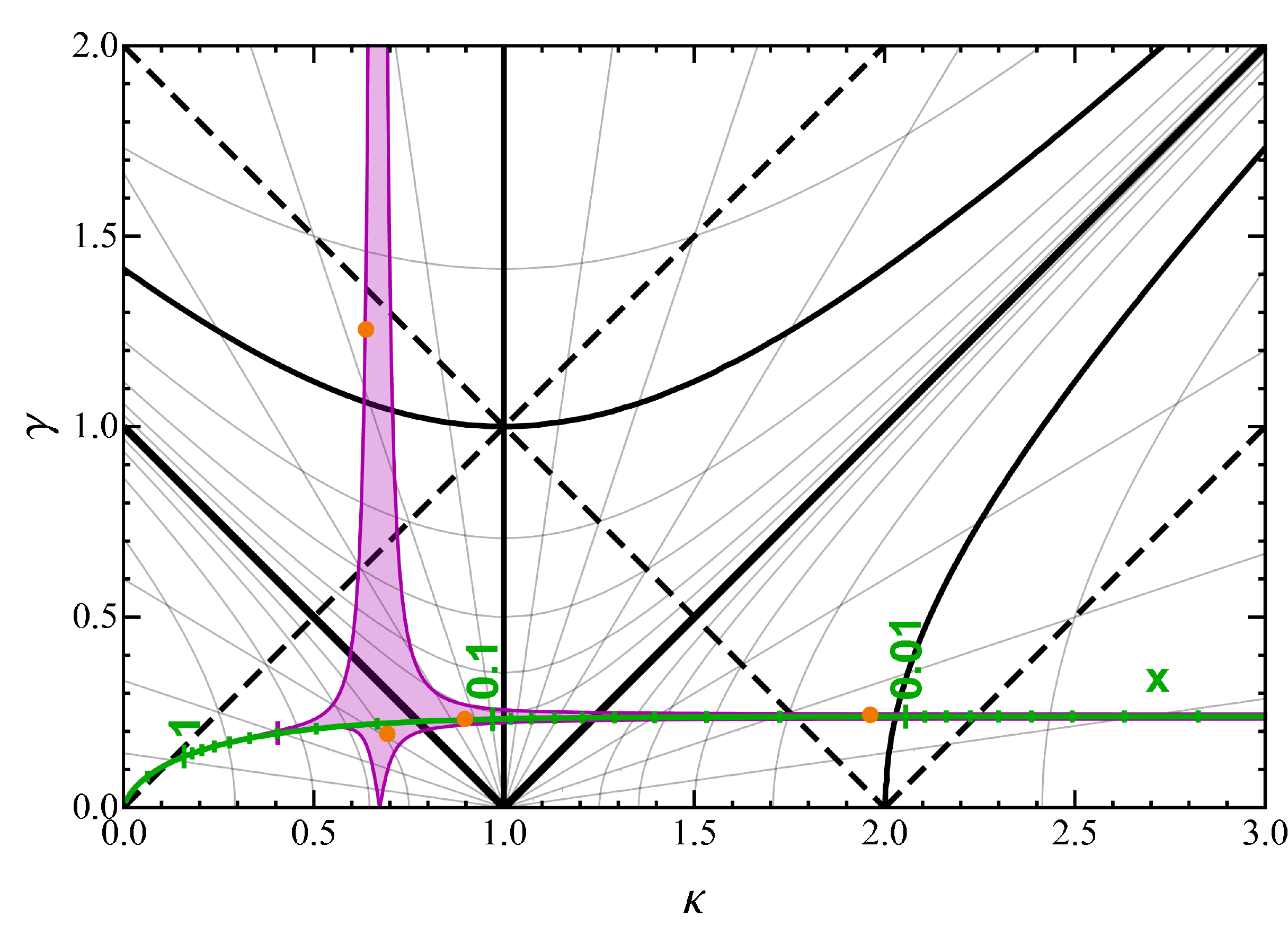}
\caption{CS diagram illustrating the geometry of images formed by the NFW halo with a $\kappa_{\text{P}}\approx 2.714\cdot 10^{-4}$ point mass at $x_{\text{P}}=0.2$, from the bottom row of Figure~\ref{fig:images}. The purple-shaded area marks the range of $\left(\kappa(\boldsymbol x),\gamma(\boldsymbol x)\right)$ combinations of the lens; the green line marks the $\left(\kappa_{\text{\tiny NFW}}(x),\gamma_{\text{\tiny NFW}}(x)\right)$ combinations of the halo-only lens from Figure~\ref{fig:CS-NFW}. The green tick marks and labels indicate the radial distance $x$ along vertical lines in this diagram. The top purple line marks the maximum shear along a circle with radius $x$ centered on the halo, which always occurs in the direction of the point mass. The bottom purple line marks the minimum shear along the circle; for larger radii $x$ (to the left of the purple tick mark) this occurs in the direction opposite the point mass; for smaller radii $x$ (to the right of the purple tick mark) this occurs at two symmetric off-axis points along the circle. The orange dots mark the $(\kappa,\gamma)$ combinations at the positions of the four full images in the bottom right panel of Figure~\ref{fig:images}.\label{fig:CS-NFWP}}}
\efi

For illustration, in Figure~\ref{fig:CS-NFWP} we present the CS diagram for a NFW halo with a $\kappa_{\text{P}}\approx 2.714\cdot 10^{-4}$ point mass located at $x_{\text{P}}=0.2$. The purple-shaded region bounded by the bold purple lines shows the range of $(\kappa,\gamma)$ combinations occurring in the image plane. Added for orientation is the green curve from Figure~\ref{fig:CS-NFW} showing the $(\kappa,\gamma)$ combinations of the NFW halo without the point mass. As in Figure~\ref{fig:CS-NFW}, the tick marks along the curve mark radial distances from the halo center. At any value of $x$ along this axis, the vertical extent between the bold purple lines indicates the range of shear values $\gamma(\boldsymbol x)$ occurring along the circle $|\boldsymbol x|=x$. The maximum shear always occurs in the direction of the point mass, i.e., for $\boldsymbol x=(x,0)$. At large distances $x$, the minimum shear occurs in the direction opposite to the point mass, i.e., for $\boldsymbol x=(-x,0)$. At lower distances, to the right of the purple vertical tick mark (in the case of Figure~\ref{fig:CS-NFWP} near $x=0.4$), minimum shear occurs at two points offset symmetrically from the axis connecting the halo center and the point mass. Close to the halo center, at the right edge of the diagram, minimum shear occurs at points offset nearly perpendicularly from the halo center, i.e., for $\boldsymbol x\approx(0,\pm\,x)$. Equation~(\ref{eq:NFWP_gamma_center}) shows that at the center of the halo the maximum shear is $\kappa_{\text{s}}+\kappa_{\text{P}}\,x_{\text{P}}^{-2}$ and the minimum shear is $|\kappa_{\text{s}}-\kappa_{\text{P}}\,x_{\text{P}}^{-2}|$.

In the example shown in Figure~\ref{fig:CS-NFWP}, the shear range at large radii does not visibly deviate from the green NFW halo shear. At radii lower than the purple tick mark near $x=0.4$, the maximum shear starts to deviate substantially from the green curve. The minimum shear starts to deviate visibly between $x=0.3$ and $x=0.2$. Along the circle with the radius of the point-mass distance, $x=x_{\text{P}}=0.2$, the maximum shear diverges at the position of the point mass. At a slightly lower radius, the minimum shear reaches $0$ at the positions of the off-axis zero-shear points. For lower radii, the shear interval shrinks back toward the NFW shear at the green curve. However, instead of reaching the central NFW shear $\gamma_{\text{\tiny{NFW}}}(0)=\kappa_s\approx 0.2390$, the limiting shear at $x=0$ varies within the interval $[0.2322,0.2458]$, as discussed in the previous paragraph.

The four orange points marked in the purple region in Figure~\ref{fig:CS-NFWP} correspond to the positions of the four images of the source center in the bottom right panel of Figure~\ref{fig:images}. Note that the fifth macro-image lying on the critical curve in Figure~\ref{fig:images} does not include an image of the source center. The point appearing at $(\kappa,\gamma)\approx (0.637,1.255)$ corresponds to the image just to the right of the point mass in Figure~\ref{fig:images}. The three remaining points in Figure~\ref{fig:CS-NFWP} correspond to perturbed versions of the three images appearing in the absence of the point mass in the top right panel of Figure~\ref{fig:images}. The point at $(\kappa,\gamma)\approx (0.694,0.194)$ corresponds to the lower right image outside the critical curve in the bottom right panel of Figure~\ref{fig:images}. The point at $(\kappa,\gamma)\approx (0.897,0.233)$ corresponds to the image just outside the unit-convergence circle at the left side of the panel in Figure~\ref{fig:images}. The fourth point at $(\kappa,\gamma)\approx (1.962,0.244)$ corresponds to the small image close to the halo center in Figure~\ref{fig:images}.

The properties of the images can be determined from the positions of the points in the diagram; the changes in the properties of the latter three due to the presence of the point mass can be studied by comparing the diagrams in Figures~\ref{fig:CS-NFWP} and \ref{fig:CS-NFW}. Note that, in this case, taking into account the full extent of each image would require marking them in the CS diagram by exact patches covering the corresponding range of $(\kappa,\gamma)$ combinations instead of by the points used in Figure~\ref{fig:CS-NFWP}. This would permit including even partial images that do not contain an image of the source center, such as the fifth macro-image in the bottom right panel of Figure~\ref{fig:images}.

CS diagrams for different masses and positions of the point mass are presented and discussed in Section~\ref{sec:plots-diagrams}.

\subsection{Weak Shear and Phase}
\label{sec:NFWP-weak}

Following the example in Section~\ref{sec:NFW-weak}, we use the geometry of image distortions to introduce the weak-lensing shear and phase estimates for the NFW halo + point-mass lens. We compute the weak shear $\gamma_{\text{w}}$ from Equation~(\ref{eq:weak_shear}), substituting the convergence from Equation~(\ref{eq:NFWP_kappa}) for $\kappa(\boldsymbol x)$, and the shear from Equation~(\ref{eq:NFWP_gamma}) for $\gamma(\boldsymbol x)$.

The angle between the major axis of the image of a small circular source and the $x_1$ axis of the image plane is equal to the phase outside the unit-convergence circle; it is perpendicular to the phase inside the unit-convergence circle. Hence, the weak phase $\varphi_{\text{w}}$ is related to the phase $\varphi$ as follows:
\beq
\varphi_{\text{w}}(\boldsymbol x)=\Bigg\{
\begin{array}{ll}
\,\varphi(\boldsymbol x) & \,\text{for $\kappa(\boldsymbol x)<1$}\,,\\
\,\varphi(\boldsymbol x)+\pi/2 & \,\text{for $\kappa(\boldsymbol x)>1$}\,,
\end{array}
\label{eq:NFWP_weak_phase}
\eeq
where $\varphi(\boldsymbol x)$ is given by Equation~(\ref{eq:NFWP_phi}) and $\kappa(\boldsymbol x)$ by Equation~(\ref{eq:NFWP_kappa}).
We use values from the interval $\left[-\pi/2,\,\pi/2\right]$ for both $\varphi$ and $\varphi_{\text{w}}$.

Image-plane maps of the weak shear $\gamma_{\text{w}}(\boldsymbol x)$ and weak phase $\varphi_{\text{w}}(\boldsymbol x)$ for different masses and positions of the point mass are presented and discussed in Section~\ref{sec:plots-weak_shear} and Section~\ref{sec:plots-weak_phase}, respectively.

\subsection{Lens Characteristics as a Function of Point-mass Parameters}
\label{sec:NFWP-plots}

\begin{deluxetable}{Lp{7.5cm}}[t]
\tabletypesize{\footnotesize}
%\tablewidth{9cm}
\tablecaption{List of Symbols\label{tab:symbols_small}}
\tablehead{\colhead{\hspace{-0.3cm}Symbol} & \colhead{\hspace{-3.9cm}Description; First Appearance}}
\startdata
 \mkern-9mu\gamma & $\mkern-8mu$Shear (combined model or general); Equation~(\ref{eq:Jacobi_matrix}) \\
 \mkern-9mu\gamma_{\text{\tiny{NFW}}} & $\mkern-8mu$Shear of NFW-halo lens; Equation~(\ref{eq:NFW_gamma}) \\
 \mkern-9mu\gamma_{\text{P}} & $\mkern-8mu$Shear of a point-mass lens; Equation~(\ref{eq:P_gamma}) \\
 \mkern-9mu\gamma_{\text{w}} & $\mkern-8mu$Weak shear (combined model or general); Equation~(\ref{eq:weak_shear}) \\
 \mkern-9mu\gamma_{\text{w,\tiny{NFW}}} & $\mkern-8mu$Weak shear of NFW-halo lens; Section~\ref{sec:NFW-weak} \\
 \mkern-9mu\delta\varphi_\text{w} & $\mkern-8mu$Weak-phase deviation due to the point mass; Section~\ref{sec:plots-weak_phase_change} \\
 \mkern-9mu\kappa & $\mkern-8mu$Convergence (combined model or general); Equation~(\ref{eq:Jacobi_matrix}) \\
 \mkern-9mu\kappa_{\text{\tiny NFW}} & $\mkern-8mu$Convergence of NFW-halo lens; Equation~(\ref{eq:NFW_kappa}) \\
 \mkern-9mu\kappa_{\text{P}},\, \kappa_{\text{PC}} & $\mkern-8mu$Mass parameter of a point mass and its critical value; \\
  & $\mkern347mu$Equation~(\ref{eq:NFWP_kappa}) \\
 \mkern-9mu\kappa_{\text{s}} & $\mkern-8mu$NFW halo convergence parameter; Equation~(\ref{eq:NFW_kappa}) \\
 \mkern-9mu\varphi & $\mkern-8mu$Phase (combined model or general); Equation~(\ref{eq:Jacobi_matrix}) \\
 \mkern-9mu\varphi_{\text{\tiny NFW}} & $\mkern-8mu$Phase of NFW-halo lens; Equation~(\ref{eq:NFW_gamma12}) \\
 \mkern-9mu\varphi_{\text{w}} & $\mkern-8mu$Weak phase (combined model or general); Equation~(\ref{eq:NFWP_weak_phase}) \\
 \mkern-9mu\varphi_{\text{w,\tiny{NFW}}} & $\mkern-8mu$Weak phase of NFW-halo lens; Equation~(\ref{eq:weak_phase}) \\
 \mkern-9mu\psi & $\mkern-8mu$Lens potential (combined model or general); Equation~(\ref{eq:NFW_gradient}) \\
 \mkern-9mu\psi_{\text{\tiny NFW}} & $\mkern-8mu$Lens potential of NFW-halo lens; Equation~(\ref{eq:NFW_psi}) \\
 \mkern-9mu\omega & $\mkern-8mu$Viewing angle of the line segment connecting the halo center and the point-mass position; Equation~(\ref{eq:cos}) \\
\enddata
\end{deluxetable}

\renewcommand{\thefigure}{7.A}
\begin{figure*}
{\centering
\vspace{0cm}
\hspace{1cm}
\includegraphics[width=15.5 cm]{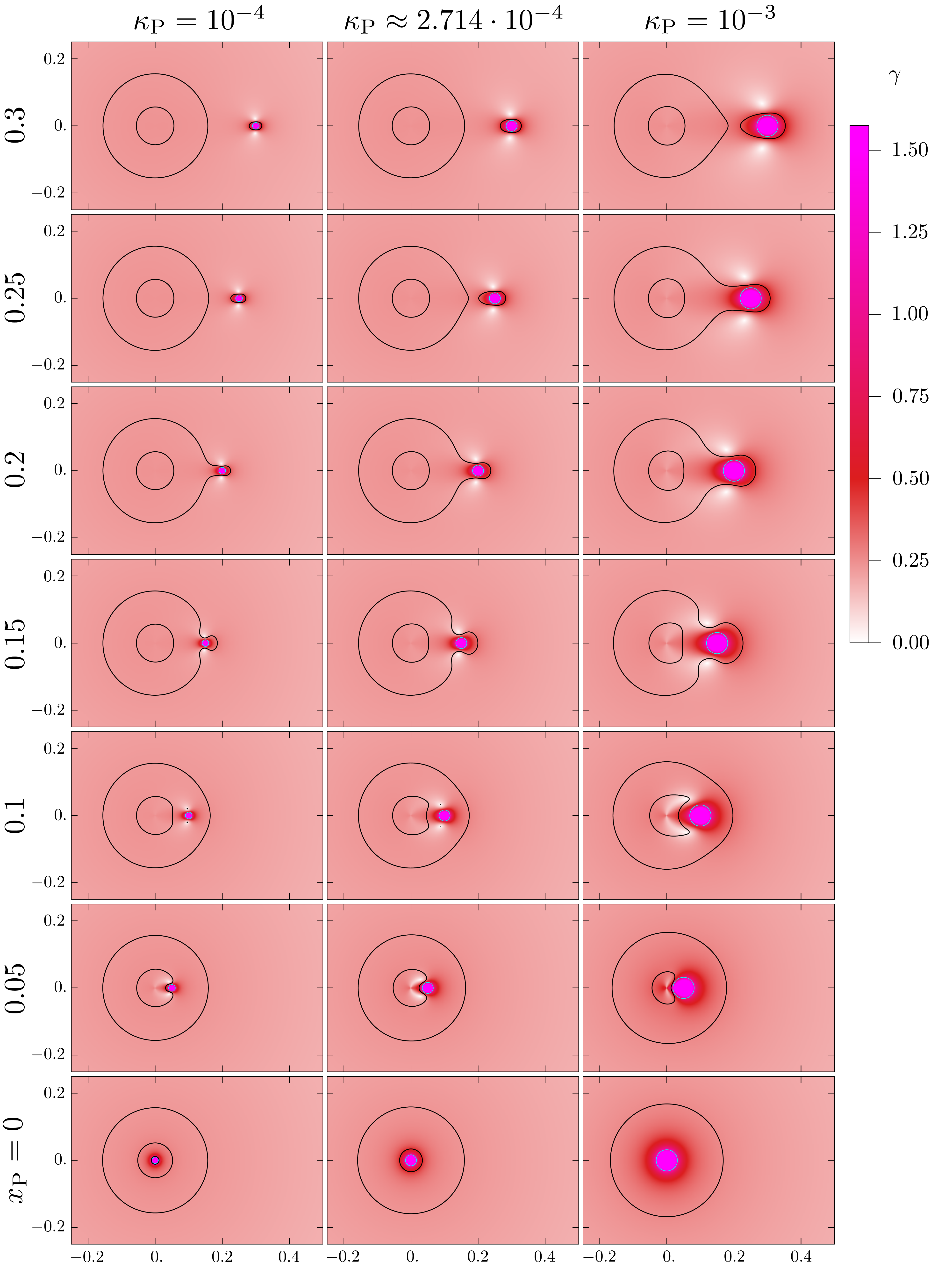}
\caption{Image-plane maps of the shear $\gamma(\boldsymbol x)$ of a NFW halo + point-mass lens, described in Section~\ref{sec:plots-shear}. Columns correspond to sub-critical, critical, and super-critical mass parameters $\kappa_{\text{P}}$ marked at the top; rows correspond to point-mass positions $x_{\text{P}}$ marked along the left side. Critical curves are plotted in black, and the point-mass location is marked by its Einstein ring (cyan). Magenta marks all positions with $\gamma\geq 1.5$.\label{fig:shear}}}
\end{figure*}

In Sections~\ref{sec:NFWP-csp}--\ref{sec:NFWP-weak} we defined the lensing quantities of interest and described their general properties. In this section we present plots illustrating these lens characteristics for different point masses embedded in a NFW halo with a fiducial convergence parameter $\kappa_{\text{s}}\approx 0.239035$. With the exception of Figure~\ref{fig:CS-diagrams} and Figure~\ref{fig:CS-diagrams-online}, all of the plots are presented as color maps in the image plane. For better orientation in these maps, we plot the critical curves (solid black) and mark the point-mass position by its Einstein ring (cyan).

In each of the following figures, the three columns of the plot grid correspond to the same three values of the mass parameter $\kappa_{\text{P}}$ of the point mass used in \citetalias{karamazov_etal21}. These differ in the number of radial critical curves they generate for $x_{\text{P}}=0$: sub-critical $\kappa_{\text{P}}=10^{-4}$ with two radial critical curves; critical $\kappa_{\text{P}}=\kappa_{\text{PC}}\approx 2.714 \cdot 10^{-4}$ with one radial critical curve; super-critical $\kappa_{\text{P}}=10^{-3}$ with no radial critical curve.

For each of the characteristics discussed in Sections~\ref{sec:plots-shear}--\ref{sec:plots-weak_phase_change} we present two plot grids. The rows in the first grid correspond to seven values of the point-mass position $x_{\text{P}}$ increasing in steps of $0.05$ from $0$ to $0.3$. These parameter combinations correspond to the critical-curve and caustic gallery in Figure~5 of \citetalias{karamazov_etal21}; they are marked by red crosses in the parameter-space plot in Figure~6 of \citetalias{karamazov_etal21}. The rows in the second grid correspond to nineteen values of $x_{\text{P}}$ increasing in steps of $0.01$ from $0$ to $0.15$, then in steps of $0.05$ to $0.3$ in the top row. These parameter combinations are marked by red and black crosses in Figure~6 of \citetalias{karamazov_etal21}.

For better orientation in the notation of the different shears, convergences, phases, and other lensing quantities, we list selected symbols together with their first appearance in the text in Table~\ref{tab:symbols_small}.

\subsubsection{Shear}
\label{sec:plots-shear}

\renewcommand{\thefigure}{7.B}
\bfi
{\centering
\vspace{0cm}
\hspace{1cm}
\includegraphics[height=22 cm]{f07_B.pdf}
\caption{Image-plane maps of the shear $\gamma(\boldsymbol x)$ of a NFW halo + point-mass lens, for a finer grid of point-mass positions than in Figure~\ref{fig:shear}. Notation same as in Figure~\ref{fig:shear}.\label{fig:shear-online}}}
\efi

Image-plane maps of the shear $\gamma(\boldsymbol x)$ are presented in Figure~\ref{fig:shear}. The shear color scale is the same as in the first two panels of Figure~\ref{fig:NFW-line}, ranging from white for $\gamma=0$ to magenta for all positions with $\gamma\geq 1.5$. In the absence of the point mass the shear varies very slowly in this region, as indicated by the featureless plot in the first panel of Figure~\ref{fig:NFW-line}.

In the bottom row ($x_{\text{P}}=0$) of Figure~\ref{fig:shear} the point mass is located at the center of the halo and the whole system thus exhibits axial symmetry. From Equations~(\ref{eq:NFWP_gamma})~and~(\ref{eq:cos}) it follows that in this case the total shear at any position is a simple sum of the NFW and point-mass shears. Near the halo center the NFW shear $\gamma_{\text{\tiny{NFW}}}(x)$, shown in the first panel of Figure~\ref{fig:NFW-line}, is surpassed by the point-mass shear $\gamma_{\text{P}}(\boldsymbol x)$ which diverges at the origin. A comparison of the bottom row in the three columns shows that the magenta high-shear region with the strongest point-mass influence naturally increases with its mass parameter $\kappa_{\text{P}}$.

Next, we focus on the left column illustrating the sub-critical case with $\kappa_{\text{P}}=10^{-4}$. Already in the second plot from the bottom ($x_{\text{P}}=0.05$), many phenomena described in detail in Section~\ref{sec:NFWP-csp} can be clearly seen. A pale circle corresponding to viewing angle $\omega=\pi/2$ connects the point mass and the halo center, marking a region with decreased shear. The shear drops to zero at two points of this circle located above and below the horizontal axis of symmetry. These zero-shear points lie inside the perturbed NFW radial critical curve close to the point mass. Their presence restricts the bright red high-shear region around the point-mass divergence to a smaller extent than in the $x_{\text{P}}=0$ case. The value of the shear in the vicinity of the halo center depends on the direction of approach. Maximum shear can be seen in the horizontal and minimum shear can be seen in the vertical direction, tangent to the $\omega=\pi/2$ circle.

Going further to $x_{\text{P}}=0.1$, we see that the paler lower-shear circle and the directional dependence near the origin become less pronounced, and the zero-shear points move even closer to the point mass, lying almost vertically above and below it. They are positioned inside the pair of tiny critical curves seen in the white low-shear areas. This hints at the fact that umbilics can only occur at zero-shear points that lie at the intersection of the $\omega=\pi/2$ circle and the $\kappa=1$ circle, as explained in Section~\ref{sec:NFWP-Jacobian}.

For even higher values of $x_{\text{P}}$ in the sub-critical case, the directional dependence at the halo center becomes indiscernible and the pattern close to the point mass becomes more regular. The region of high shear around the point mass has a horizontally elongated oval shape and the zero-shear points lie above and below it in the white spots outside the critical curve. With increasing $x_{\text{P}}$, this shear pattern around the point mass resembles the total shear of the Chang--Refsdal model, which consists of a point mass and a constant external shear \citep{chang_refsdal84}.

The preceding discussion made for the sub-critical case holds also for the critical case \mbox{($\kappa_{\text{P}}=\kappa_{\text{PC}}\approx 2.714 \cdot 10^{-4}$)} in the central column. In terms of shear, the corresponding plots portray qualitatively the same sequence of situations as in the left column; the difference is merely quantitative. More specifically, the pattern around the point mass is considerably larger and the directional dependence around the halo center is more pronounced, observable even for higher values of $x_{\text{P}}$.

These patterns are even larger and more distinct in the right column illustrating the super-critical case with $\kappa_{\text{P}}=10^{-3}$. However, there are some important differences. In the second plot from the bottom, \mbox{$x_{\text{P}}=0.05$} does not exceed the threshold value of \mbox{$\sqrt{\kappa_{\text{P}}/\kappa_{\text{s}}}\approx 0.0647$}. Hence, there are no zero-shear points. Minimum shear, which is now non-zero in the central region, can be found at the halo center when approached vertically. For point-mass positions \mbox{$x_{\text{P}}\gtrsim 0.0647$} this minimum shear drops to zero and its position detaches from the origin, moving along the $\omega=\pi/2$ circle. In addition, the third plot from the bottom now depicts the situation before the detachment of the two small critical curves and, thus, the zero-shear points still lie inside the perturbed NFW radial critical curve. Note that in this case the directional dependence of the shear at the halo center can be seen up to the top row.

The dependence of the shear $\gamma(\boldsymbol x)$ on the point-mass position $x_{\text{P}}$ can be examined in more detail in Figure~\ref{fig:shear-online}, which includes plots for a finer grid in terms of $x_{\text{P}}$. Zero-shear points appear at the halo center at point-mass positions $\{0.0205, 0.0337, 0.0647\}$ in the sub-critical, critical, and super-critical cases. Other key values of $x_{\text{P}}$ correspond to changes in the critical-curve topology, as indicated by the color boundaries in Figure~6 of \citetalias{karamazov_etal21}.

Notice that the zero-shear points always occur in a positive-Jacobian region, as indicated by Equation~(\ref{eq:Jacobian}): inside the perturbed NFW radial critical curve, inside the symmetric pair of small critical-curve loops, or outside all critical-curve loops. Apart from this, there is hardly any correlation between the shear pattern and the critical-curve geometry.

\subsubsection{Shear Deviation Due to the Point Mass}
\label{sec:plots-shear_deviation}

In Figure~\ref{fig:shear_deviation_perturbation}, we present image-plane maps of the shear deviation $\gamma/\gamma_{\text{\tiny{NFW}}}-1$ caused by the presence of the point mass. As indicated by the formula, this quantity represents the relative difference between the shear $\gamma(\boldsymbol x)$ of the NFW halo + point-mass lens (shown in Figure~\ref{fig:shear}) and the shear $\gamma_{\text{\tiny{NFW}}}(x)$ of the NFW halo alone (shown in the first panel of Figure~\ref{fig:NFW-line}). As the deviation falls rather quickly with increasing distance from the point mass, we introduce a semi-logarithmic color scale to visualize even minor changes in the deviation. In the positive yellow- and orange-hued regions the shear is increased, while in the negative blue regions it is decreased by the point mass. Darkest blue is used for $-1$, the lowest possible deviation. From $-1$ to $-10^{-3}$, blue saturation decreases logarithmically, and then to $0$ linearly, where it reaches white. From $0$ to $10^{-3}$, yellow saturation increases linearly, and then to $10^{-1}$ logarithmically. The logarithmic scale then continues to color red at deviation $10$, beyond which the color is kept constant even though the shear deviation can reach arbitrarily large values near the point mass.

For better orientation, we also include contours for a few specific values of the shear deviation. The dot-dashed lines represent the zero-deviation contour, along which the shears are equal. Paler and darker shades of orange are used for positive-deviation contours with values $10^{-2}$ and $10^{-1}$, respectively. Similarly, paler and darker shades of blue indicate negative deviations $-10^{-2}$ and $-10^{-1}$, respectively.

We first inspect the deviation map for a centrally positioned sub-critical point mass (bottom left plot). In this case the deviation is equal to $\gamma_{\text{P}}/\gamma_{\text{\tiny{NFW}}}$, which is positive in the entire image plane, i.e., the shear is globally increased by the point mass. The deviation diverges at the halo center, since the point-mass shear increases to $\infty$ while the halo shear tends to the constant $\kappa_\text{s}$. Further from the halo center the deviation approaches zero, as the point-mass shear given by Equation~(\ref{eq:P_gamma}) falls quickly with distance. Contours representing deviations $10^{-1}$ and $10^{-2}$ are slightly larger than the outer radial and tangential critical curves, respectively.

In the second row ($x_{\text{P}}=0.05$) the point mass is displaced from the center and a pair of blue regions with negative deviation appears. These regions reach the halo center from the vertical direction, while the deviation is positive along the full horizontal axis, as indicated by the orange and yellow color and by the dot-dashed zero contour pinched at the halo center. In fact, the deviation along the horizontal axis is always positive for any $x_\text{P}$ and $\kappa_\text{P}$, since here $\gamma/\gamma_{\text{\tiny{NFW}}}-1=\gamma_\text{P}/\gamma_{\text{\tiny{NFW}}}$ according to the discussion in the paragraphs following Equation~(\ref{eq:NFWP_gamma}). The pattern near the halo center arises from the directional dependence of the shear shown in Equation~(\ref{eq:NFWP_gamma_center}). Places with the darkest blue color lie above and below the point mass, with the lowest shear deviation $-1$ occurring at the zero-shear points. The region with a deviation larger than $10\%$ in absolute value is now roughly centered on the point mass, while the region with a deviation lower than $1\%$ in absolute value lies outside the near-circular pale orange contour and in narrow bands along the dot-dashed zero-deviation contour.

At $x_{\text{P}}=0.1$ on the third row we see that the affected area becomes more asymmetric, with the pale orange contour with $10^{-2}$ deviation now broken into two lobes extending to the left of the center and to the right of the point mass. In this case there is a single region with deviation lower than $1\%$ in absolute value, reaching inside the critical curves, including the zero-deviation contour, and reaching the halo center along it.

For higher values of $x_{\text{P}}$, the blue regions of negative deviation expand as the point mass shifts to the right. Their borders indicated by the zero-deviation contour become more and more circular except for the vicinity of the point mass, where they enclose the zero-shear points but avoid the vicinity of the point mass. These dot-dashed contours intersect at the halo center at a right angle and the deviation remains positive in the spindle-shaped region along the horizontal axis from the halo center to the point mass. The orange and blue contours gradually detach from the halo center and for $x_{\text{P}}\geq 0.25$ they form a four-lobed structure around the point mass, with positive lobes extending horizontally from the point mass and negative lobes separated vertically from the point mass. The single region with deviation lower than $1\%$ in absolute value includes the halo center as well as a progressively larger area around it, including the entire image plane except the four lobes around the point mass.

In the critical and super-critical cases in the two right columns, the plots look similar to those in the sub-critical case, with the colors getting progressively more saturated indicating higher shears $\gamma_\text{P}$ from heavier point masses. Naturally, the orange and blue contours also expand with increasing mass. On the other hand, the blue regions of negative deviation inside the dot-dashed contours do not expand with increasing $\kappa_{\text{P}}$. On the contrary, they shrink as they recede from a heavier point mass. Away from the point mass, the dot-dashed zero contours are almost circular. They intersect at the halo center at a right angle and reach nearly to the point mass before avoiding it.

Comparing the columns in the different rows, we see that the geometry of the zero-deviation contour is generic, with increasing $\kappa_{\text{P}}$ affecting only the vicinity of the point mass, and increasing $x_{\text{P}}$ only enlarging the scale. The pattern arises naturally from Equation~(\ref{eq:NFWP_gamma}) in the regime $\gamma_{\text{P}}\ll\gamma_{\text{\tiny{NFW}}}$ valid anywhere except in the immediate vicinity of the point mass. In this case we expand the shear and get the simple result $\gamma/\gamma_{\text{\tiny{NFW}}}-1\simeq (\gamma_\text{P}/\gamma_{\text{\tiny{NFW}}})\,\cos{2\,\omega}$. The ratio in the parentheses is always positive, hence, the zero-deviation contour is purely given by the condition on the viewing angle requiring $\cos{2\,\omega}=0$. Figure~\ref{fig:omega} shows that the corresponding $\omega=\pi/4$ and $\omega=3\pi/4$ circles describe the dot-dashed contours seen in Figure~\ref{fig:shear_deviation_perturbation} practically exactly, except in the point-mass vicinity where $\gamma_{\text{P}}\gtrsim\gamma_{\text{\tiny{NFW}}}$.

In the bottom row of the critical and super-critical columns we see the same pattern of globally positive shear deviation as in the sub-critical column. What differs is the larger extent of the orange contours. In fact, in the super-critical case the entire paler-orange contour of deviation $10^{-2}$ lies outside the plotted area. Within the plotted area in all other panels in the right column, the regions with a deviation lower than $1\%$ in absolute value are limited to a band along the zero-deviation contour. This band expands as $x_{\text{P}}$ increases, and eventually connects with the outer low-deviation region. For $x_{\text{P}}=0.05$ in the super-critical case we can see very small blue regions of negative deviation without zero-shear points inside, as these appear at a higher separation, for $x_{\text{P}}\geq 0.0645$. Moreover, here the angle of intersection of the dot-dashed contours is very different from a right angle. In this case the influence of the point mass at the halo center is too strong ($\gamma_\text{P}/\gamma_{\text{\tiny{NFW}}}\approx 1.7$), so that the expansion illustrating the generic shape of the zero-deviation contour is not valid here.

\renewcommand{\thefigure}{8.A}
\begin{figure*}
{\centering
\vspace{0cm}
\hspace{1cm}
\includegraphics[width=15.5 cm]{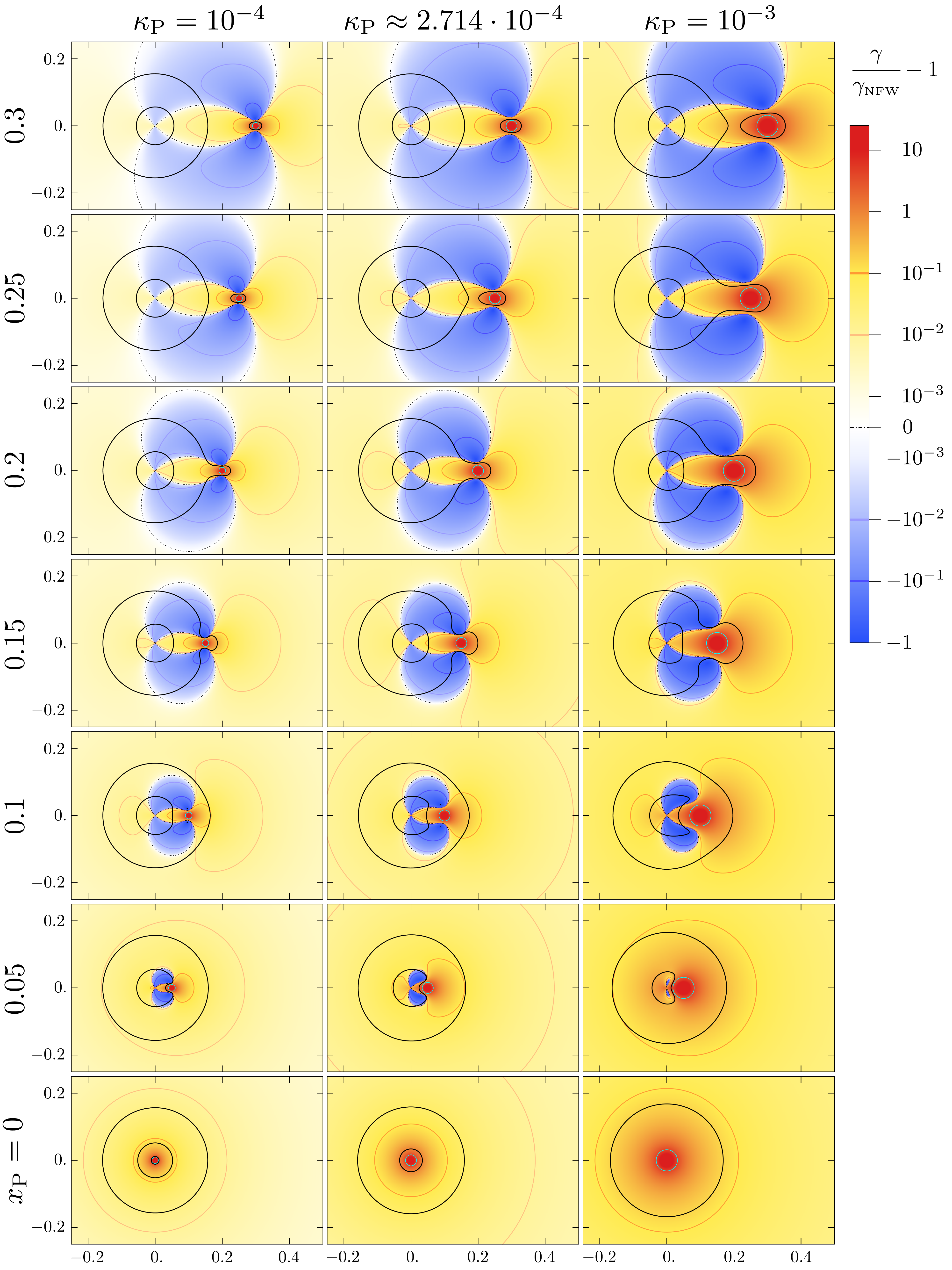}
\caption{Image-plane maps of the relative shear deviation, $\gamma/\gamma_{\text{\tiny{NFW}}}-1$, caused by the presence of the point mass, described in Section~\ref{sec:plots-shear_deviation}. The color scale changes from logarithmic to linear in the interval $[-10^{-3},10^{-3}]$. All positions with deviation greater than $10$ are marked in red. Contours are plotted for five deviation values indicated in the color bar. Remaining notation as in Figure~\ref{fig:shear}. \label{fig:shear_deviation_perturbation}}}
\end{figure*}

A more detailed view of the changing deviation patterns with point-mass position can be seen in Figure~\ref{fig:shear_deviation_perturbation-online}. Its closer inspection reveals that the deviation is globally positive not only for centrally-positioned point masses, but also for plots up to $x_{\text{P}}=0.01$ in the sub-critical case, up to $0.02$ in the critical case, and up to $0.04$ in the super-critical case. Imposing the condition $\gamma\leq\gamma_{\text{\tiny{NFW}}}$ on Equation~(\ref{eq:NFWP_gamma}) reveals that negative deviation first appears at the halo center in the $\omega=\pi/2$ vertical direction once the point-mass shear at the center decreases to $\gamma_\text{P}=2\,\gamma_{\text{\tiny{NFW}}}$. Using Equation~(\ref{eq:NFW_gamma_origin}) and Equation~(\ref{eq:P_gamma}) with $\boldsymbol{x}=(0,0)$ then yields the condition for the existence of negative-deviation regions: $x_{\text{P}}\geq\sqrt{\kappa_{\text{P}}/(2\,\kappa_{\text{s}})}$. In the sub-critical case we find $x_{\text{P}}\geq 0.0145$, in the critical $x_{\text{P}}\geq 0.0238$, and in the super-critical $x_{\text{P}}\geq 0.0457$, in agreement with the deviation maps.

The sizes of the contours can be used to estimate the areas with a strong effect on the shear due to the presence of the point mass. As an example, for the three different masses we find that at the moment of separation of the critical curve surrounding the point mass from the perturbed NFW tangential critical curve, the darker contours of deviation $\pm\,10^{-1}$ extend roughly seven Einstein radii from the point mass.

\subsubsection{Convergence--Shear Diagrams}
\label{sec:plots-diagrams}

\renewcommand{\thefigure}{8.B}
\bfi
{\centering
\vspace{0cm}
\hspace{1cm}
\includegraphics[height=22 cm]{f08_B.pdf}
\caption{Image-plane maps of the relative shear deviation, $\gamma/\gamma_{\text{\tiny{NFW}}}-1$, caused by the presence of the point mass, for a finer grid of point-mass positions than in Figure~\ref{fig:shear_deviation_perturbation}. Notation same as in Figure~\ref{fig:shear_deviation_perturbation}.\label{fig:shear_deviation_perturbation-online}}}
\efi

In Figure~\ref{fig:CS-diagrams} we present a grid of CS diagrams, which provide a description complementary to the image-plane plots of the shear, its deviation due to the point mass, and the quantities discussed in the following sections. For a general understanding of CS diagrams see Appendix~\ref{sec:Appendix-images} with Figure~\ref{fig:CS-diagram}, Section~\ref{sec:NFW-images} with Figure~\ref{fig:CS-NFW}, and, in particular, Section~\ref{sec:NFWP-images} with Figure~\ref{fig:CS-NFWP}.

The purple-shaded area marks the full range of $(\kappa,\gamma)$ combinations of the NFW halo + point-mass lens. Its intersection with the green curve corresponds to the dot-dashed zero-deviation curve in Figure~\ref{fig:shear_deviation_perturbation}. The part of the area above the green curve then corresponds to the yellow- and orange-hued positive-deviation regions, and the part below the green curve corresponds to the blue negative-deviation regions in Figure~\ref{fig:shear_deviation_perturbation}.

We start by describing the sub-critical case shown in the left column. For $x_{\text{P}}=0$ the point mass lies at the center of the halo and the system therefore has axial symmetry. In addition, the radial dependence of the convergence is monotonic. This implies that only one value of the shear $\gamma(x)$ can occur for any value of the convergence $\kappa(x)$. These combinations $\left(\kappa(x),\gamma(x)\right)$ are plotted here as the bold purple curve. For positions far from the halo center (at the left side of the plot), this curve closely follows the unperturbed-halo green curve, which starts at the origin of the plot. Proceeding to the right (closer to the halo center), the purple curve reaches the bold black line with slope $-1$ representing the tangential critical curve. Here the magnification is infinite and the flattening reaches $1$.

\renewcommand{\thefigure}{9.A}
\begin{figure*}
{\centering
\vspace{0cm}
\hspace{1cm}
\includegraphics[width=14 cm]{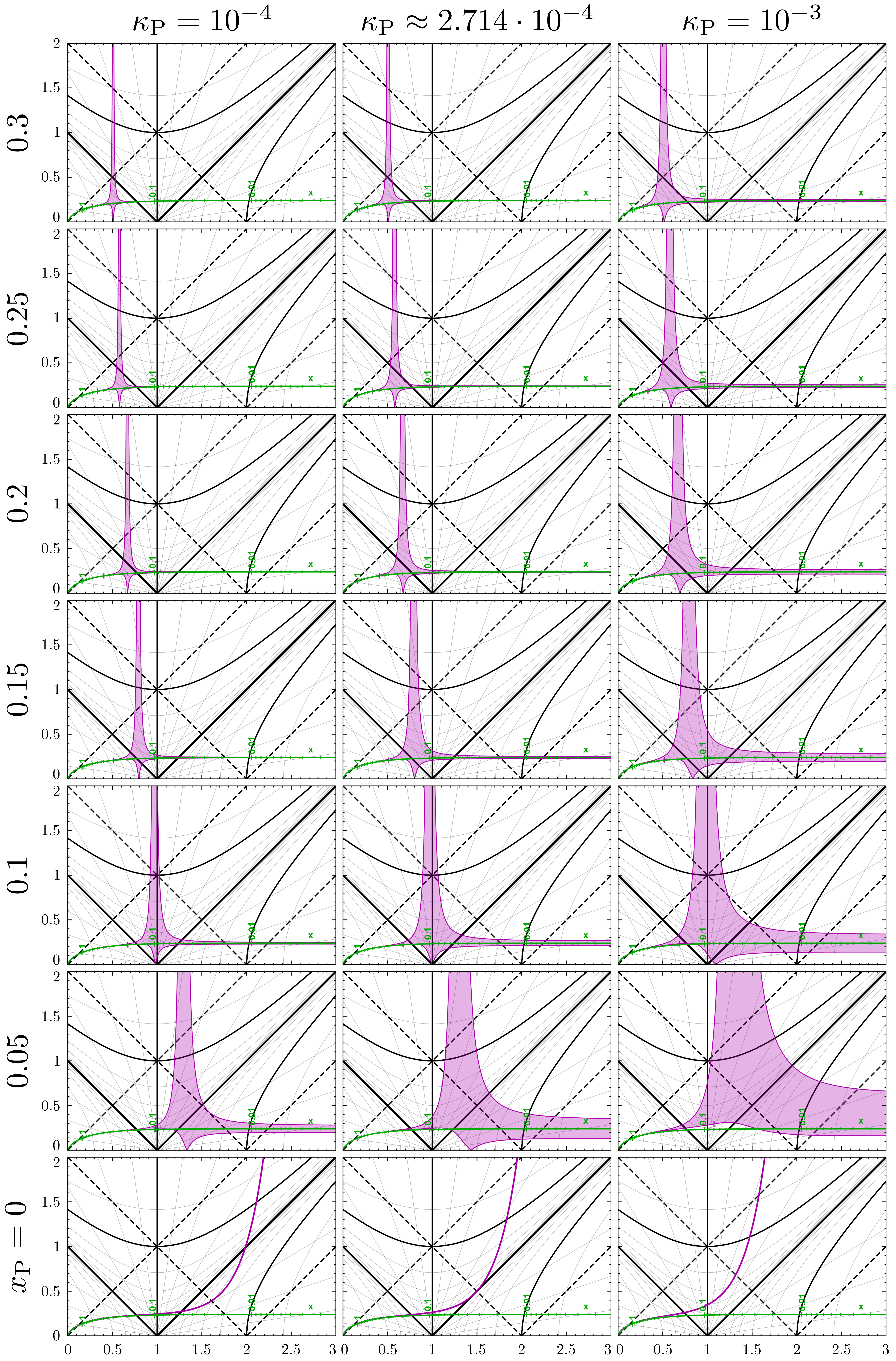}
\caption{Convergence--shear (CS) diagrams of a NFW halo + point-mass lens, described in Section~\ref{sec:plots-diagrams}. Combinations $\left(\kappa(\boldsymbol x),\gamma(\boldsymbol x)\right)$ occurring in each of the lens configurations are marked by the purple regions or curves. The green curve corresponds to the NFW-halo lens from Figure~\ref{fig:CS-NFW}. For further details on the notation see Figure~\ref{fig:CS-NFWP}; for the interpretation of CS diagrams see Figure~\ref{fig:CS-diagram}.\label{fig:CS-diagrams}}}
\end{figure*}

Further to the right, the curve enters an area of negative parity, where the magnification decreases and the flattening drops to $0$ at the bold vertical $\kappa=1$ line corresponding to the unit-convergence radius $x_0\approx0.0936$. To the right of this line, images are elongated perpendicularly to the phase. Roughly here, the purple curve of $\left(\kappa(x),\gamma(x)\right)$ combinations starts to deviate significantly from the green curve. It rises rapidly and eventually leaves the plot, as the shear diverges at the location of the point mass. Close to the point mass the parity is always negative and both magnification and flattening approach zero. Before this happens, the purple curve intersects the bold black line with slope $1$ twice. The first intersection represents the outer radial critical curve and the second intersection represents the inner radial critical curve. Between them, images have positive parity.

Increasing the point-mass position from $x_{\text{P}}=0$ to $x_{\text{P}}=0.05$ brings about several important changes. As the system loses its axial symmetry, for each value of convergence $\kappa(x)$ there is a continuous interval of shear values $\gamma(\boldsymbol x)$ in the image plane and the set of convergence--shear combinations is represented by a two-dimensional region. At the left side of the plot, far from the halo center, these combinations remain limited to the close vicinity of the green curve of the unperturbed halo. At the right side, close to the halo center, the set of combinations forms a horizontal band, with shear values $\gamma\in[\gamma_{\text{\tiny{NFW}}}-\gamma_{\text{P}}, \gamma_{\text{\tiny{NFW}}}+\gamma_{\text{P}}]$ corresponding to its directional dependence at the halo center, demonstrated by Equations~(\ref{eq:NFWP_gamma}) and (\ref{eq:NFWP_gamma_center}). In the vicinity of $x=x_\text{P}$ we see the broadest range of shear values. On the one hand, the shear diverges at the position of the point mass while on the other hand, the shear drops to $0$ at the zero-shear points occurring here at a radius slightly lower than $x_\text{P}$. Overall, the set of $\left(\kappa(\boldsymbol x),\gamma(\boldsymbol x)\right)$ combinations looks similar to the case illustrated in Figure~\ref{fig:CS-NFWP} and described in detail in Section~\ref{sec:NFWP-images}, with one important difference. The shear divergence and the zero-shear points both lie in the area of $\kappa>1$, meaning that nearby images would now be elongated perpendicularly to the phase.

\renewcommand{\thefigure}{9.B}
\bfi
{\centering
\vspace{0cm}
\hspace{1cm}
\includegraphics[height=22 cm]{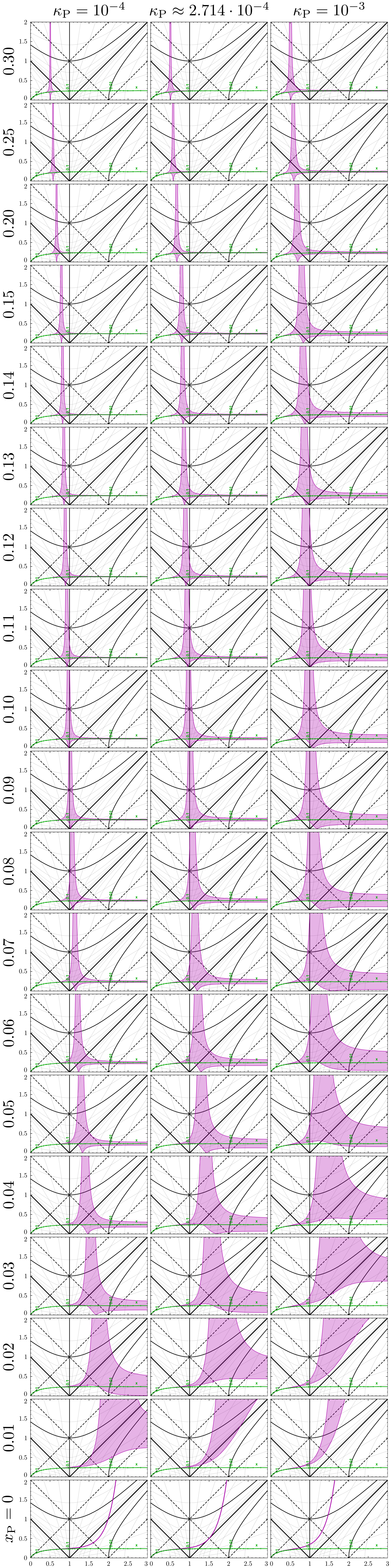}
\caption{Convergence--shear (CS) diagrams of a NFW halo + point-mass lens, for a finer grid of point-mass positions than in Figure~\ref{fig:CS-diagrams}. Notation same as in Figure~\ref{fig:CS-diagrams}.\label{fig:CS-diagrams-online}}}
\efi

\renewcommand{\thefigure}{10.A}
\begin{figure*}
{\centering
\vspace{0cm}
\hspace{1cm}
\includegraphics[width=15.5 cm]{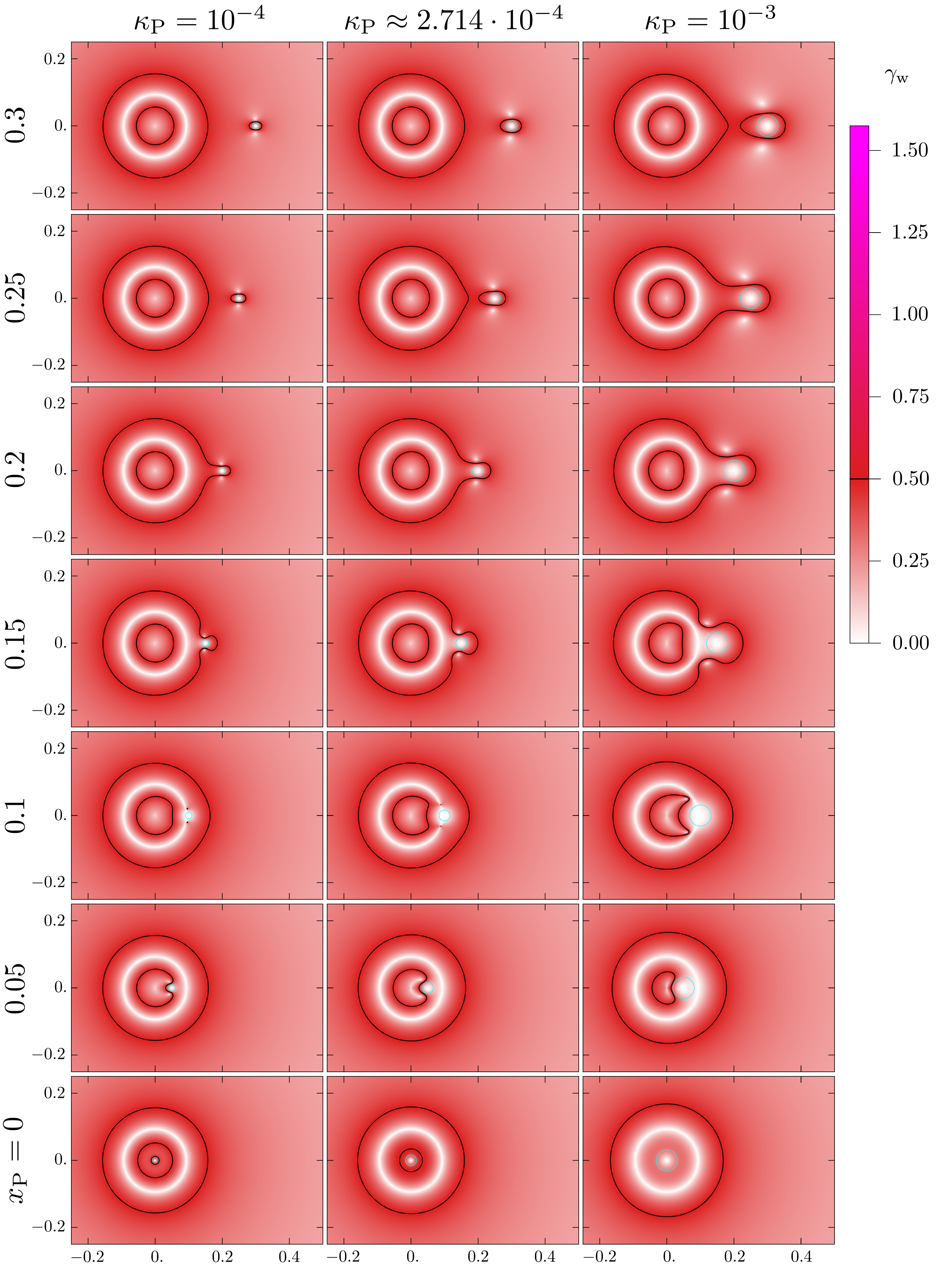}
\caption{Image-plane maps of the weak shear $\gamma_{\text{w}}(\boldsymbol x)$ of a NFW halo + point-mass lens, described in Section~\ref{sec:plots-weak_shear}. These maps also illustrate image flattening with values $f(\boldsymbol x)=2\,\gamma_{\text{w}}(\boldsymbol x)$. Maximum weak shear $\gamma_{\text{w}}=0.5$ occurs exclusively along the critical curves. Remaining notation as in Figure~\ref{fig:shear}. \label{fig:weak_shear}}}
\end{figure*}

As $x_{\text{P}}$ increases in the following rows, both the shear divergence and the zero-shear point shift to the left, indicating that the point mass moves to locations with progressively lower halo convergence. By the third row from the bottom ($x_{\text{P}}=0.1$), the purple region touches the horizontal axis to the left of the vertical $\kappa=1$ line, indicating that the lens underwent an umbilic transition at a slightly lower $x_{\text{P}}$ value. The surroundings of the shear divergence and the zero-shear point now lie in the region $\kappa<1$, where images are elongated parallel to the phase. The purple regions in the diagram also become narrower with increasing $x_{\text{P}}$. In the case of the divergence this is due to the changing scale of CS diagrams in terms of image-plane positions, as indicated by the green ticks. In the case of the horizontal band this is due to the decreasing value of $\gamma_{\text{P}}(0)$, i.e., the shear due to the point mass at the halo center.

There are a few differences to notice in the critical case, which is shown in the central column of Figure~\ref{fig:CS-diagrams}. For a centrally positioned point mass ($x_{\text{P}}=0$), the purple curve of the $\left(\kappa(x),\gamma(x)\right)$ combinations deviates from the green curve of halo combinations similarly as it does in the sub-critical case. However, instead of intersecting the bold black line with slope $1$, the curve merely touches it at a single point. This indicates the disappearance of the inner positive-parity region and the presence of a degenerate radial critical curve, described in detail in Appendix~B of \citetalias{karamazov_etal21}. The plot in the third row from the bottom ($x_{\text{P}}=0.1$) depicts a configuration extremely close to the elliptic umbilic, as the zero-shear points now occur almost precisely at $(\kappa,\gamma)=(1,0)$. Generally speaking, the purple regions of convergence--shear combinations are broader than in the sub-critical case, meaning that at a given distance from the halo center a greater range of shear values occurs.

For the super-critical mass in the right column, in the axially symmetric $x_{\text{P}}=0$ case the purple curve of the $\left(\kappa(x),\gamma(x)\right)$ combinations does not touch the black line with slope $1$ at all. This means that there are no radial critical curves. The second plot from the bottom ($x_{\text{P}}=0.05$) is now profoundly different than in previous cases. Here, the purple area does not touch the horizontal axis of the plot. As discussed in Section~\ref{sec:plots-shear}, in this case there are no zero-shear points and minimum shear can be found at the halo center when approached vertically. Moreover, unlike in the lower-mass cases, at the halo center the horizontal band does not spread symmetrically around the green line. In this case, at the center the shear due to the point mass is higher than the shear due to the halo, so that $\gamma\in[\gamma_{\text{P}}-\gamma_{\text{\tiny{NFW}}}, \gamma_{\text{P}}+\gamma_{\text{\tiny{NFW}}}]$ as shown by Equation~(\ref{eq:NFWP_gamma_center}) and discussed in the following paragraphs and in Section~\ref{sec:NFWP-images}.

\renewcommand{\thefigure}{10.B}
\bfi
{\centering
\vspace{0cm}
\hspace{1cm}
\includegraphics[height=22 cm]{f10_B.pdf}
\caption{Image-plane maps of the weak shear $\gamma_{\text{w}}(\boldsymbol x)$ of a NFW halo + point-mass lens, for a finer grid of point-mass positions than in Figure~\ref{fig:weak_shear}. Notation same as in Figure~\ref{fig:weak_shear}.\label{fig:weak_shear-online}}}
\efi

The variation of the CS diagrams with point-mass position can be inspected in more detail in Figure~\ref{fig:CS-diagrams-online}. Note that the shear interval at the halo center is centered on $\gamma_{\text{P}}$ rather than on $\gamma_{\text{\tiny{NFW}}}$ in all cases with $x_{\text{P}}<\sqrt{\kappa_{\text{P}}/\kappa_{\text{s}}}$, namely: from $x_\text{P}=0$ to $x_\text{P}=0.02$ in the sub-critical case; to $x_\text{P}=0.03$ in the critical case; to $x_\text{P}=0.06$ in the super-critical case. The $x_\text{P}=0.01$ diagrams illustrate the nature of the transition from the axially symmetric lens configurations at $x_\text{P}=0$, best seen in the right panel. The bold purple curve from $x_\text{P}=0$ gradually expands to a broader band at lower radii for $x_\text{P}=0.01$. The remaining structure of the purple region lies outside the plotted area in the right panel, but it has a similar nature to the plot in the left panel (also similar to the right panels for higher $x_\text{P}$ values): shear divergence at the lens position $(x_{\text{P}}=0.01)$, and the shear interval shrinking to a horizontal band centered on $\gamma_{\text{P}}$.

Another feature to notice for the lower $x_\text{P}$ values is that the entire purple region lies above the green curve, which means that for such configurations the shear is higher than in the absence of the point mass everywhere in the image plane. As shown in Section~\ref{sec:plots-shear_deviation}, this is the case for $x_{\text{P}}<\sqrt{\kappa_{\text{P}}/(2\,\kappa_{\text{s}})}$, namely: plots up to $x_\text{P}=0.01$ in the sub-critical case; up to $x_\text{P}=0.02$ in the critical case; up to $x_\text{P}=0.04$ in the super-critical case.
For higher values of $x_{\text{P}}$ the shear may be lower than in the absence of the point mass, but only in limited parts of the image plane. For example, along the horizontal axis of the lens the shear always stays higher than in the absence of the point mass, as shown in Section~\ref{sec:plots-shear_deviation}. For lower masses at higher separations $x_\text{P}$, the generic geometry of the zero-deviation contour discussed in Section~\ref{sec:plots-shear_deviation} shows that the regions of lower shear are limited to radial distances $x<x_\text{P}\,\sqrt{2}$.

\subsubsection{Weak Shear}
\label{sec:plots-weak_shear}

The plots in Figure~\ref{fig:weak_shear} show image-plane color maps of the weak shear $\gamma_{\text{w}}(\boldsymbol x)$, which we defined in Section~\ref{sec:NFW-weak} as the shear that would be measured from image deformations using weak-lensing analysis. At the same time, these maps illustrate image flattening, since $f(\boldsymbol x)=2\,\gamma_{\text{w}}(\boldsymbol x)$, as shown in Equation~(\ref{eq:weak_shear}). This equality also implies that weak shear values range from $0$ (corresponding to white color in the maps) to $0.5$ (bright red color), with the maximum value occurring exclusively along the full length of the critical curve. In spite of this limited range, for purposes of comparison we retain the same color scale as in the first two panels of Figure~\ref{fig:NFW-line} and in Figure~\ref{fig:shear}.

We begin our description in the left column of Figure~\ref{fig:weak_shear} with a sub-critical point mass, starting from the bottom row corresponding to its central position in the NFW halo. Directly at the center the weak shear is equal to zero, since the image cannot be flattened in any direction due to the axial symmetry of the lens configuration. Going further from the halo center, the weak shear reaches $0.5$ at the small inner radial critical curve, then it decreases slightly before returning to $0.5$ at the outer radial critical curve. The weak shear then drops to zero at the unit-convergence circle, along which images are undistorted, increasing again to $0.5$ at the tangential critical curve, beyond which it drops asymptotically to zero.

In the second row ($x_{\text{P}}=0.05$), the outer red ring of high weak shear along the perturbed NFW tangential critical curve is preserved. The same holds for the white $\gamma_{\text{w}}=0$ ring along the unit-convergence circle, which is in fact preserved exactly in all the configurations for all point masses. However, several changes can be seen closer to the halo center, where there are now four points with zero weak shear. These points coincide with the halo center, the point-mass position, and the pair of zero-shear points. This can be understood by inspecting Equation~(\ref{eq:flattening}) and taking its limit at these points. At the halo center, the difference of signs before $\gamma$ is suppressed by diverging $\kappa$, which also suppresses the directional dependence of the shear close to the center. Both fractions in Equation~(\ref{eq:flattening}) tend to $1$, which results in zero flattening. At the position of the point mass, both numerators and denominators are dominated by the diverging shear and the fractions thus again approach unity. At the zero-shear points, the fractions for $\gamma=0$ are directly equal to $1$. It is worth noting here that zero shear $\gamma$ implies zero weak shear $\gamma_{\text{w}}$, but not vice versa. All four points are connected by a paler low-weak-shear region corresponding to the $\omega=\pi/2$ circle, which is interrupted between the zero-shear points and the point-mass position by the perturbed NFW radial critical curve, along which the weak shear reaches $0.5$.

At $x_{\text{P}}=0.1$ the region of low weak shear surrounding the point mass is superimposed over the white ring of the unit-convergence circle, while the pair of $\gamma_{\text{w}}=0$ zero-shear points is now trapped inside tiny critical curves, here very close to the elliptic umbilic transition. For higher values of $x_{\text{P}}$, a lobe forms on the perturbed NFW tangential critical curve with the point mass inside and the zero-shear points outside. By $x_{\text{P}}=0.25$ the curve splits and the point mass is surrounded by a small oval critical-curve loop elongated toward the halo center. Three points with zero weak shear remain associated with this loop: the point-mass position inside, and the two zero-shear points directly above and below the loop. Similar but larger weak-shear patterns near the tangential-critical-curve lobes and the detached ovals can be seen in the two right columns showing the critical and super-critical cases, in the rows with $x_{\text{P}}\geq 0.15$.

In the critical case with a centrally located point mass (bottom plot in central column), there is only one $\gamma_{\text{w}}=0.5$ circle along the single radial critical curve between the white center and the white $\kappa=1$ circle. The third row ($x_{\text{P}}=0.1$) illustrates the peculiar situation in the presence of elliptic umbilic points (technically, these occur at $x_{\text{P}}\approx0.0996$, but the plot is visually identical). These are zero-shear points lying directly on the unit-convergence circle. At these points the weak shear (and flattening) is undefined, as can be seen by substituting $\gamma=0$ and $\kappa=1$ in Equation~(\ref{eq:flattening}). In the plot we can see them as point-like interruptions of the white unit-convergence circle.

In the super-critical case with $x_{\text{P}}=0$ (bottom right plot) there is no radial critical curve. Hence, the weak shear increases only slightly between the center and the $\kappa=1$ circle without reaching $0.5$. In the $x_{\text{P}}=0.05$ plot the pair of vertically offset zero-weak-shear points is missing, since the condition $x_{\text{P}}>\sqrt{\kappa_{\text{P}}/\kappa_{\text{s}}}$ for the existence of zero-shear points is not fulfilled here. Note in the same plot that the weak-shear pattern near the halo center resembles the central directionally dependent shear pattern. However, while the shear is undefined at the center, the weak shear reaches $0$ from any direction, with only the rate of convergence depending on the direction. The rate is slowest in the horizontal direction with $\gamma_{\text{w}}\sim(\gamma_{\text{P}}+\gamma_{\text{\tiny{NFW}}})/(\kappa-1)$, and fastest in the vertical direction with $\gamma_{\text{w}}\sim|\gamma_{\text{P}}-\gamma_{\text{\tiny{NFW}}}|/(\kappa-1)$ to first order in $1/(\kappa-1)$. Clearly, the directionality will be most pronounced when $\gamma_{\text{P}}=\gamma_{\text{\tiny{NFW}}}$ at the halo center, i.e., at the appearance of the zero-shear points. The directionality will be least pronounced when either of the component shears $\gamma_{\text{\tiny{NFW}}}, \gamma_{\text{P}}$ dominates over the other at the halo center.

The emergence and evolution of the features and structures discussed above can be studied in more detail in Figure~\ref{fig:weak_shear-online}. For example, the formation of the tiny critical-curve loops around the zero-(weak)-shear points is well visible in the sub-critical case at $x_{\text{P}}=0.08$ or in the critical case at $x_{\text{P}}=0.09$. The location of the white zero-(weak)-shear points inside these loops is best visible for $x_{\text{P}}=0.12$ for all three masses, as well as for $x_{\text{P}}=0.13$ in the sub-critical case.

\subsubsection{Weak-shear Deviation Due to the Point Mass}
\label{sec:plots-weak_shear_deviation}

The plots in Figure~\ref{fig:weak-shear_deviation_perturbation} depict the weak-shear deviation caused by the presence of the point mass, given by \mbox{$\gamma_{\text{w}}/\gamma_{\text{w,\tiny{NFW}}}-1$}. As indicated by the formula, it is defined as the relative difference between the weak shear $\gamma_{\text{w}}(\boldsymbol x)$ of a NFW halo with a point mass (see Figure~\ref{fig:weak_shear}) and the weak shear $\gamma_{\text{w,\tiny{NFW}}}(\boldsymbol x)$ of a NFW halo alone (see the second panel of Figure~\ref{fig:NFW-line}). This means that the plots are the weak-shear equivalents of the plots from Figure~\ref{fig:shear_deviation_perturbation} described in Section~\ref{sec:plots-shear_deviation}. Therefore, we use the same color scale and the same set of contours as in Figure~\ref{fig:shear_deviation_perturbation}. Note that due to Equation~(\ref{eq:weak_shear}), the plots in Figure~\ref{fig:weak-shear_deviation_perturbation} also exactly portray the relative image-flattening deviation, including the color bar and the values of the contours. The yellow- and orange-hued regions thus correspond to higher flattening and weak shear, while the blue regions correspond to lower flattening and weak shear than in the absence of the point mass.

The striking patterns of the colored areas in Figure~\ref{fig:weak-shear_deviation_perturbation} look remarkably complex at first. However, especially in the top rows, away from the halo center and from the vicinity of the point mass, the plots are very similar to those in Figure~\ref{fig:shear_deviation_perturbation}. This should be expected, since in these regions with low values of $\gamma$ and $\kappa$ the weak shear is a good approximation of the shear. Closest to the point mass, the plots differ from those in Figure~\ref{fig:shear_deviation_perturbation} fundamentally. In Figure~\ref{fig:shear_deviation_perturbation}, the relative shear deviation diverges to $\infty$ at the position of the point mass. In Figure~\ref{fig:weak-shear_deviation_perturbation}, even inside the positive weak-shear deviation contours in the top rows, there is a blue negative area in which the deviation falls to the minimum possible value of $-1$ at the point-mass position. This is a consequence of the weak shear $\gamma_{\text{w}}$ converging to $0$, while the shear $\gamma$ diverges there. Interestingly, this result holds even for $x_{\text{P}}=0$ at the halo center, where even $\gamma_{\text{w,\tiny{NFW}}}$ converges to $0$. Nevertheless, in the immediate vicinity of the center the weak-shear ratio $\gamma_{\text{w}}/\gamma_{\text{w,\tiny{NFW}}}\simeq 4\,(x\,\ln{x})^2\,\kappa_\text{s}/\kappa_\text{P}\to 0$, so that even in this case the weak-shear deviation at the point-mass position is $-1$.

In order to decipher the alternating positive and negative regions in Figure~\ref{fig:weak-shear_deviation_perturbation}, we concentrate on their boundaries which are indicated by the dot-dashed zero-weak-shear-deviation contour. Especially in the top rows, for higher $x_{\text{P}}$, we can see that parts of the contours look identical to the zero-shear-deviation contours from Figure~\ref{fig:shear_deviation_perturbation}. However, here there is an additional strong effect closely associated with the critical curve, unlike in the case of the shear-deviation patterns which show very little influence of the critical curve. More specifically, this additional effect reflects the relative deformations of the critical curve caused by the point mass.

\renewcommand{\thefigure}{11.A}
\begin{figure*}
{\centering
\vspace{0cm}
\hspace{1cm}
\includegraphics[width=15.5 cm]{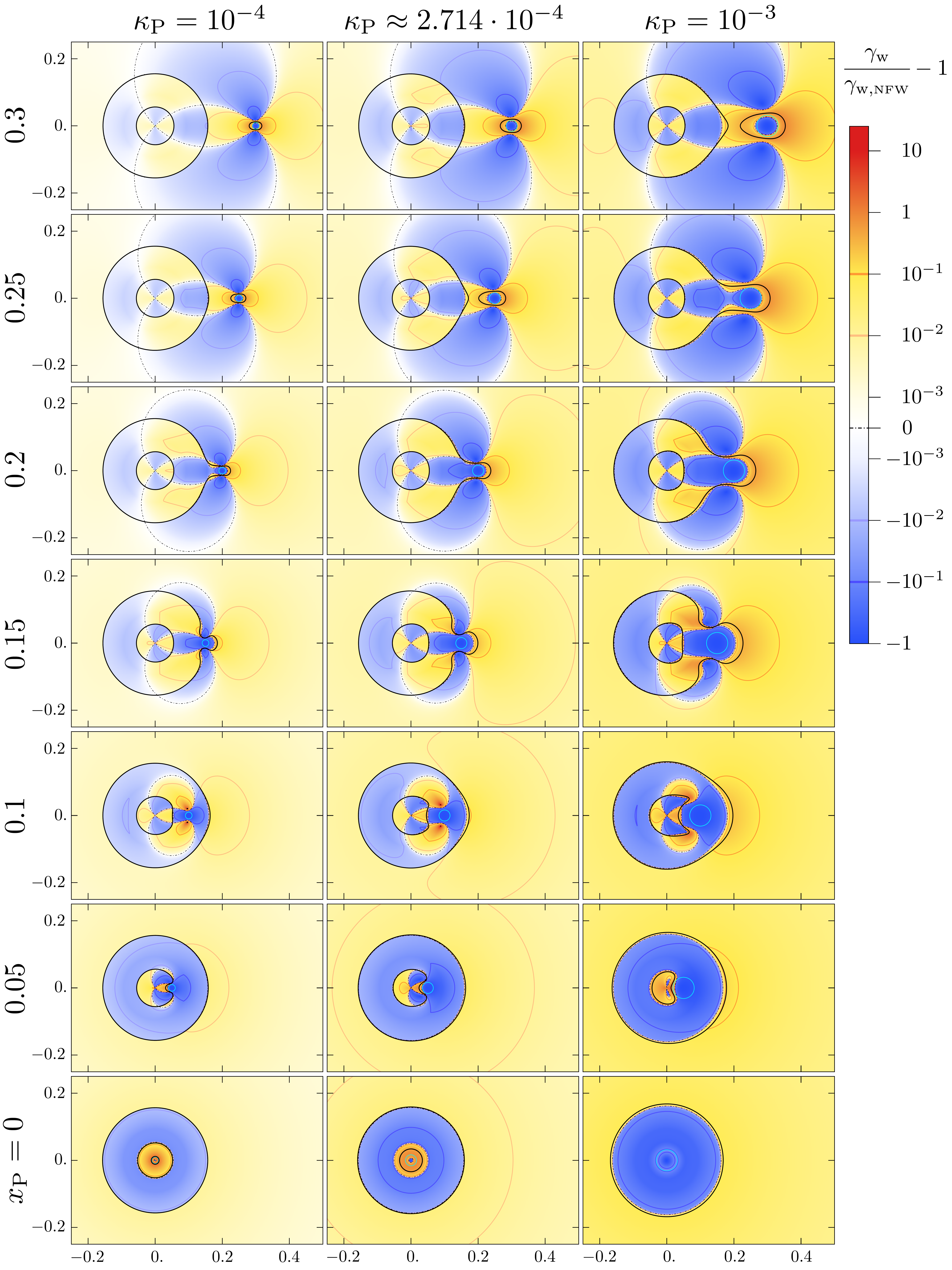}
\caption{Image-plane maps of the relative weak-shear deviation, $\gamma_{\text{w}}/\gamma_{\text{w,\tiny{NFW}}}-1$, caused by the presence of the point mass, described in Section~\ref{sec:plots-weak_shear_deviation}. The maps also exactly portray the relative deviation in image flattening due to the point mass. Color scale, contours, and remaining notation as in Figure~\ref{fig:shear_deviation_perturbation}. \label{fig:weak-shear_deviation_perturbation}}}
\end{figure*}

The zero-weak-shear-deviation contour can be defined using the flattening from Equation~(\ref{eq:flattening}) by setting $f(\kappa,\gamma)=f(\kappa,\gamma_{\text{\tiny{NFW}}})$. Since the convergence is the same with or without the point mass (except at the point-mass position), we immediately see that the zero-shear-deviation contour
\beq
\gamma(\boldsymbol x)=\gamma_{\text{\tiny{NFW}}}(x)
\label{eq:contour_pt1}
\eeq
automatically also forms a component of the zero-weak-shear-deviation contour. All dot-dashed contours from Figure~\ref{fig:shear_deviation_perturbation} thus appear also in Figure~\ref{fig:weak-shear_deviation_perturbation}. The remaining components can be obtained by solving the flattening equality, which yields the additional non-trivial solution
\beq
\gamma(\boldsymbol x)\,\gamma_{\text{\tiny{NFW}}}(x)=\left[\,1-\kappa(x)\,\right]^2\,.
\label{eq:contour_pt2}
\eeq
This equation describes all components of the zero-weak-shear-deviation contour that do not appear in Figure~\ref{fig:shear_deviation_perturbation}.

In particular, these include components closely associated with the critical curves, which is best illustrated when the NFW critical curves are only weakly perturbed by the point mass. Along the NFW tangential critical curve we recall that $\gamma_{\text{\tiny{NFW}}}(x)=1-\kappa(x)$, as discussed in Section~\ref{sec:NFW-Jacobian}. Similarly, along the perturbed NFW tangential critical curve which lies outside the unit-convergence circle, the shear satisfies $\gamma(\boldsymbol x)=1-\kappa(x)$, as discussed in Section~\ref{sec:NFWP-Jacobian}. It is straightforward to show that the product of the two shears is lower than $\left[\,1-\kappa(x)\,\right]^2$ along a section of one of these critical curves and at the same time it is higher than $\left[\,1-\kappa(x)\,\right]^2$ along the corresponding section of the other critical curve. Due to continuity, there is a contour between the critical curves of the two models along which Equation~(\ref{eq:contour_pt2}) is satisfied. When the point-mass perturbation is weak, such as in the top left plot in Figure~\ref{fig:weak-shear_deviation_perturbation}, the corresponding component of the zero-weak-shear-deviation contour is indistinguishable from the perturbed NFW tangential critical curve. The contour can be distinguished for example in the top right plot near the horizontal axis in the direction of the point mass.

A similar argument can be made for the NFW radial critical curve, along which $\gamma_{\text{\tiny{NFW}}}(x)=\kappa(x)-1$, as discussed in Section~\ref{sec:NFW-Jacobian}, and for the perturbed NFW radial critical curve inside the unit-convergence circle with $\gamma(\boldsymbol x)=\kappa(x)-1$, as discussed in Section~\ref{sec:NFWP-Jacobian}. Even in this case there is a contour between the two critical curves along which Equation~(\ref{eq:contour_pt2}) is satisfied.  The contour can be distinguished from the perturbed NFW radial critical curve in Figure~\ref{fig:weak-shear_deviation_perturbation} for example in the super-critical $x_\text{P}=0.15$ plot near the horizontal axis in the direction of the point mass. Note that Equation~(\ref{eq:contour_pt2}) also accounts for other components of the zero-weak-shear-deviation contour, such as the loop around the point mass in the top row of Figure~\ref{fig:weak-shear_deviation_perturbation}.

At the mutual intersections of the components of the zero-weak-shear-deviation contour, Equation~(\ref{eq:contour_pt1}) and Equation~(\ref{eq:contour_pt2}) have to be satisfied simultaneously. This implies that these points also represent the intersections of the critical curves of the NFW halo with and without the point mass. The combined geometry of the components with their mutual intersections explains the partitioning of the image plane into the color patterns seen in Figure~\ref{fig:weak-shear_deviation_perturbation}.

\renewcommand{\thefigure}{11.B}
\bfi
{\centering
\vspace{0cm}
\hspace{1cm}
\includegraphics[height=22 cm]{f11_B.pdf}
\caption{Image-plane maps of the relative weak-shear deviation, $\gamma_{\text{w}}/\gamma_{\text{w,\tiny{NFW}}}-1$, caused by the presence of the point mass, for a finer grid of point-mass positions than in Figure~\ref{fig:weak-shear_deviation_perturbation}. Notation same as in Figure~\ref{fig:weak-shear_deviation_perturbation}. \label{fig:weak-shear_deviation_perturbation-online}}}
\efi

\renewcommand{\thefigure}{12.A}
\begin{figure*}
{\centering
\vspace{0cm}
\hspace{1cm}
\includegraphics[width=15.5 cm]{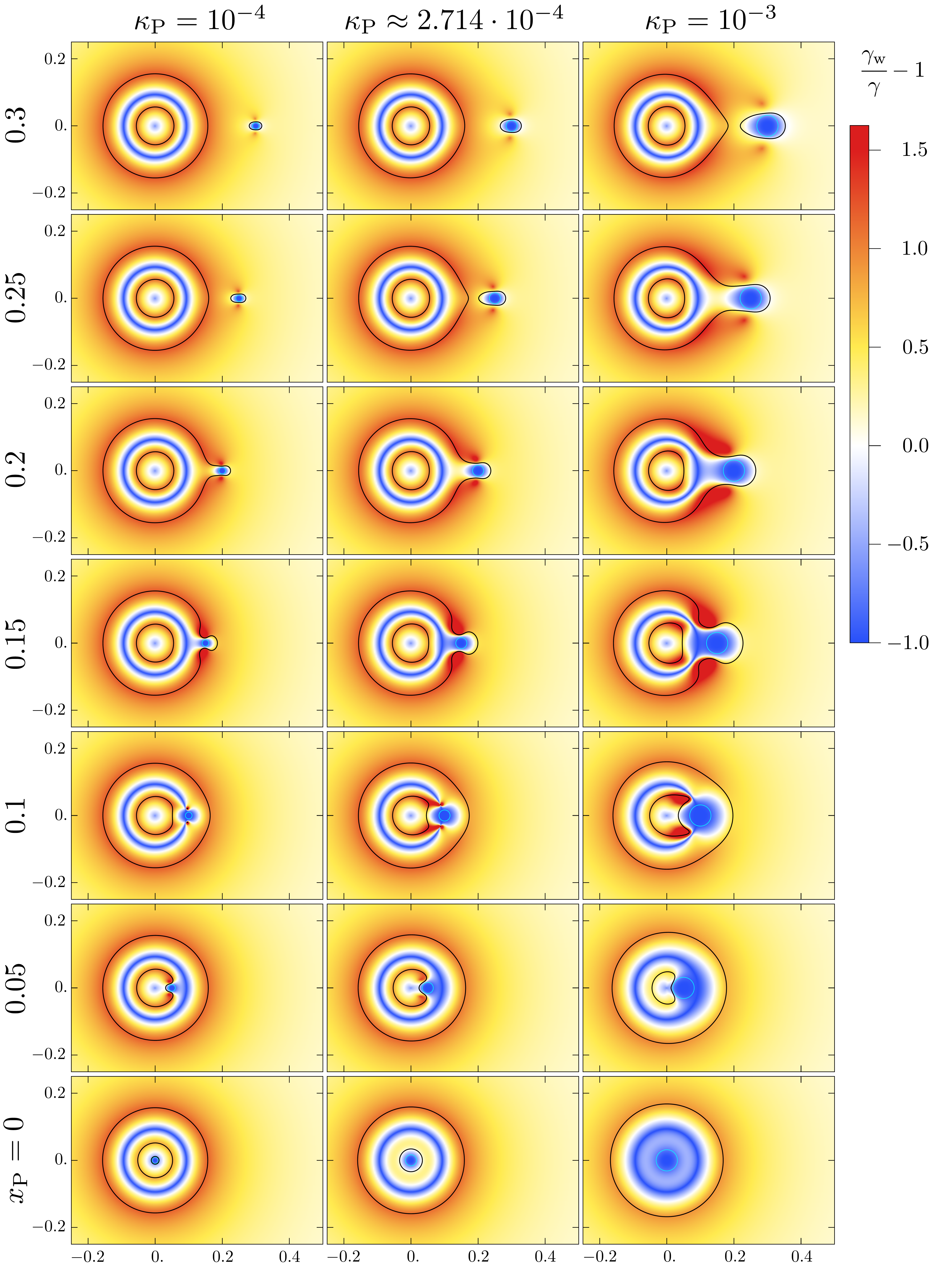}
\caption{Image-plane maps of the relative weak-shear deviation from the shear, $\gamma_{\text{w}}/\gamma-1$, of a NFW halo + point-mass lens, described in Section~\ref{sec:plots-weak_shear_deviation_from_shear}. The maps illustrate the relative error of the weak-lensing shear estimate. All positions with deviation greater than $1.5$ are marked in red. Remaining notation as in Figure~\ref{fig:shear}.\label{fig:weak-shear_deviation}}}
\end{figure*}

Starting with a centrally positioned sub-critical point mass (bottom left plot), the deviation shows a very small negative region around the point mass inside the inner radial critical curve, followed by a positive annulus reaching just beyond the outer radial critical curve, a negative annulus almost to the tangential critical curve, and a positive outer region. All three boundaries separating these regions are described by Equation~(\ref{eq:contour_pt2}). At $x_\text{P}=0.05$, the negative region around the point mass is connected with the larger negative annulus. In addition, the zero-shear-deviation boundary given by Equation~(\ref{eq:contour_pt1}) can be seen, introducing negative areas above and below the halo center, with positive crescents where it reaches beyond the zero-weak-shear-deviation contour associated with the perturbed NFW radial critical curve. At $x_\text{P}=0.1$, strong positive deviation can be seen close to the two tiny critical-curve loops, along which $\gamma_{\text{w}}=0.5$ while $\gamma_{\text{w,\tiny{NFW}}}\approx 0$. However, note that at the zero-shear points inside the loops the weak-shear deviation equals $-1$, with the indiscernible negative regions around them bordered by contour loops obeying Equation~(\ref{eq:contour_pt2}). At $x_\text{P}=0.15$, the region inside the zero-shear-deviation contour flips color again as it crosses the zero-weak-shear-deviation contour associated with the perturbed NFW tangential critical curve. At $x_\text{P}=0.25$, the point-mass critical-curve loop is detached from the NFW-halo critical curve, showing the characteristic four-lobed contour pattern seen in Figure~\ref{fig:shear_deviation_perturbation}, here with the central negative region described above.

A similar sequence can be seen in the two right columns, in which the structures are larger and the deviations more prominent. In particular, the zero-weak-shear-deviation contours near the critical curves are better visible in some of the plots here. Note also the change with increasing mass for a central position of the point mass in the bottom row. In the critical case there is only a single degenerate radial critical curve, with the accompanying zero-weak-shear-deviation contour well separated from it. The positive-deviation annulus is narrower as the central negative region is larger, and for a higher mass it disappears entirely. This can be seen in the super-critical case, where there is only a single large negative-deviation region reaching almost to the tangential critical curve.

\renewcommand{\thefigure}{12.B}
\bfi
{\centering
\vspace{0cm}
\hspace{1cm}
\includegraphics[height=22 cm]{f12_B.pdf}
\caption{Image-plane maps of the relative weak-shear deviation from the shear, $\gamma_{\text{w}}/\gamma-1$, of a NFW halo + point-mass lens, for a finer grid of point-mass positions than in Figure~\ref{fig:weak-shear_deviation}. Notation same as in Figure~\ref{fig:weak-shear_deviation}.\label{fig:weak-shear_deviation-online}}}
\efi

The transitions of the patterns with increasing point-mass position $x_\text{P}$ can be studied in more detail in Figure~\ref{fig:weak-shear_deviation_perturbation-online}. For example, note the appearance of a positive-deviation region in the super-critical case at $x_\text{P}=0.01$, as soon as the point mass is displaced from the halo center. For larger separations the boundary of this region forms the zero-weak-shear-deviation contour associated with the perturbed NFW radial critical curve. The small negative regions inside the tiny critical-curve loops can be seen here at $x_\text{P}=0.13$ in the sub-critical case, or at $x_\text{P}=0.12$ in all cases. In the critical and super-critical cases at $x_\text{P}=0.13$, we see that these regions persist even after the loops merge with the outer critical curve.

Studying the colored contours, we see that the regions with the strongest negative weak-shear deviation occur close to the point mass and near the zero-shear points (except when they lie between the perturbed NFW radial and tangential critical curves). The strongest positive deviation occurs close to the point mass along the horizontal axis outside its Einstein radius, and just outside the tiny critical-curve loops enclosing the zero-shear points. A closer inspection of the contours reveals that some of them display sharp bends at specific positions. These may appear at three types of locations. First, at the unit-convergence circle, as can be seen for example in the critical case at $x_\text{P}=0.15$ on the pale blue, pale orange, and dark orange contours. These kinks are caused by the switching of the minimum fractions in Equation~(\ref{eq:flattening}) at $\kappa=1$, which causes a discontinuity in the derivatives.

Second, at the critical curves, as can be seen for example in the super-critical case at $x_\text{P}=0.25$ on the pale orange and pale blue contours at the left side of the plot, or in the sub-critical case at $x_\text{P}=0.15$ on the pale orange and pale blue contours above and below the halo center. These kinks are caused by crossing the zero points of the absolute values in Equation~(\ref{eq:flattening}), which also causes a discontinuity in the derivatives. Third, at the NFW-halo critical curves, as can be seen for example in the super-critical case at $x_\text{P}=0.15$ on the dark blue contours around the zero-shear points above and below the point mass, or in the critical case at $x_\text{P}=0.20$ on the dark blue contour extending from the point mass toward the halo center. These kinks are similar to the previous ones, being caused by crossing the zero points of the absolute values in Equation~(\ref{eq:flattening}) when evaluating the NFW halo weak shear.

Regarding the extent of the contours, they are generally slightly smaller than those in Figure~\ref{fig:shear_deviation_perturbation}, indicating that the weak-shear deviation falls more quickly with the distance from the point mass than the shear deviation. As a rough estimate, at the moment of separation of the critical curve surrounding the point mass from the perturbed NFW tangential critical curve, the $\gamma_{\text{w}}/\gamma_{\text{w,\tiny{NFW}}}-1 = \pm\,10^{-1}$ contours extend roughly six Einstein radii from the point mass.

\subsubsection{Weak-shear Deviation from the Shear}
\label{sec:plots-weak_shear_deviation_from_shear}

The relative deviation $\gamma_{\text{w}}/\gamma-1$ of the weak shear (shown in Figure~\ref{fig:weak_shear}) from the shear (shown in Figure~\ref{fig:shear}) is plotted in the image-plane color maps in Figure~\ref{fig:weak-shear_deviation}. We use the same color scale as in the third panel of Figure~\ref{fig:NFW-line}, which shows the same quantity plotted for the NFW halo alone. By its definition, this deviation shows the relative error of shear estimation using the weak-lensing approximation.

In the case of the sub-critical mass (left column), the plots are very similar to the plot in the third panel of Figure~\ref{fig:NFW-line}, with the point mass affecting only its nearby surroundings. Note that the weak shear is always zero at the position of the point mass, so that the deviation reaches its minimum value of $-1$ there, leading to a dark blue spot similar to the one associated with the point mass in Figure~\ref{fig:weak-shear_deviation_perturbation}. In the bottom row, the central negative spot is thus more prominent than in the absence of the point mass and it includes even the inner radial critical curve. At $x_\text{P}=0.05$, there is a pair of red spots with a strong positive deviation close to the zero-shear points above and below the point mass. At these points, both $\gamma_{\text{w}}$ and $\gamma$ drop to $0$, but their ratio $\gamma_{\text{w}}/\gamma$ converges to $1/|\kappa-1|$. The deviation is thus positive as long as $\kappa<2$, corresponding to zero-shear points located more than circa $0.011$ from the center of our fiducial halo.

As the point mass crosses the unit-convergence circle near $x_\text{P}=0.1$, the red spots shrink but their peak deviation increases. When the zero-shear points approach the perturbed NFW tangential critical curve at $x_\text{P}=0.15$, the spots expand to their largest as they merge with the positive-deviation band along the critical curve. When the point-mass critical-curve loop is detached from the perturbed tangential critical curve of the NFW halo as seen in the two top rows, the positive-deviation spots remain associated with the zero-shear points above and below the point mass, even though their peak deviation declines with increasing point-mass position $x_\text{P}$. The pattern around the halo center at $x_\text{P}=0.3$ is virtually identical to the unperturbed-halo pattern in the third panel of Figure~\ref{fig:NFW-line}.

The plots for a critical mass in the central column follow a similar sequence, with larger affected regions around the point mass. When it is positioned at the halo center (bottom row), we see that the single radial critical curve shows practically zero deviation, and between it and the blue unit-convergence circle the positive deviation reaches lower values. At $x_{\text{P}}=0.1$, the red areas of high deviation extend from the perturbed NFW radial critical curve past the zero-shear points and beyond the unit-convergence circle. The pinched pattern at its intersection with the circle is indicative of the elliptic umbilic at which the weak shear is undefined, as discussed in Section~\ref{sec:plots-weak_shear}.

The regions strongly influenced by the super-critical mass in the right column are much larger, with nearly half of the pattern around the NFW halo critical curve affected for $x_{\text{P}}=0.1, 0.15,$ and $0.2$. For $x_{\text{P}}=0$ there is no region of positive deviation between the halo center and the unit-convergence circle. For $x_{\text{P}}=0.05$ a positive region is present but there are no red spots, as there are no zero-shear points at this configuration. At $x_{\text{P}}=0.1$ the red spots are very large and prominent. At \mbox{$x_{\text{P}}=0.15$}, in addition to the zero-shear-point red spots there are two adjacent smaller red spots along the perturbed NFW radial critical curve.

The changing patterns can be studied with a finer step in point-mass positions in Figure~\ref{fig:weak-shear_deviation-online}. For example, the pattern around the tiny critical-curve loops can be seen at $x_{\text{P}}=0.12$ and in a few neighboring panels. We also point out the $x_{\text{P}}=0.03$ panel in the sub-critical case and the $x_{\text{P}}=0.04$ panel in the critical case. In these panels there are no red spots close to the zero-shear points, which lie still too close to the halo center. However, in the super-critical case at $x_{\text{P}}=0.06$ we see red (or rather orange) spots at the tips of the perturbed NFW radial critical curve even though the zero-shear points are not present yet. One panel higher, at $x_{\text{P}}=0.07$, these red spots are more prominent while the zero-shear points are located near their edge closer to the halo center. Clearly, the red spots develop at the perturbed NFW radial critical curve even for lower point-mass positions at which there are no zero-shear points. With a slight increase in $x_\text{P}$, the points reach the red spots, which then remain associated with them at more distant point-mass positions.

Overall, the pattern of the blue areas shows that the weak shear underestimates the shear near the halo center, along the unit-convergence circle, and roughly within an Einstein radius of the point mass (extending further when it overlaps with the unit-convergence circle). Practically everywhere else the weak shear overestimates the shear, most prominently in the red and orange areas: close to the zero-shear points, along the perturbed NFW tangential critical curve, and along the perturbed NFW radial critical curve (except when a super-critical mass is positioned close to the halo center). Further from the halo center, e.g., close to the right edge of the plots in Figure~\ref{fig:weak-shear_deviation} for our fiducial halo, the positive deviation is very low, so that the weak shear may serve as a good approximation of the shear. However, the agreement fails in the vicinity of the point mass. This occurs primarily inside its Einstein radius, but also further to the tangentially offset zero-shear points. Note that for $\kappa_{\text{\tiny{NFW}}}(x_{\text{P}})\ll 1$ the deviation at the zero-shear points reaches a value $\gamma_{\text{w}}/\gamma-1\approx\kappa_{\text{\tiny{NFW}}}(x_{\text{P}})$, so that their influence decreases as the halo convergence declines for higher $x_{\text{P}}$.

\subsubsection{Weak Phase}
\label{sec:plots-weak_phase}

In Figure~\ref{fig:weak_phase} we present image-plane plots of the weak phase $\varphi_{\text{w}}$ of the NFW halo + point-mass lens, given by Equation~(\ref{eq:NFWP_weak_phase}). We defined the weak phase in Section~\ref{sec:NFW-weak} as the phase that would be measured from the orientation of image deformations using weak-lensing analysis. For an elliptical image of a small circular source the weak phase is simply the angle between its major axis and the horizontal axis in the plots. Thus, $\varphi_{\text{w}}=0$ corresponds to horizontally and $|\varphi_{\text{w}}|=\pi/2$ to vertically elongated images. The weak phase of the NFW-halo lens is described in Section~\ref{sec:NFW-weak} and shown in the fourth panel of Figure~\ref{fig:NFW-line}. Without the point mass, outside the unit-convergence circle the weak phase is equal to the phase and images are oriented tangentially. Inside the unit-convergence circle the weak phase is perpendicular to the phase and images are oriented radially.

In Figure~\ref{fig:weak_phase} we use the same weak-shear color scale as in the fourth panel of Figure~\ref{fig:NFW-line}, with white corresponding to horizontal images, orange to images oriented counterclockwise from the horizontal, and blue to images oriented clockwise from the horizontal. The saturation of both colors increases with the angle from the horizontal, from zero saturation for angle $0$ to maximum saturation for angle $\pi/2$. The dot-dashed contour marks several special orientations: $\varphi_{\text{w}}=0$, $|\varphi_{\text{w}}|=\pi/2$, and undefined $\varphi_{\text{w}}$. The weak phase is undefined at all points with zero weak shear, as described in Section~\ref{sec:plots-weak_shear}. These include the halo center, the point-mass position, the zero-shear points, and the unit-convergence circle. All of these points thus play an important role in the patterns seen in the color maps in Figure~\ref{fig:weak_phase}.

Thanks to the axial symmetry of the lens configurations with $x_{\text{P}}=0$ in the bottom row of Figure~\ref{fig:weak_phase}, the color maps are identical to that of the unperturbed halo, as seen in the fourth panel of Figure~\ref{fig:NFW-line}. This means that outside the $\kappa=1$ circle images are oriented tangentially, while inside the circle the weak phase is flipped by $\pi/2$ and images are oriented radially. Exactly at the halo center the weak phase is undefined. Note the colors alternating around the center with orange in the first quadrant and the saturation varying from zero along the horizontal to maximum along the vertical axis. Such a pattern corresponds to radial orientation of images around the center.

\renewcommand{\thefigure}{13.A}
\begin{figure*}
{\centering
\vspace{0cm}
\hspace{1cm}
\includegraphics[width=15.5 cm]{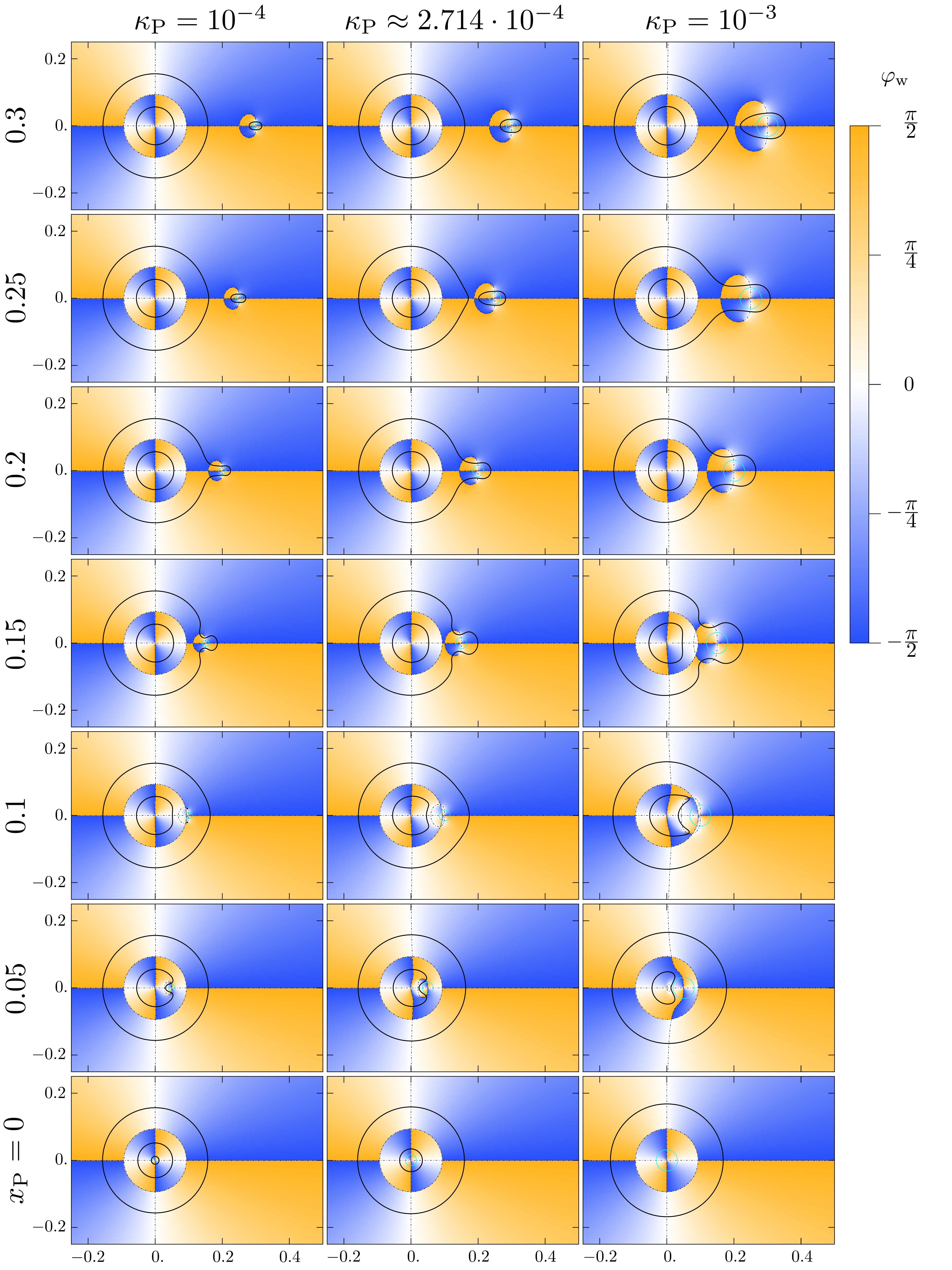}
\caption{Image-plane maps of the weak phase $\varphi_{\text{w}}(\boldsymbol x)$ of a NFW halo + point-mass lens, described in Section~\ref{sec:plots-weak_phase}. Orange and blue correspond to images oriented counterclockwise and clockwise, respectively, from the horizontal. Dot-dashed curves indicate exactly horizontal, exactly vertical, or undefined image orientation. Remaining notation as in Figure~\ref{fig:shear}.\label{fig:weak_phase}}}
\end{figure*}

The variation of the weak-phase maps with point-mass position $x_\text{P}$ is best seen in the super-critical case (right column). In the top row with $x_\text{P}=0.3$ the point mass is well separated from the halo center. The left part of the plot closer to the halo center is similar to the unperturbed pattern seen also in the bottom row. Note here the slight shift toward the point mass of the dot-dashed contour extending vertically outward from the unit-convergence circle, which corresponds to horizontal images. The main new feature is the oval region with inverted colors bordered by a loop of the dot-dashed contour, which passes through the position of the point mass at its right side and extends in the direction of the halo center on its left side. The pattern seen around the point mass also has colors alternating around the center, however, here with blue in the first quadrant. In addition, the saturation varies from zero along the vertical to maximum along the horizontal axis. This pattern corresponds to tangential orientation of images around the point mass.

The left part of the dot-dashed border of the oval separates high-saturation regions and, thus, corresponds to vertically oriented images. The right part of the border lies in the white band separating low-saturation regions and, thus, corresponds to horizontally oriented images. The two parts of the boundary meet at the zero-shear points that lie above and below the point mass. In this panel, horizontal images occur along the horizontal axis inside the unit-convergence circle, along the vertical dot-dashed contour outside the unit-convergence circle, and along the right boundary of the oval region connecting the zero-shear points vertically through the point-mass position. Vertical images occur along the horizontal axis outside the unit-convergence circle, along the vertical dot-dashed contour inside the unit-convergence circle, and along the left boundary of the oval region connecting the zero-shear points and passing vertically through a point between the halo center and the point mass.

Going down in the right column, the inverted-color oval shrinks slightly as it moves with the point mass closer to the halo center. At $x_{\text{P}}=0.15$, the oval partly overlaps the unit-convergence circle. The very pale color in the region of their overlap indicates near-horizontal orientation of images there. At the boundary of the oval in this plot, images are oriented horizontally along its left part inside the unit-convergence circle and along its right part from the zero-shear points to the point-mass position. Along the segments extending from the zero-shear points to the left till the unit-convergence circle, images are oriented vertically. Image orientations along the other parts of the dot-dashed contour remain unchanged.

\renewcommand{\thefigure}{13.B}
\bfi
{\centering
\vspace{0cm}
\hspace{1cm}
\includegraphics[height=22 cm]{f13_B.pdf}
\caption{Image-plane maps of the weak phase $\varphi_{\text{w}}(\boldsymbol x)$ of a NFW halo + point-mass lens, for a finer grid of point-mass positions than in Figure~\ref{fig:weak_phase}. Notation same as in Figure~\ref{fig:weak_phase}.\label{fig:weak_phase-online}}}
\efi

\renewcommand{\thefigure}{14.A}
\begin{figure*}
{\centering
\vspace{0cm}
\hspace{1cm}
\includegraphics[width=15.5 cm]{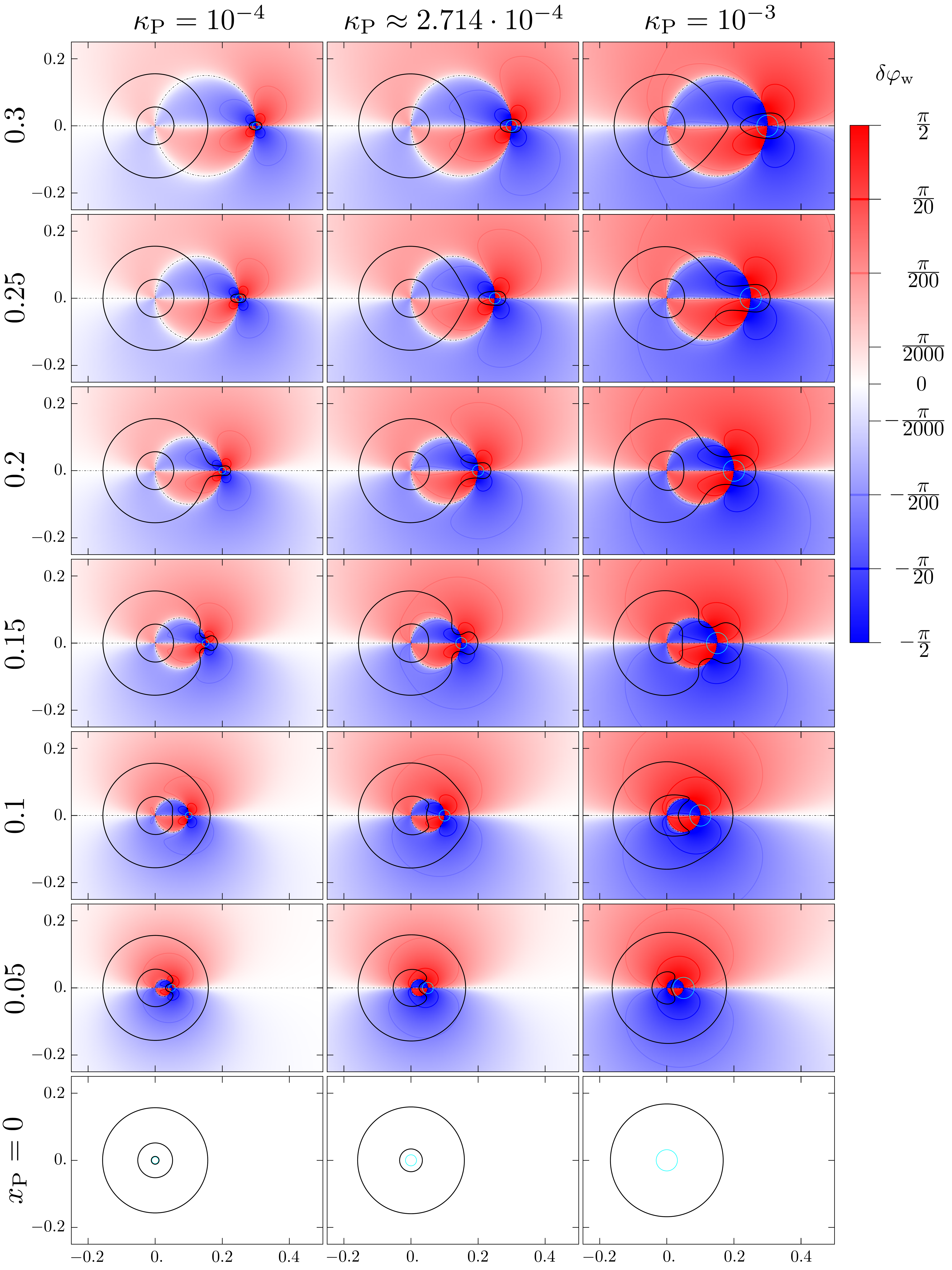}
\caption{Image-plane maps of the weak-phase deviation $\delta\varphi_\text{w}(\boldsymbol x)$ caused by the presence of the point mass, described in Section~\ref{sec:plots-weak_phase_change}. The maps also exactly portray the change in image orientation due to the point mass. Red and blue indicate counterclockwise and clockwise deviation, respectively. The color scale changes from logarithmic to linear in the interval $[-\pi/2000,\pi/2000]$. Four colored contours correspond to $\delta\varphi_\text{w}$ values indicated in the color bar; the dot-dashed contour corresponds to $\delta\varphi_\text{w}=0$ and $\delta\varphi_\text{w}=\pm \pi/2$. Remaining notation as in Figure~\ref{fig:shear}.\label{fig:weak-phase_change}}}
\end{figure*}

At $x_{\text{P}}=0.1$, the zero-shear points now lie inside the unit-convergence circle, so that horizontally oriented images now occur along the left boundary of the oval up to the zero-shear points and along the part of the right boundary outside the unit-convergence circle passing through the point-mass position. Vertically oriented images occur along the segments extending from the zero-shear points to the right till the unit-convergence circle. The pattern inside the overlap of the oval and the unit-convergence circle reflects two general properties of its boundaries. Crossing the boundary of the oval inverts only the color at $\varphi_{\text{w}}=0$ or $|\varphi_{\text{w}}|=\pi/2$, which corresponds to a continuous change in orientation. Crossing the unit-convergence circle inverts the color as well as the saturation, which corresponds to a $\pi/2$ flip in orientation. Note here also the more prominent bulging of the vertical dot-dashed contour passing through the halo center as the point mass lies closer.

Between $x_{\text{P}}=0.1$ and $x_{\text{P}}=0.05$, the boundary of the inverted-color oval undergoes reconnection with the bulging vertical dot-dashed contour. At $x_{\text{P}}=0.05$ this contour passes through the point-mass position, while a small inverted-color oval lies to the left of it, with its left boundary passing through the halo center. Note that at $x_{\text{P}}=0.05$ there are no zero-shear points, so that here the entire boundary of the small oval corresponds to horizontal images. What is more striking in this plot is the pattern around the point mass, which now lies inside the unit-convergence circle. Orange in the first quadrant and zero saturation along the horizontal axis indicate that images are now oriented radially around the point mass. As the point mass moves closer to the halo center, at a certain value $0\leq x_{\text{P}} < 0.05$ the small oval disappears and the dot-dashed vertical curve passes through the origin for $x_{\text{P}}=0$.

In the critical and sub-critical cases (the central and left columns, respectively) the same sequence of changes occurs as in the super-critical case, although for slightly different values of $x_{\text{P}}$. Note that the inverted-color oval increases in size with increasing mass parameter $\kappa_{\text{P}}$. In the top rows, the point at which the left side of its boundary crosses the horizontal axis marks a boundary of influence of the halo and the point mass (at least as far as image orientation is concerned). Both boundaries passing through this point correspond to vertically oriented images. However, images located off the horizontal axis to the left of this point are tilted tangentially to the halo, while those to the right of this point are tilted tangentially to the point mass.

Details of the changing patterns and contours can be studied in more detail in Figure~\ref{fig:weak_phase-online}. For example, it can be seen that the reconnection of the vertical and oval dot-dashed contours occurs at the zero-shear points. At higher values of $x_{\text{P}}$ the points lie on the larger oval passing through the point-mass position, while at lower values they lie on the smaller oval passing through the halo center. For progressively lower point-mass distances, the zero-shear points move along the smaller oval to the center, where they disappear at $x_{\text{P}}=\sqrt{\kappa_{\text{P}}/\kappa_{\text{s}}}$.

\renewcommand{\thefigure}{14.B}
\bfi
{\centering
\vspace{0cm}
\hspace{1cm}
\includegraphics[height=22 cm]{f14_B.pdf}
\caption{Image-plane maps of the weak-phase deviation $\delta\varphi_\text{w}(\boldsymbol x)$ caused by the presence of the point mass, for a finer grid of point-mass positions than in Figure~\ref{fig:weak-phase_change}. Notation same as in Figure~\ref{fig:weak-phase_change}.\label{fig:weak-phase_change-online}}}
\efi

The plot for the critical case at $x_{\text{P}}=0.1$ illustrates the situation at the elliptic umbilics, which occur for a slightly lower point-mass distance. In this case, the full length of the oval contour corresponds to horizontal image orientations (except the point-mass position and the zero-shear points with undefined orientation). Similarly, the $x_{\text{P}}=0.09$ plots for the critical and super-critical cases indicate the interesting pattern near the point mass when it lies on the unit-convergence circle (in both cases for a slightly higher point-mass distance). To the right of the point mass images are oriented vertically (tangentially), to the left horizontally (radially), and above and below the orientation is undefined (i.e., the images are circular).

The weak-phase plots presented in Figure~\ref{fig:weak_phase} can also be used to visualize the phase, which we do not present in a separate plot. The reason is indicated by Equation~(\ref{eq:NFWP_weak_phase}), which shows that outside the unit-convergence circle the phase is equal to the weak phase, while inside it differs by $\pi/2$. The phase plots would thus differ from the weak-phase plots in Figure~\ref{fig:weak_phase} by having the colors and saturations flipped inside the unit-convergence circle around the halo center. The circle would disappear in such plots, and the color and saturation outside would extend continuously inside all the way to the halo center. The dot-dashed contour would then consist only of the horizontal axis, the perturbed vertical line through the halo center, and the large oval associated with the point mass reconnecting to the small oval associated with the halo center.

Overall, note that unlike the weak shear in Figure~\ref{fig:weak_shear}, the weak phase in Figure~\ref{fig:weak_phase} shows very little correlation with the geometry of the critical curve. Finally, we point out that placing the point mass off the horizontal axis would not lead to a simple rotation of the patterns as in the other presented plot grids. Instead, the overall halo pattern and the pattern close to the point mass would remain unchanged. A point mass in a blue region of the halo would thus locally generate a pair of orange lobes, while in an orange region it would generate a pair of blue lobes.

\subsubsection{Weak-phase Deviation Due to the Point Mass}
\label{sec:plots-weak_phase_change}

The image-plane color maps presented in Figure~\ref{fig:weak-phase_change} depict the weak-phase deviation due to the point mass, defined as the difference between the weak shear of the NFW halo + point-mass lens (shown in Figure~\ref{fig:weak_phase}) and the weak shear of the NFW halo alone (shown in the fourth panel of Figure~\ref{fig:NFW-line}), $\delta\varphi_\text{w}(\boldsymbol x)= \varphi_\text{w}-\varphi_{\text{w,\tiny{NFW}}}$. We correct the difference if necessary by adding or subtracting $\pi$ to keep $\delta\varphi_\text{w}$ in the interval $\left[-\pi/2,\,\pi/2\right]$. The deviation is also equal to the angle by which the orientation of an image changes due to the presence of the point mass. Red positive values of $\delta\varphi_\text{w}$ correspond to a counterclockwise change, blue negative values to a clockwise change in orientation. The color saturation is scaled linearly for $|\delta\varphi_\text{w}|\leq\pi/2000$ and logarithmically for $|\delta\varphi_\text{w}|\geq\pi/2000$. Contours are plotted for $|\delta\varphi_\text{w}|=\pi/20=9^{\circ}$ (dark red and blue), and for $|\delta\varphi_\text{w}|=\pi/200=0.9^{\circ}$ (light red and blue).

When the point mass is located exactly at the center of the halo (bottom row of Figure~\ref{fig:weak-phase_change}), the lens has axial symmetry and the weak-phase deviation is zero everywhere, which explains the completely white plots. Note that this result also reflects the fact that images are oriented radially around a point mass lying inside the unit-convergence circle, as shown in Section~\ref{sec:plots-weak_phase}.

All the other plots with an off-center point mass share the same characteristic color pattern; they differ only in its scale and in color saturation. The blue and red regions are separated by the dot-dashed contour, which marks all positions with zero deviation $\delta\varphi_\text{w}=0$ or maximum deviation $|\delta\varphi_\text{w}|=\pi/2$. The geometry of the contour is simple and independent of the mass parameter $\kappa_\text{P}$: it includes the horizontal axis and the $\omega=\pi/2$ circle reaching from the halo center to the position of the point mass (see Figure~\ref{fig:omega}). Along the horizontal axis, images are always horizontal inside and vertical outside the unit-convergence circle, unaffected by the presence of a point mass, as shown in Section~\ref{sec:plots-weak_phase}. Hence, this part of the dot-dashed contour corresponds to $\delta\varphi_\text{w}=0$.

Along the $\omega=\pi/2$ circle, there is a right angle between the direction to the halo center and the direction to the point mass. The shear from the halo and the shear from the point mass thus act in perpendicular directions. Going from the halo center along the $\omega=\pi/2$ circle, the halo shear decreases and the point-mass shear increases, as described in Section~\ref{sec:NFWP-csp}. Hence, from the halo center to the zero-shear points the halo shear dominates and the image orientation remains unchanged, so that $\delta\varphi_\text{w}=0$ along this part of the circle. From the zero-shear points to the point mass, the point-mass shear dominates and the images are oriented perpendicular to the orientation they would have in absence of the point mass, so that $|\delta\varphi_\text{w}|=\pi/2$ along the remaining part of the circle. For point-mass positions $x_{\text{P}}<\sqrt{\kappa_{\text{P}}/\kappa_{\text{s}}}$, there are no zero-shear points and the entire $\omega=\pi/2$ circle corresponds to $|\delta\varphi_\text{w}|=\pi/2$.

Above the horizontal axis and outside the $\omega=\pi/2$ circle, the weak-shear deviation is positive meaning that in this region the image orientation changes counterclockwise. Conversely, under the axis and outside the circle, the deviation is negative and the image orientation changes clockwise. Inside the $\omega=\pi/2$ circle, the sign of the deviation in either half-plane is switched and image orientations change in the opposite sense. At the center of the halo and at the location of the point mass, four regions of alternating positive and negative deviation meet.

Deviations $|\delta\varphi_\text{w}|$ peak close to the point mass (particularly along the  $\omega=\pi/2$ circle), and fall rapidly with increasing distance from it. The color pattern near the point mass indicates that images are oriented tangentially around it (when it lies outside the unit-convergence circle) or radially around it (when it lies inside the unit-convergence circle). The color pattern around the halo center indicates that the point mass orients the images more horizontally there. When zero-shear points are present, zero deviations occur in the horizontal and vertical directions from the center and strongest deviations occur along the diagonals. In the absence of zero-shear points, for $x_{\text{P}}<\sqrt{\kappa_{\text{P}}/\kappa_{\text{s}}}$, zero deviations occur in the horizontal directions and strongest $|\delta\varphi_\text{w}|=\pi/2$ deviations occur in the vertical directions.

The areas with high deviation $|\delta\varphi_\text{w}|$ extend further from the point mass for greater values of $\kappa_{\text{P}}$, as seen in the central and right columns of Figure~\ref{fig:weak-phase_change}. For point masses far from the halo center (in the top rows), all contours form four-lobed butterfly-like shapes. For point masses closer to the halo center, the left lobes of the contours become pointy (e.g., $x_{\text{P}}=0.2$ in the central column) and eventually they extend to the halo center (e.g., $x_{\text{P}}=0.15$ in the central column). To illustrate the area affected by the point mass, we estimate the extent of the contours at the moment of separation of the critical curve surrounding the point mass from the perturbed NFW tangential critical curve, i.e., between $x_{\text{P}}=0.20$ and $x_{\text{P}}=0.25$ for the sub-critical and critical cases, and between $x_{\text{P}}=0.25$ and $x_{\text{P}}=0.30$ in the super-critical case. In all three cases, the inner \mbox{$|\delta\varphi_\text{w}|=\pi/20$} contours extend about four Einstein radii, while the outer $|\delta\varphi_\text{w}|=\pi/200$ contours extend as far as twelve Einstein radii from the point mass.

The changes in the contours for different point-mass positions can be studied in more detail in Figure~\ref{fig:weak-phase_change-online}. The abrupt change from zero deviation to $|\delta\varphi_\text{w}|=\pi/2$ along the $\omega=\pi/2$ circle at the zero-shear points can be best seen in the presence of the small critical-curve loops surrounding them, e.g., for $x_{\text{P}}=0.12$. Note also the very small extent of the colored regions for the lowest separations $x_\text{P}$.

Similarly to Figure~\ref{fig:weak_phase}, the weak-phase deviation in Figure~\ref{fig:weak-phase_change} shows very little correlation with the geometry of the critical curve. However, unlike Figure~\ref{fig:weak_phase}, it also shows no influence of the unit-convergence circle.

\section{Discussion}
\label{sec:discussion}

The examples of images formed by the NFW halo + point-mass lens in the bottom right panel of Figure~\ref{fig:images} can be compared with the results presented in Section~\ref{sec:NFWP-plots}, specifically with the corresponding $x_\text{P}=0.2$, $\kappa_\text{P}\approx 2.714\cdot 10^{-4}$ panels of the plot grids. In particular, the weak-shear map in Figure~\ref{fig:weak_shear} shows the image flattening, the weak-phase map in Figure~\ref{fig:weak_phase} shows the image orientation, and the weak-shear-deviation-from-shear map in Figure~\ref{fig:weak-shear_deviation} shows the relative shear error that would be incurred by assuming the weak-lensing relation between image distortion and shear.

A few points should be noted regarding the interpretation of such comparisons. First, Figures~\ref{fig:shear}--\ref{fig:weak-phase_change} present maps of their respective quantities for point-like sources. For images of extended sources such as those shown in Figure~\ref{fig:images}, one has to consider the full variation of the quantities within the area of the image. For example, an image lying on the unit-convergence circle will have its inner part extended radially and its outer part tangentially, as seen from Figure~\ref{fig:weak_phase}.

Second, when studying the changes in image shape and orientation due to the point mass using Figure~\ref{fig:images} with Figure~\ref{fig:weak-shear_deviation_perturbation} or Figure~\ref{fig:weak-phase_change}, a direct comparison can be made when there is at least a partial overlap of the images formed by halos with and without the point mass. In the bottom right panel of Figure~\ref{fig:images}, the left and central images have a large overlap with the images in the top right panel, the lower right image has a smaller overlap, and the two images close to the point mass have no overlap. For these two images, Figure~\ref{fig:weak-shear_deviation_perturbation} and Figure~\ref{fig:weak-phase_change} show the deviations from images formed at the same positions by the halo-only lens, but for different source positions. In this particular case, the sources would lie above the horizontal axis at different radial distances in the top left panel of Figure~\ref{fig:images}. In fact, even in the case of overlapping images, the overlapping parts may be images of different parts of the source. Thus, Figure~\ref{fig:weak-shear_deviation_perturbation} and Figure~\ref{fig:weak-phase_change} compare the properties of images formed by the two lens models at a same position in the image plane of sources at different positions in the source plane. They do not compare the properties of images of a fixed source formed by a halo with and without a point mass.

Third, in this work we do not explore the changes in image positions due to the point mass. This would be difficult to present in general, if only due to the change in the number of images. However, it could be done in a perturbative regime, by studying the displacement and distortion of particular images of a fixed source due to the presence of a point mass. This approach was taken by \cite{wagner18}, who studied general perturbations of large arc-like images along the tangential critical curve of axisymmetric lenses. One provided example shows the influence of a point-mass perturber on the radii and lengths of arcs formed by a singular isothermal sphere lens. Even though NFW halos have different lensing properties, the generic pattern of angular distortions seen for example in the $x_\text{P}=0.25$, $\kappa_\text{P} = 10^{-4}$ panel of Figure~\ref{fig:weak-phase_change} indicates that an arc-like image between the tangential critical curve and the point mass inside the $\omega=\pi/2$ circle would be straightened by its influence. Hence, its radius of curvature would be increased, in agreement with the example in \cite{wagner18}. Overall, image-plane maps such as those in Figures~\ref{fig:weak-shear_deviation_perturbation} and \ref{fig:weak-phase_change} are suitable for assessing the influence of compact masses in observed lensing clusters or galaxies where the image positions are fixed.

The results presented in this work correspond to one fiducial value of the halo convergence parameter $\kappa_\text{s}$. The variation of the critical curves and the unit-convergence circle with $\kappa_\text{s}$, as well as the variation of $\kappa_\text{s}$ and $\kappa_\text{P}$ with source redshift are discussed in \citetalias{karamazov_etal21}. The plots presented above in Sections~\ref{sec:plots-shear}--\ref{sec:plots-weak_phase_change} can be expected to follow the combined geometry of the critical curve, the unit-convergence circle, and the $\omega=\pi/2$ circle. Note that the last mentioned circle is given purely by the value of $x_\text{P}$, which is independent of the source redshift and the halo convergence parameter.

Some of the obtained results have more general relevance than just for a point mass embedded in a spherically symmetric NFW halo. These include some of the analytic results, such as the shear of a combination of two mass distributions described in Section~\ref{sec:NFWP-csp}. Among the numerical results shown in Figures~\ref{fig:shear}--\ref{fig:weak-phase_change}, the patterns seen around the point mass at sufficient separation from the halo critical curves will be very similar for other halo mass distributions with low spatial variation on the scale of the point-mass Einstein radius \citep[e.g.,][]{chang_refsdal84}, as discussed further below. Best seen in the top left plots for $x_\text{P}=0.30$, $\kappa_\text{P} = 10^{-4}$, the patterns are relevant not only for point masses separated from other mass concentrations, but also for extended bodies with compact mass distributions that do not extend significantly beyond their Einstein radius.

The applicability of the studied lens model to the astrophysical scenarios of a galaxy within a galaxy cluster, of a satellite galaxy within a galactic halo, and of a massive black hole in a galactic halo is discussed in \citetalias{karamazov_etal21}. In addition, possible extensions toward more advanced models are pointed out there, such as replacing the point mass by an extended mass distribution. For distributions that do not extend significantly beyond their total-mass Einstein radius, the lensing quantities in the surroundings will not differ significantly from the patterns seen in Figures~\ref{fig:shear}--\ref{fig:weak-phase_change}. However, for more extended distributions the patterns will be affected more substantially. Nevertheless, at large separations the lensing impact of any compact object may be approximated by that of a point mass.

The explored model can be extended also by altering the properties of the NFW halo. On the one hand, one may change its central properties by adding a core radius, or by changing its density divergence \citep{evans_wilkinson98}. Such changes would alter the radii of the halo critical curves and caustics, the reference plots in Figure~\ref{fig:NFW-line}, and the critical value $\kappa_{\text{PC}}$ of the mass parameter of the point mass. This would impact the presented results primarily for point-mass positions close to the halo center. Regardless of the nature of the alterations, for sufficiently large distances the point-mass critical-curve loop disconnects from the halo critical curves. In this distant regime the patterns in the vicinity of the point mass will have the same character as seen in the figures.

On the other hand, one may abandon spherical symmetry and study the effect of a massive object embedded in a more realistic elliptical NFW halo. The lensing properties of an NFW halo with an elliptically symmetric mass distribution are poorly studied due to the lack of simple analytic expressions to describe them \citep{oguri21}. Nevertheless, for low values of the halo ellipticity the properties can be approximated by those of the pseudo-elliptical model, which has an elliptically symmetric lens potential \citep{golse_kneib02,meneghetti_etal03,dumet-montoya_etal12}. In this model the critical curve typically consists of two nested oval loops instead of the two circles of the spherical model. The point-like tangential caustic of the spherical model is replaced by a four-cusped loop; the circular radial caustic by an oval loop. In this model one may expect an even more complex dependence of the critical curves and caustics of the combined lens on the mass and two-dimensional position of the point mass than in the spherical case described in \citetalias{karamazov_etal21}.

In the elliptical model, the unit-convergence circle which plays a key role in the spherical model when studying the geometry of images is replaced by a unit-convergence ellipse. The grids of the plots corresponding to those in Figures~\ref{fig:shear}--\ref{fig:weak-phase_change} would be complicated by an additional parameter, the angular position of the point mass with respect to the axes of the elliptical halo. Nevertheless, the main factors driving the patterns described in the text, such as the existence and location of zero-shear points or the geometry of the perturbations of the critical curve, would remain the same. The asymptotic patterns would have the same nature as seen in the spherical model, as mentioned above. However, they would be less symmetric and the extent of the contours would additionally depend on the angular position within the halo.

The presented single-point-mass results will be useful for interpreting the properties of cluster lens models with multiple (point) masses embedded in an NFW halo. Masses that are sufficiently separated from the perturbed NFW halo critical curves as well as from other masses should display similar patterns in their vicinity, as discussed above. Each of these masses will also produce deviation patterns near the halo center similar to those seen in the top left plots in Figures~\ref{fig:shear_deviation_perturbation}, \ref{fig:weak-shear_deviation_perturbation}, or \ref{fig:weak-phase_change}. Due to the directional dependence of these patterns, their superposition for a sufficient number of isotropically distributed masses would drive the amplitude of the central deviations to zero. Strong differences can be expected when one or more of the masses are positioned close to the halo center, or when two or more neighboring masses are mutually separated by less than a few Einstein radii. These situations cannot be simply extrapolated from the results presented in this work and their study requires direct simulations.

\section{Summary}
\label{sec:summary}

In this paper we proceeded in our study of gravitational lensing by a compact massive object in a dark matter halo. In \citetalias{karamazov_etal21} we analyzed the critical curves and caustics of a lens consisting of a point mass embedded in a spherical NFW halo. Here we concentrated on the shear and phase of the same lens model, focusing on their relation to the geometry of images formed by the lens.

In Section~\ref{sec:NFW-csp}, we described the properties of the shear and phase of a lens consisting only of a NFW halo. In order to study the images, we used the eigenvalue decomposition of the inverse matrix of a general lens-equation Jacobian matrix presented in Equation~(\ref{eq:A_matrix}). Based on it, we introduced the convergence--shear diagram (CS diagram) in Appendix~\ref{sec:Appendix-images}, which illustrates concisely the connection between arbitrary convergence and shear values and the geometry of an image corresponding to them. Specific lens models occupy characteristic regions in the diagram, which then define the properties of all images that could be formed by the lens. We described the properties of images formed by the NFW halo by reading them off the CS diagram in Section~\ref{sec:NFW-images}. In Section~\ref{sec:NFW-weak} we defined the weak shear and weak phase as the shear and phase values obtained by assuming weak-lensing relations involving the semiaxes of images and their orientation.

We followed the same outline for the NFW halo + point-mass lens in Section~\ref{sec:NFWP}. In particular, we derived the formula for the shear in Equation~(\ref{eq:NFWP_gamma}) which provides a geometric interpretation in terms of the halo and point-mass shears and the viewing angle $\omega$ of the line segment separating them. The formula which is valid for combinations of other axisymmetric lenses is a special case of the more general formula in Equation~(\ref{eq:shear_combination}) for the shear of a combination of two arbitrary lenses. For the NFW halo + point-mass lens, we discuss the appearance and location of zero-shear points in Section~\ref{sec:NFWP-csp} and describe the conditions under which they form umbilic points in Section~\ref{sec:NFWP-Jacobian}.

Figures~\ref{fig:shear}--\ref{fig:weak-phase_change} illustrate the main results in terms of image-plane maps of different lens characteristics and CS diagrams, all presented for the same grid of point-mass parameter combinations as the grid used in \citetalias{karamazov_etal21}. Important features and trends seen in the figures are described in the corresponding Sections~\ref{sec:plots-shear}--\ref{sec:plots-weak_phase_change}. As discussed in Section~\ref{sec:discussion}, the obtained results have broader implications beyond the specific properties of the studied lens model.

%\begin{acknowledgments}

\vskip 8mm

We thank the anonymous referee for comments and suggestions that helped improve the manuscript. Work on this project was supported by Charles University Grant Agency project GA UK 1000218.

%\end{acknowledgments}

\clearpage

\appendix

\section{IMAGE GEOMETRY AS A FUNCTION OF CONVERGENCE AND SHEAR}
\label{sec:Appendix-images}

The mapping from a source to its image is described locally by the inverse of the lens-equation Jacobian matrix. The eigenvalue decomposition of this inverse matrix $\mathbb{A}$ presented in Equation~(\ref{eq:A_matrix}) shows that the full geometry of an image of a small source, including its shape, size, and orientation with respect to the phase, is given by the lens convergence and shear at the position of the image. We demonstrate here the connection between the geometry of an image and its position in a general convergence--shear diagram.

The properties of an image appearing at position $\boldsymbol x$ in the image plane can be determined from the convergence--shear diagram (CS diagram) in Figure~\ref{fig:CS-diagram} using the combination of the local convergence $\kappa(\bm x)$ and shear $\gamma(\bm x)$, and the eigenvalues introduced in Equation~(\ref{eq:A_eigenvalues}) computed from them. The eigenvalue $\lambda_\parallel$, which defines the scaling factor in the direction of the phase, is constant in the diagram along straight lines with slope $-1$. It starts at $1$ at the origin of the plot, and increases to $\infty$ at the solid red line, which corresponds to critical curves (tangential in the case of axially symmetric lenses). Above it $\lambda_\parallel$ changes discontinuously to $-\infty$, with the negative sign indicating that the image is flipped in the direction of the phase. The value of $\lambda_\parallel$ increases to $-1$ along the dashed red line, above which the flipped image is contracted rather than expanded in the direction of the phase. Further beyond the top right corner of the diagram, $\lambda_\parallel$ increases asymptotically to $0$. Overall, below the dashed red line the image is expanded in the direction of the phase, while above it the image is contracted in the direction of the phase.

\renewcommand{\thefigure}{15}
\bfi[b]
{\centering
\vspace{0cm}
\hspace{0cm}
\includegraphics[width=12 cm]{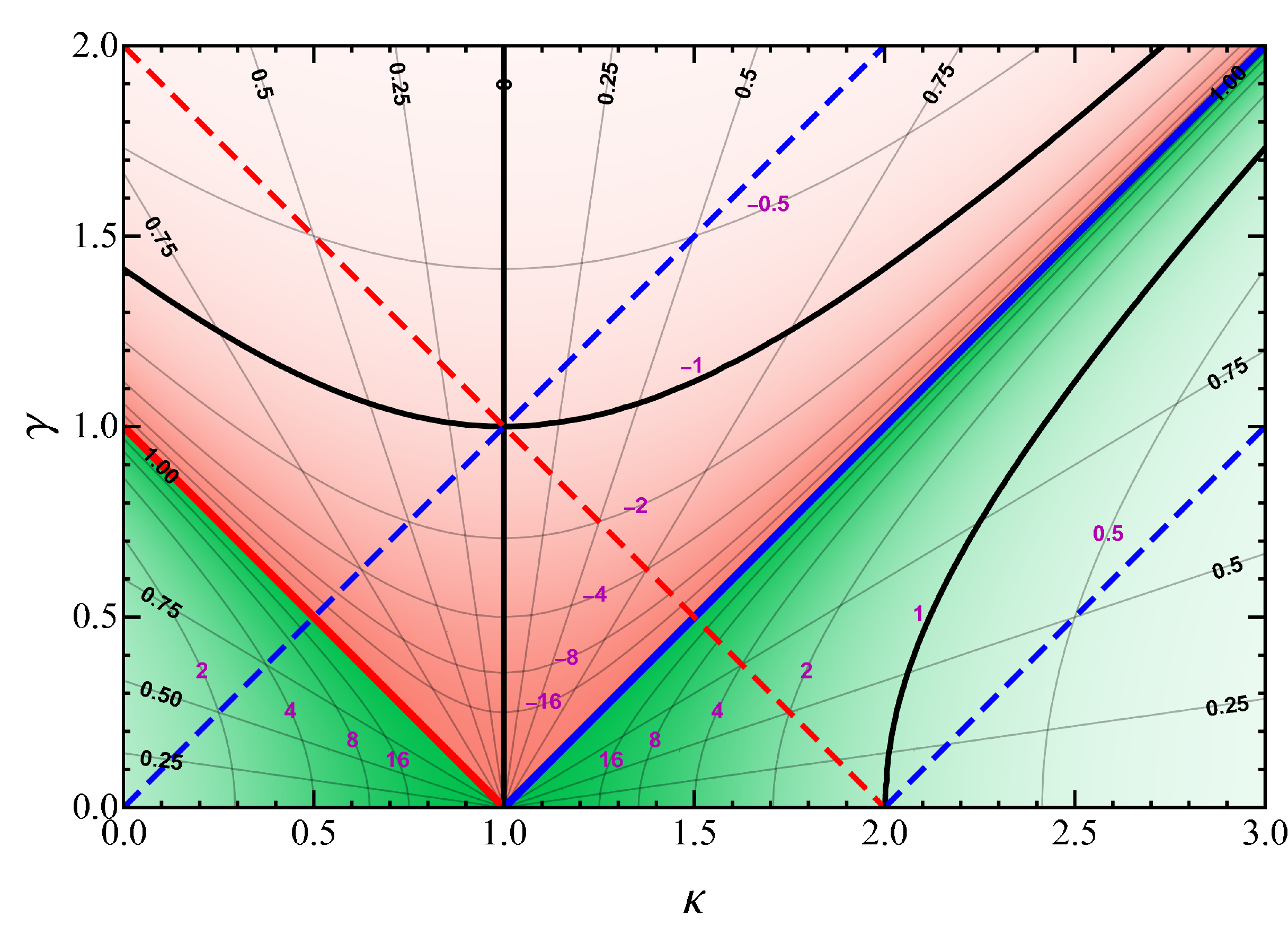}
\caption{Convergence--shear diagram (CS diagram) illustrating the geometry of lensed images as determined by the $(\kappa,\gamma)$ combination at their position. The color map and the purple-labeled hyperbolic contours show the values of $\mathrm{det}\,\mathbb{A}$; its absolute value yields the magnification and its sign yields the parity of the image (positive in green, negative in pink areas). Unit magnification occurs at the origin of the diagram and along the bold black hyperbolae; outside the region delimited by them images are demagnified. The black-labeled straight-line contours correspond to constant flattening of the image, given by Equation~(\ref{eq:flattening}). Undistorted images occur along the $\gamma=0$ axis and the bold black $\kappa=1$ line. The solid red and solid blue lines correspond to critical-curve points; the combination $(\kappa,\gamma)=(1,0)$ specifically to umbilic points. Unit absolute values of the eigenvalues of $\mathbb{A}$ occur along the dashed diagonals: $\lambda_\parallel=-1$ (red); $\lambda_\perp=1$ (blue from origin); $\lambda_\perp=-1$ (blue from $\kappa=2$). For $\kappa<1$ images are elongated in the direction of the phase; for $\kappa>1$ perpendicular to the phase. For more details, see Appendix~\ref{sec:Appendix-images}.\label{fig:CS-diagram}}}
\efi

The scaling factor in the direction perpendicular to the phase is defined by the eigenvalue $\lambda_\perp$, which is constant in the diagram along straight lines with slope $1$. Along the dashed blue line passing through the origin we find $\lambda_\perp=1$. Above it the image is contracted in the direction perpendicular to the phase, and $\lambda_\perp$ decreases asymptotically to $0$ above the top left of the diagram. Below the dashed blue line passing through the origin $\lambda_\perp$ increases to $\infty$ at the solid blue line, which corresponds to critical curves (radial in the case of axially symmetric lenses). Below it $\lambda_\perp$ changes discontinuously to $-\infty$, with the negative sign indicating that the image is flipped in the direction perpendicular to the phase. The value of $\lambda_\perp$ increases to $-1$ along the right dashed blue line, below which the flipped image is contracted in the direction perpendicular to the phase. Further beyond the bottom right corner of the diagram, $\lambda_\perp$ increases asymptotically to $0$. Overall, in the band between the dashed blue lines the image is expanded in the direction perpendicular to the phase, while outside it the image is contracted in the direction perpendicular to the phase.

The described scalings in the two perpendicular directions can be combined to yield information about the orientation, shape, and size of the image. Comparing their absolute values for a non-zero shear, $|\lambda_\parallel|$ is larger for $\kappa<1$ and $|\lambda_\perp|$ is larger for $\kappa>1$. Hence, the vertical solid black line at $\kappa=1$ divides images by the orientation of their distortion, i.e., the orientation of the major axis of an elliptical image of a small circular source. To the left of the line, images are oriented in the direction $\varphi$ (parallel to the phase), while to the right of the line, images are oriented in the direction $\varphi+\pi/2$ (perpendicular to the phase).

The distortion of the shape can be quantified by the flattening, computed from Equation~(\ref{eq:flattening}). In the diagram in Figure~\ref{fig:CS-diagram}, the flattening $f$ is constant along the straight lines radiating from the point $(\kappa,\gamma)=(1,0)$, labeled by their $f$ value at the outer edge of the plot. The horizontal axis corresponds to $f=0$, i.e., there is no distortion for zero shear. Clockwise from the direction to the origin of the plot, $f$ values along the lines increase in steps of $0.25$ to $1$ at the solid red line, corresponding to maximum flattening in the direction of the phase at the tangential critical curve. For the following lines, $f$ decreases in steps of $0.25$ to $0$ at the vertical $\kappa=1$ line, along which there is no distortion either. Continuing clockwise, $f$ increases to $1$ at the solid blue line, corresponding to maximum flattening in the direction perpendicular to the phase at the radial critical curve. The flattening along the following lines decreases back to $0$ along the horizontal axis.

The change in size of the image is given by the absolute value of the product of the two scale factors. Since they are eigenvalues of $\mathbb{A}$, their product is equal to its determinant,
\beq
\mathrm{det}\,\mathbb{A}(\boldsymbol x)=\lambda_\parallel(\boldsymbol x)\,\lambda_\perp(\boldsymbol x)=\left\{\,\left[1-\kappa(\boldsymbol x)\right]^2-\gamma^2(\boldsymbol x)\,\right\}^{-1}\,,
\label{eq:A_determinant}
\eeq
which is the inverse of the Jacobian $\mathrm{det}\,J(\boldsymbol x)$ from Equation~(\ref{eq:Jacobian}). For an image at $\boldsymbol x$, the sign of $\mathrm{det}\,\mathbb{A}(\boldsymbol x)$ yields the parity and its absolute value $|\mathrm{det}\,\mathbb{A}(\boldsymbol x)|$ yields the (point-source) magnification, the ratio of solid angles subtended by the image and by the source. The values of $\mathrm{det}\,\mathbb{A}$ are indicated by the purple-labeled hyperbolic contours and the color map in Figure~\ref{fig:CS-diagram}. Shades of pink above the critical-curve lines indicate negative $\mathrm{det}\,\mathbb{A}$, i.e., all images here are mirror images with negative parity. Positive-parity images lie in the green regions below the critical-curve lines.

The origin of the diagram with zero convergence and shear has unit magnification and positive parity, $\mathrm{det}\,\mathbb{A}=1$. Proceeding from the origin, contours are plotted for magnifications increasing in powers of two, corresponding to $\mathrm{det}\,\mathbb{A}\in\left\{2, 4, 8, 16\right\}$. Along the solid red and solid blue lines the magnification is infinite; above them the sign of $\mathrm{det}\,\mathbb{A}$ flips to negative. Going upward from the critical-curve lines, the magnifications along the plotted hyperbolas decrease in powers of two, corresponding to $\mathrm{det}\,\mathbb{A}\in\left\{-16, -8, -4, -2, -1, -0.5\right\}$. The bold black hyperbola passing through the point $(\kappa,\gamma)=(1,1)$ thus corresponds to unit magnification. All images above it are demagnified. Below the solid blue radial-critical-curve line the sign of $\mathrm{det}\,\mathbb{A}$ flips back to positive. Proceeding from it to the right, contours are plotted for $\mathrm{det}\,\mathbb{A}\in\left\{16, 8, 4, 2, 1, 0.5\right\}$. The bold black hyperbola passing through the point $(\kappa,\gamma)=(2,0)$ thus also corresponds to unit magnification. All images to its right are demagnified.

Note that the values of $\lambda_\parallel$ along diagonal lines with slope $-1$ are equal to the value of $\mathrm{det}\,\mathbb{A}$ at their intersection with the dashed blue line starting from the origin. The values of $\lambda_\perp$ along diagonal lines with slope $1$ are equal to $-\mathrm{det}\,\mathbb{A}$ at their intersection with the dashed red line.

The two perpendicular lines in the diagram corresponding to undistorted images differ by the sign of $\mathrm{det}\,\mathbb{A}$. Images along the horizontal axis (with $\gamma=0$) have positive parity; they are magnified for $\kappa<2$ and demagnified for $\kappa>2$. Images along the vertical bold line (with $\kappa=1$) are mirror images with negative parity; they are magnified for $\gamma<1$ and demagnified for $\gamma>1$. Clearly, the point $(\kappa,\gamma)=(1,0)$ lying at the intersection of these lines has special significance. At this point, the entire Jacobian matrix given by Equation~(\ref{eq:Jacobi_matrix}) is equal to zero, and its inverse $\mathbb{A}$ is thus undefined. If such points exist in the image plane of a gravitational lens, they define the position of critical-curve umbilic points (\citealt{schneider_etal92}; \citetalias{karamazov_etal21}). The properties of images in their vicinity depend on higher-order derivatives of the lens equation.

The structure of the CS diagram shows that in the general non-critical case there are four different $(\kappa,\gamma)$ combinations that lead to the same combination of $(f,|\mathrm{det}\,\mathbb{A}|)$, i.e., an image of the same shape and size. Two of these have positive and two have negative parity, as illustrated in Figure~\ref{fig:small-source-images}. In the case of zero flattening, there are three different $(\kappa,\gamma)$ combinations with only one negative-parity image.

For any specific gravitational lens, the range of $(\kappa(\boldsymbol x),\gamma(\boldsymbol x))$ combinations occurring in its image plane defines a region in the diagram which demonstrates the properties of all possible images formed by the lens. For a lens with an axially symmetric mass distribution the region is one-dimensional, described by the curve $\left(\kappa(x),\gamma(x)\right)$ with the radial position $x$ varying from $0$ to $\infty$. For a spherical NFW halo this case is described in Section~\ref{sec:NFW-images} and illustrated in Figure~\ref{fig:CS-NFW}; for a NFW halo with a centrally positioned point mass see the bottom row of Figure~\ref{fig:CS-diagrams}. For more asymmetric lenses with a (non-constant) continuous mass distribution, the region in the diagram is two-dimensional, as shown for the spherical NFW halo with an off-center point mass in Figure~\ref{fig:CS-NFWP} and \ref{fig:CS-diagrams} (except the bottom row).

For lenses consisting of point masses without continuous matter (e.g., stars and stellar systems in Galactic microlensing), the corresponding one-dimensional region is the $\kappa=0$ vertical axis, with $\gamma\to\infty$ at the positions of the masses and $\gamma\to 0$ far from them. For quasar microlensing, in which point masses are combined with a constant background convergence $\kappa_0$ and shear $\gamma_0$, the one-dimensional region lies along the $\kappa=\kappa_0$ vertical line. Finally, the regime of weak lensing with $\kappa\ll1$ and $\gamma\ll1$ is confined to the vicinity of the origin of the diagram in Figure~\ref{fig:CS-diagram}.

\end{document}